\documentclass[11pt]{article}
\usepackage{eurosym}
\usepackage{amsmath}
\usepackage{amsfonts}
\usepackage{amssymb}
\usepackage{graphicx}
\usepackage{epstopdf}
\usepackage{cite}
\usepackage[usenames]{color}
\usepackage{multirow}

\providecommand{\U}[1]{\protect\rule{.1in}{.1in}}
\newcommand{\ba}{\begin{array}}
\newcommand{\ea}{\end{array}}
\newcommand{\lab}{\label}
\newcommand{\Dsl}[1] { \setbox0=\hbox{$#1$}     
\dimen0=\wd0   \setbox1=\hbox{/} \dimen1=\wd1  \ifdim\dimen0>\dimen1        
 \rlap{\hbox to \dimen0{\hfil/\hfil}}  #1 \else \rlap{\hbox to \dimen1{\hfil$#1$\hfil}}  /  \fi  }
\newcommand{\bea}{\begin{eqnarray}}
\newcommand{\eea}{\end{eqnarray}}

\newcommand{\ns}{\Dsl{n}}

\newcommand {\nbs}{\Dsl{\bar n}}
\newcommand{\nbn}{\frac{\nbs\ns}{4}}
\newcommand{\nnb}{\frac{\ns\nbs}{4}}

\begin{document}
\title{ {\Large A description of  charmonia decays $J/\psi\rightarrow B\bar{B}$
within the QCD factorisation framework }}
\author{ Nikolay Kivel \\
\textit{\ Physik-Department, Technische Universit\"at M\"unchen,}\\
\textit{James-Franck-Str. 1, 85748 Garching, Germany } }
\maketitle

\begin{abstract}
Decays  $J/\psi\to B\bar B$  into octet baryon-antibaryons pairs are studied within the QCD factorisation framework.
The next-to-leading power correction  to the leading-order  decay amplitude is  calculated  for the first time.  
The phenomenological analysis also includes the interference with the electromagnetic decay  $J/\psi\to \gamma^*\to B\bar B$. 
The branching fractions and the angular coefficients $\alpha_B$  are obtained with an accuracy of about $20\%$.  The obtained results allow us 
to conclude that the calculated amplitudes provide a dominant contribution and therefore QCD factorisation provides a suitable basis for understanding of these decays.

 \end{abstract}

\noindent

\vspace*{1cm}

\newpage

\section{Introduction}
\label{int}

A systematic description of exclusive charmonia decays  can be performed  using QCD factorisation framework, see {\it e.g.} Refs.\cite{Brodsky:1981kj, Chernyak:1983ej} and references therein.  Within this approach one constructs  the expansion of an amplitude in inverse powers of $1/m_c$ combining non-relativistic QCD and hard-collinear effective field theory.  As a rule,  most of these calculations are performed with  leading power accuracy. In many cases this gives a reliable description, but in many cases this approximation is insufficient even for a good qualitative description of the data. Potentially such  discrepancies can be associated  with various large corrections and with a complicated underlying QCD dynamics.  Therefore,  a description  of exclusive charmonia decays still includes many open questions, see, for example,  reviews in  Refs.\cite{Brambilla:2004wf, Brambilla:2010cs}.  

Decays of $S$-wave charmonia  into proton-antiproton  pairs have been discussed  already long time ago within  the QCD factorisation framework in Refs\cite{Brodsky:1981kj, Chernyak:1987nv}.    It was found  that  with the appropriate models of baryon matrix elements one can obtain a reasonable description of the branching ratio for $J/\psi\to p \bar p$.   This observation was further used   in order to constrain the  nucleon DAs  in Refs. \cite{Stefanis:1992nw, Bolz:1997as }

In recent years, the BESII and BESIII collaborations have provided a lot of new and accurate  data in Refs.$\ $\cite{BES:2008hwe, BESIII:2012ion,BESIII:2016ssr,BESIII:2016nix,BESIII:2017kqw,BESIII:2020fqg}. 
The description of these data is a non-trivial task, since requires calculations beyond the leading-power approximation, which are often are very problematic.  In addition, the phenomenological consideration suggested  in Refs.\cite{Alekseev:2018qjg, BaldiniFerroli:2019abd}  indicates potential problems for the description based on QCD factorisation.
  
 The new data provide not only the values of the cross sections, but also interesting information about the angular behaviour of the $e^+e^-\to J/\psi\to B\bar B$ cross section.   This  behaviour is sensitive to the  amplitude, which is  power suppressed  within the systematic QCD description. The contribution of this amplitude to the decay width  is suppressed by additional factor  $1/m_c^2$   and therefore is expected to be a small correction. On the other hand, new data show that the effect of this amplitude on the description of the angular behaviour is very significant. 
 
  The various  attempts to describe  the angular coefficient 
 $\alpha_B$ have been considered  in Refs.$\ $\cite{Claudson:1981fj,Carimalo:1985mw,Murgia:1994dh} using different  phenomenological models and crude approximations.   For the first time  the power suppressed amplitude was calculated in the QCD factorisation framework in Refs.\cite{Kivel:2019wjh, Kivel:2021uzl}.  The qualitative phenomenological analysis carried out in these works indicates that  the corresponding  description works sufficiently well.  However, in order to better understand the role of the  power corrections  some theoretical uncertainties need to be clarified. 
 
 In this paper we continue the investigation of the decays into  baryon-antibaryon pairs  within the QCD factorisation framework.  Two additional effects are included into phenomenological consideration, which was ignored  in Ref.\cite{Kivel:2021uzl}.  First, we calculate the next-to-leading power correction $\sim \Lambda^2/m^2_c$ to the dominant leading-power amplitude, where $\Lambda$ is a typical hadronic scale.  This correction is interesting because  the mass of  charm quark  is not very large, and such contribution could  provide rather large numerical effect. In addition, this correction is described by the same non-perturbative parameters  as the subleading amplitude, which gives them better phenomenological constraints. 
 
  This power correction is closely associated with the higher Fock states of the baryon wave function  and  do not involve relativistic corrections with respect to the small heavy quark velocity in NRQCD.  The decay mechanism associated with  annihilation into the three hard gluons suggests that such  contributions can also be described within the collinear factorisation framework that  gives us  an opportunity of  a more detailed  understanding  of the systematic expansion in the inverse mass $1/m_c$. The higher twist baryon  light-cone distribution amplitudes (DAs), associated with  the non-perturbative  collinear matrix elements,  have already been  investigated  in  Refs. \cite{Braun:2000kw,Braun:1999te, Braun:2008ia,Krankl:2011gch, Anikin:2013aka,Wein:2015oqa,Anikin:2015qos,Bali:2019ecy}.  These results allows one to reduce  theoretical uncertainties associated with the non-perturbative QCD input.  
  
In this work we also discuss the effect that occurs due to the contribution of the  electromagnetic amplitude $J/\psi\to \gamma^*\to B\bar B$.  The given amplitude is sensitive to the baryon time-like electromagnetic form factors. For a given energy scale these form factors can not be computed within the QCD framework because of relatively  large effect from the soft-overlap rescattering (or Feynman)  mechanism.  Time-like baryonic FFs are also poorly studied experimentally.   Different cross sections  for the $e^+e^-\to B\bar B$  channels  have been measured by BESIII recent years, and the so-called effective FFs are obtained  from  data fitting, 
see Refs.$\,$\cite{BaBar:2007fsu,BESIII:2017hyw,BESIII:2019hdp,BESIII:2020ktn,BESIII:2020uqk,BESIII:2021dfy,BESIII:2021aer,BESIII:2021rkn}.  We use these results in order to estimate the numerical influence of the electromagnetic admixture  in the angular distribution $\alpha_B$.

Our  paper is organised as follows. In Sec.~\ref{BDA} we discuss the higher twist baryon DAs
and describe the models, which are used in our calculations. The calculation and  the analytical
expressions for the hard kernels and for the total amplitudes  are discussed in the
Sec.~\ref{calc}.  The phenomenological analysis is presented in
Sec.~\ref{phen}. In Sec.~\ref{conc}  we summarise our findings. In Appendix  we provide additional useful details and a more detailed discussion of some  technical issues.

\section{ Baryon light-cone distribution amplitudes}
\label{BDA}
\subsection{The basic set of the twist-5 DAs and their properties}
\label{T5DA}

In this section we describe the properties of  twist-5 DAs and define the
twist-5 auxiliary matrix elements, which are used in our calculation of the
power corrections.  We  consider contributions with only  three quark operators, which greatly simplifies the calculations. 
 The  total power correction is given by the sum of various contributions with the different combinations of DAs.
 The leading-power (lp) contribution to amplitude $A_{1}$ only includes
 the integral with the twist-3 DAs $\varphi_{3}$, schematically:
 $A_{1}^{(\text{lp})}\sim\varphi_{3}\ast H\ast\varphi_{3}$, where the
asterisk is used as a short notation for the convolution integral  with respect
to quark momentum fractions.  The next-to-leading power (nlp) correction is more
complicated, it depends on  twist-4 $\varphi_{4}$ and twist-5 $\varphi
_{5}$ DAs  and can be represented as 
\begin{equation}
A_{1}^{(\text{nlp})}\sim\varphi_{3}\ast H_{35}\ast\varphi_{5}+\varphi_{4}\ast
H_{44}\ast\varphi_{4}+\frac{m_{B}^{2}}{m_{Q}^{2}}\varphi_{3}\ast H_{33}%
\ast\varphi_{3},
\end{equation}
where the last term can be associated with the kinematical  contribution.
 The various higher twist DAs and their properties have been discussed in
Refs.\cite{Braun:2008ia, Anikin:2013aka,Wein:2015oqa,Anikin:2015qos}  and we
use these results in the present work.  

Let us  define the  light-cone  matrix elements, which define the light-cone DAs. The analysis of the twist-4 auxiliary DAs are given in Refs.\cite{Kivel:2019wjh,Kivel:2021uzl,Anikin:2013aka}.  Therefore, here we will focus only  on the discussion of the twist-5 matrix elements.

We use the same notations as in Refs.\cite{Kivel:2019wjh,Kivel:2021uzl}. The
light-like vectors $n,\bar{n}$ ( $n^{2}=\bar{n}^{2}=0$, $(n\bar{n})=2$ )  define the longitudinal and transverse
 components of a Lorentz vector $V^{\mu}$
\begin{equation}
V_{+}=(kn),\quad V_{-}=(V\bar{n}),\ \ \ V^{\mu}=V_{+}\frac{\bar{n}^{\mu}}%
{2}+V_{-}\frac{n^{\mu}}{2}+V_{\bot}^{\mu}.
\end{equation}

In the following we also assume that the baryon momentum $k$ is directed along
$z$-axis and can be expanded as
\begin{equation}
k=k_{-}\frac{{n}}{2}+k_{+}\frac{\bar n}{2},\ \ k_{+}\gg k_{-}=\frac{m_{B}^{2}%
}{k_{+}},~~k^{2}=m_{B}^{2}. \label{def:k}%
\end{equation}
For the baryon spinor $N(k,s)$ we define  the large and small components $N_{\bar{n}%
}$ and $N_{{n}}$, respectively:
\begin{equation}
N_{\bar{n}}=\nbn N(k,s),~\ ~N_{{n}}=\nnb N(k,s)=\frac{m_{B}}{k_{+}}\frac{\Dsl{n}}{2}N_{\bar{n}}.~
\end{equation}
The similar relations are also used for the antibaryon spinors.

The relevant three-quark light-cone operators are constructed from the QCD
quark fields $q=\{{u,d,s\}}$ and from the  light-cone the Wilson lines
\begin{equation}
W_{\bar{n}}[x_{-},z_{-}]=\text{P}\exp\left\{  ig\int_{(z_{-}-x_{-})/2}%
^{0}ds\ A_{+}(x_{-}n/2+sn)\right\}  .
\end{equation}
The basic operator can be written as
\begin{equation}
\mathcal{O}_{\alpha_{1}\alpha_{2}\alpha_{3}}(x_{-},y_{-},z_{-})=\varepsilon
^{ijk}\ q_{\alpha_{1}}^{i^{\prime}}(x_{-})W_{\bar{n}}[x_{-},v_{-}]_{i^{\prime
}i}\ q_{\alpha_{2}}^{j^{\prime}}(y_{-})W_{\bar{n}}[y_{-},v_{-}]_{j^{\prime}%
j}\ q_{\alpha_{3}}^{k^{\prime}}(z_{-})W_{\bar{n}}[z_{-},v_{-}]_{k^{\prime}k},
\label{calO123}%
\end{equation}
where we use the short notation $q(x_{-})\equiv q(x_{-}n/2)$.  
Below, for brevity,  we also simplify the notation of the Dirac indices  setting
$\{\alpha_{1},\alpha_{2},\alpha_{3}\}\equiv\{1,2,3\}$ and we also will not show
the colour structure. Following to Ref.~\cite{Wein:2015oqa} we assume the
following flavour content of the operators $\langle0|q_{\alpha_{1}}%
q_{\alpha_{2}}q_{\alpha_{3}}|B\rangle\equiv\langle0|q_{1}q_{2}q_{3}|B\rangle
$:
\begin{align}
&  \langle0|u_{1}u_{2}d_{3}|p\rangle,\,\langle0|u_{1}d_{2}s_{3}|\Lambda
\rangle,\label{qqq1}\\
&  \langle0|u_{1}d_{2}s_{3}|\Sigma^{0}\rangle,\,\langle0|u_{1}u_{2}%
s_{3}|\Sigma^{+}\rangle,\label{qqq2}\\
&  \langle0|s_{1}s_{2}u_{3}|\Xi^{0}\rangle,\,\langle0|s_{1}s_{2}d_{3}|\Xi
^{-}\rangle. \label{qqq3}%
\end{align}
For simplicity, this will not be shown explicitly.

Let us write the light-cone matrix element for the baryon state $B$ as
\begin{align}
\left\langle 0\right\vert \mathcal{O}_{123}(x_{-},y_{-},z_{-})\left\vert
B(k)\right\rangle &=\left\langle 0\right\vert \mathcal{O}_{123}\left\vert
B(k)\right\rangle _{\text{tw3}}+\left\langle 0\right\vert \mathcal{O}%
_{123}\left\vert B(k)\right\rangle _{\text{tw4}}
\nonumber\\
&+\left\langle 0\right\vert
\mathcal{O}_{123}\left\vert B(k)\right\rangle _{\text{tw5}}+\ldots,
\label{meO123}%
\end{align}
where the subscript \textquotedblleft tw3\textquotedblright\ indicate that we
only count in this term the contributions of twist-3 $\ $\ and so on. The
leading twist-3 contribution read%
\begin{align}
\left\langle 0\right\vert \mathcal{O}_{123}\left\vert B(k)\right\rangle
_{\text{tw3}}  &  =~\frac{1}{8}k_{+}\left[  \nbs C\right]  _{12}~\left[
\gamma_{5}N_{\bar{n}}\right]  _{3}\text{FT}\left[  V_{1}^{B}\right]
+~\frac{1}{8}k_{+}\left[  \nbs\gamma_{5}C\right]  _{12}~\left[  N_{\bar{n}%
}\right]  _{3}\text{FT}\left[  A_{1}^{B}\right]
\nonumber\\
&  +\frac{1}{8}k_{+}\left[  \nbs\gamma_{\bot}^{\alpha}C\right]
_{12}~\left[  \gamma_{\bot}^{\alpha}\gamma_{5}N_{\bar{n}}\right]
_{3}\text{FT}\left[  T_{1}^{B}\right]  .
 \label{Metw3}
\end{align}
The twist-4 contribution  is defined as
\begin{eqnarray}
\left\langle 0\right\vert \mathcal{O}_{123}\left\vert B(k)\right\rangle
_{\text{tw4}}=~\frac{m_{B}}{16}\left[  \nbs C\right]  _{12}~\left[
\gamma_{5}\ns  N_{\bar{n}}\right]  _{3}\text{FT}\left[  V_{2}^{B}\right]
+\frac{m_{B}}{8}\left[  \gamma_{\bot}C\right]  _{12}~\left[  \gamma_{\bot
}\gamma_{5}N_{\bar{n}}\right]  _{3}~\text{FT}\left[  V_{3}^{B}\right]
\nonumber \\
+~\frac{m_{B}}{16}\left[  \nbs\gamma_{5}C\right]  _{12}~\left[  \ns
N_{\bar{n}}\right]  _{3}\text{FT}\left[  A_{2}^{B}\right]  +\frac{m_{B}}%
{8}\left[  \gamma_{\bot}\gamma_{5}C\right]  _{12}~\left[  \gamma_{\bot}%
N_{\bar{n}}\right]  _{3}~\text{FT}\left[  A_{3}^{B}\right] %
\nonumber \\
+\frac{m_{B}}{4}\left[  C\right]  _{12}~\left[  \gamma_{5}N_{\bar{n}}\right]
_{3}\text{FT}\left[  S_{1}^{B}\right]  +\frac{m_{B}}{4}\left[  \gamma
_{5}C\right]  _{12}~\left[  N_{\bar{n}}\right]  _{3}\text{FT}\left[  P_{1}%
^{B}\right]
\nonumber \\
+\frac{m_{B}}{16}\left[  \nbs\gamma_{\bot}^{\alpha}C\right]  _{12}~\left[
\ns  \gamma_{\bot\alpha}\gamma_{5}N_{\bar{n}}\right]  _{3}%
\text{FT}\left[  T_{2}^{B}\right]  +\frac{m_{B}}{8}k_{+}\left[  i\sigma
^{\bar{n}n}C\right]  _{12}~\left[  \gamma_{5}N_{\bar{n}}\right]  _{3}%
\text{FT}\left[  T_{3}^{B}\right]
\nonumber \\
+\frac{m_{B}}{8}\left[  \sigma_{\bot\bot}^{\alpha\beta}C\right]  _{12}~\left[
\sigma_{\alpha\beta}\gamma_{5}N_{\bar{n}}\right]  _{3}%
\text{FT}\left[  T_{7}^{B}\right]  .
 \label{Metw4}
\end{eqnarray}

The twist-5 contribution read%
\begin{eqnarray}
\left\langle 0\right\vert \mathcal{O}_{123}\left\vert B(k)\right\rangle
_{\text{tw5}}=   ~\frac{1}{4}\frac{m_{B}^{2}}{4k_{+}}\left[  \gamma_{\bot
}^{\alpha}C\right]  _{12}~\left[  \ns  \gamma_{\bot\alpha}\gamma
_{5}N_{\bar{n}}\right]  _{3}~\text{FT}\left[  V_{4}^{B}\right]  
+\frac{1}%
{4}~\frac{m_{B}^{2}}{4k_{+}}\left[  \gamma_{\bot}^{\alpha}\gamma_{5}C\right]
_{12}~\left[  \gamma_{\bot\alpha}\ns  N_{\bar{n}}\right]  _{3}%
\text{FT}\left[  A_{4}^{B}\right]
\nonumber \\
  +~~\frac{1}{4}\frac{m_{B}^{2}}{2k_{+}}\left[  \ns  C\right]
_{12}~\left[  \gamma_{5}N_{\bar{n}}\right]  _{3}\text{FT}\left[  V_{5}%
^{B}\right]  +~\frac{1}{4}\frac{m_{B}^{2}}{2k_{+}}\left[  \ns  \gamma
_{5}C\right]  _{12}~\left[  N_{\bar{n}}\right]  _{3}\text{FT}\left[  A_{5}%
^{B}\right]
\nonumber \\
-\frac{1}{4}\frac{m_{B}^{2}}{2k_{+}}~\left[  C\right]  _{12}\left[  \ns
\gamma_{5}N_{\bar{n}}\right]  _{3}~\text{FT}\left[  S_{2}^{B}\right]
+\frac{1}{4}\frac{m_{B}^{2}}{2k_{+}}~\left[  \gamma_{5}C\right]  _{12}\left[
\ns  N_{\bar{n}}\right]  _{3}~\text{FT}\left[  P_{2}^{B}\right]
\nonumber \\
+\frac{1}{4}\frac{m_{B}^{2}}{4k_{+}}\left[  i\sigma_{\bar{n}n}C\right]
_{12}\left[  \ns  \gamma_{5}N_{\bar{n}}\right]  _{3}~\text{FT}\left[
T_{4}^{B}\right]  +\frac{1}{4}\frac{m_{B}^{2}}{2k_{+}}\left[  i\sigma_{\alpha
n}C\right]  _{12}\left[  \gamma_{\bot}^{\alpha}\gamma_{5}N_{\bar{n}}\right]
_{3}~\text{FT}\left[  T_{5}^{B}\right]
\nonumber \\
-\frac{1}{4}\frac{m_{B}^{2}}{4k_{+}}\left[  \sigma_{\bot\bot}^{\alpha\beta
}C\right]  _{12}\left[  \ns  \sigma_{\alpha\beta}\gamma_{5}N_{\bar{n}%
}\right]  _{3}~\text{FT}\left[  T_{8}^{B}\right] , 
\label{<O123>}%
\end{eqnarray}
where we always assume that $\sigma_{\bar{n}n}=\sigma_{\alpha\beta}\bar
{n}^{\alpha}n^{\beta}$ .  The symbol \textquotedblleft FT" denotes the
Fourier transformation with respect to the quark momentum fractions
\begin{equation}
\text{FT}[f]=\int Du_{i}~e^{-i(k_{1}x)-i(k_{2}y)-i(k_{3}z)}f(u_{1},u_{2}%
,u_{3}),~\ \label{FT}%
\end{equation}
with the measure $Du_{i}=du_{1}du_{2}du_{3}\delta(1-u_{1}-u_{2}-u_{3})$, the
quark momenta in (\ref{FT}) are defined as
\begin{equation}
k_{i}=u_{i}k_{+}\frac{\bar{n}}{2}.
\end{equation}
The DAs defined in Eqs.(\ref{Metw3})-(\ref{<O123>})  are symmetric or antisymmetric with respect to
interchange $u_{1}\leftrightarrow u_{2}$%
\begin{equation}
F_{i}^{B}(x_{2},x_{1},x_{3})=-(-1)_{B}F_{i}^{B}(x_{1},x_{2},x_{3}),\ \text{for
}\ \ F=\{S,P,A\}, \label{AsDA}%
\end{equation}%
\begin{equation}
F_{i}^{B}(x_{2},x_{1},x_{3})=+(-1)_{B}F_{i}^{B}(x_{1},x_{2},x_{3}),\ \text{for
}\ \ F=\{V,T\}, \label{SymDA}%
\end{equation}
with%
\begin{equation}
(-1)_{B}=\left\{
\begin{array}
[c]{c}%
+1\ \ \ B\neq\Lambda\\
-1\ \ \ B=\Lambda
\end{array}
\right.  .\text{ }%
\end{equation}

The  DAs   in Eqs(\ref{Metw3})-(\ref{<O123>})  can be rewritten  in
terms of the \textit{basis} light-cone DAs as \cite{Wein:2015oqa,
Anikin:2015qos} ( below we  use  short notation $(x_i)\equiv(x_{123})\equiv(x_{1},x_{2},x_{3})$)%
\begin{equation}
f_{B}\varphi_{3\pm}^{B}(x_{123})=\frac{c_{B}^{\pm}}{2}\left\{  \left(
V_{1}-A_{1}\right)  ^{B}(x_{123})\pm\left(  V_{1}-A_{1}\right)  ^{B}%
(x_{321})\right\}  , \label{phi3Bpm}%
\end{equation}%
\begin{equation}
f_{B}\Pi_{3}^{B}(x_{123})=c_{B}^{-}(-1)_{B}T_{1}^{B}(x_{132}),
\label{pi3B}
\end{equation}%
\begin{equation}
\Phi_{4\pm}^{B}(x_{123})=c_{B}^{\pm}\left\{  \left(  V_{2}-A_{2}\right)
^{B}(x_{123})\pm(-1)_{B}\left(  V_{3}-A_{3}\right)  ^{B}(x_{231})\right\}  ,
\label{phi4pm}%
\end{equation}%
\begin{equation}
\Xi_{4\pm}^{B}(x_{123})=(-1)_{B}3c_{B}^{\pm}\left[  ~\left(  T_{3}-T_{7}%
+S_{1}+P_{1}\right)  ^{B}(x_{123})\pm\left(  T_{3}-T_{7}+S_{1}+P_{1}\right)
^{B}(x_{132})\right]  , \label{Ksi4pm}%
\end{equation}%
\begin{equation}
\Pi_{4}^{B}(x_{123})=c_{B}^{-}\left(  T_{3}+T_{7}+S_{1}-P_{1}\right)
^{B}(x_{312}), \label{Pi4}%
\end{equation}%
\begin{equation}
\Upsilon_{4}^{B}(x_{123})=6c_{B}^{-}T_{2}^{B}(x_{321}), \label{Ups4}%
\end{equation}%
\begin{equation}
\Phi_{5\pm}^{B}(x_{123})=c_{B}^{\pm}\left\{  \left(  V_{2}-A_{2}\right)
^{B}(x_{123})\pm(-1)_{B}\left(  V_{3}-A_{3}\right)  ^{B}(x_{231})\right\}  ,
\label{PhiB5pm}%
\end{equation}%
\begin{equation}
\Xi_{5\pm}^{B}(x_{123})=(-1)_{B}3c_{B}^{\pm}\left[  ~\left(  T_{4}-T_{8}%
+S_{2}+P_{2}\right)  ^{B}(x_{123})\pm\left(  T_{4}-T_{8}+S_{2}+P_{2}\right)
^{B}(x_{132})\right]  , \label{KsiB5pm}%
\end{equation}%
\begin{equation}
\Pi_{5}^{B}(x_{123})=c_{B}^{-}\left(  T_{4}+T_{8}+S_{2}-P_{2}\right)
^{B}(x_{312}), \label{PiB5}%
\end{equation}%
\begin{equation}
\Upsilon_{5}^{B}(x_{123})=6c_{B}^{-}T_{5}^{B}(x_{321}), \label{UpsB5}%
\end{equation}
where 
\begin{equation}
c_{B}^{+}=\left\{
\begin{array}
[c]{c}%
1,~B\neq\Lambda\\
\sqrt{\frac{2}{3}},B=\Lambda
\end{array}
\right.  ,~~\ c_{B}^{-}=\left\{
\begin{array}
[c]{c}%
1,~B\neq\Lambda\\
-\sqrt{6},B=\Lambda
\end{array}
\right.  , \label{cBpm}%
\end{equation}
The coefficients $c_{B}^{\pm}$ are chosen in such a way that these DAs satisfy
the simple relations in the $SU(3)$-flavour symmetry limit, see \textit{e.g. }
Ref.\cite{Wein:2015oqa}. 

For the nucleon state, the basic set of DAs can be further simplified by the
isospin relationships, namely, it is sufficient to determine%
\begin{equation}
f_{N~}\varphi_{3}(x_{123})=\phi_{3+}^{N}(x_{123})+\phi_{3-}^{N}(x_{123}%
)=\left(  V_{1}-A_{1}\right)  ^{N}(x_{123}), \label{phi3}%
\end{equation}%
\begin{equation}
\frac{1}{2}\Phi_{4}(x_{123})=\frac{1}{2}\left(  \Phi_{4+}^{N}+\Phi_{4-}%
^{N}\right)  (x_{123})=\left(  V_{2}-A_{2}\right)  ^{N}(x_{123}), \label{Phi4}%
\end{equation}%
\begin{equation}
\frac{1}{2}\Psi_{4}(x_{123})=\frac{1}{2}\left(  \Phi_{4+}^{N}-\Phi_{4-}%
^{N}\right)  (x_{312})=\left(  V_{3}-A_{3}\right)  ^{N}(x_{231}), \label{Psi4}%
\end{equation}%
\begin{equation}
\frac{\lambda_{2}}{6}\Xi_{4}(x_{123})=\left(  \Xi_{4+}^{N}+\Xi_{4-}%
^{N}\right)  (x_{123})=\left(  T_{3}-T_{7}+S_{1}+P_{1}\right)  ^{N}(x_{123}).
\label{Ksi4}%
\end{equation}%
\begin{equation}
\frac{1}{2}\Phi_{5}(x_{123})=\frac{1}{2}\left(  \Phi_{5+}^{N}+\Phi_{5-}%
^{N}\right)  (x_{123})=\left(  V_{5}-A_{5}\right)  ^{N}(x_{123}), \label{Phi5}%
\end{equation}%
\begin{equation}
\frac{1}{2}\Psi_{5}(x_{123})=\frac{1}{2}\left(  \Phi_{5+}^{N}-\Phi_{5-}%
^{N}\right)  (x_{312})=\left(  V_{4}-A_{4}\right)  ^{N}(x_{231}), \label{Psi5}%
\end{equation}%
\begin{equation}
\frac{1}{6}\Xi_{5}(x_{123})=\frac{1}{6}\left(  \Xi_{5+}^{N}+\Xi_{5-}%
^{N}\right)  (x_{123})=\left(  T_{4}-T_{8}+S_{2}+P_{2}\right)  ^{N}(x_{123}).
\label{Ksi5}%
\end{equation}

The twist-4 and twist-5 DAs can be decomposed into components corresponding
to the contributions of the light-cone operators of \textit{geometrical}
twist-3, twist-4 and twist-5. Let us write such decompositions in the following
form%
\begin{equation}
\Phi_{4\pm}^{B}(x_{123})=f_{B}\Phi_{4\pm}^{B(3)}(x_{123})+\lambda_{1}^{B}%
\bar{\Phi}_{4\pm}^{B}(x_{123}), \label{WWPhi4pm}%
\end{equation}%
\begin{equation}
\Pi_{4}^{B}(x_{i})=f_{\bot}^{B}\Pi_{4}^{B(3)}(x_{123})+\lambda^B_1\bar{\Pi}_{4}^{B}(x_{i}),\ B\neq\Lambda, 
\label{WWPi4B}%
\end{equation}%
\begin{equation}
\Pi_{4}^{\Lambda}(x_{i})=f_{\Lambda}\Pi_{4}^{\Lambda(3)}(x_{123}%
)+\lambda^\Lambda_{\bot}\bar{\Pi}_{4}^{\Lambda}(x_{i}),\ \label{WWPi4L}%
\end{equation}%
\begin{equation}
\Xi_{4\pm}^{B}(x_{i})=\lambda_{2}^{B}\bar{\Xi}_{4}^{B}(x_{i}), \label{WWKsi4}%
\end{equation}%
\begin{equation}
\Upsilon_{4}^{B}(x_{i})=\lambda_{2}^{B}\bar{\Upsilon}_{4}^{B}(x_{i}).
\label{WWUps4}%
\end{equation}%
\begin{equation}
\Phi_{5\pm}^{B}(x_{i})=f_{B}~\Phi_{5\pm}^{B(3)}(x_{i})+\lambda_{1}^{B}%
\Phi_{5\pm}^{B(4)}(x_{i})+\bar{\Phi}_{5\pm}^{B}(x_{i}), \label{WWPhi5}%
\end{equation}%
\begin{equation}
\Xi_{5\pm}^{B}(x_{i})=\lambda_{2}^{B}\Xi_{5}^{B(4)}(x_{i})+\bar{\Xi}_{5\pm
}^{B}(x_{i}), \label{WWKsi5}%
\end{equation}%
\begin{equation}
\Pi_{5}^{B}(x_{i})=f_{\bot}^{B}\Pi_{5}^{B(3)}(x_{i})+f_{\bot}^{B}\Pi
_{5}^{B(4)}(x_{i})+\bar{\Pi}_{5}^{B}(x_{i}), \label{WWPi5B}%
\end{equation}%
\begin{equation}
\Pi_{5}^{\Lambda}(x_{i})=\lambda_{\bot}^{\Lambda}\Pi_{5}^{\Lambda(3)}%
(x_{i})+\lambda_{\bot}^{\Lambda}\Pi_{5}^{\Lambda(4)}(x_{i})+\bar{\Pi}%
_{5}^{\Lambda}(x_{i}), \label{WWPi5L}%
\end{equation}%
\begin{equation}
\Upsilon_{5}^{B}(x_{i})=\lambda_{2}^{B}\Upsilon_{5}^{B(4)}(x_{i}%
)+\bar{\Upsilon}_{5}^{B}(x_{i}). \label{WWUps5}%
\end{equation}
Here the superscript $(i)$ indicates that the corresponding function is completely
defined in terms of the moments of the light-cone operators of geometrical
twist-$i$. The functions with the bar denote the genuine twist-4/5
contributions. The moments of genuine twist-5 DAs are not known.  In the present work we assume
that the corresponding matrix elements are sufficiently small and therefore  be neglected that yields
\begin{equation}
\bar{\Phi}_{5\pm}^{B}(x_{i})=\bar{\Xi}_{5\pm}^{B}(x_{i})=\bar{\Pi}_{5}%
^{B}(x_{i})=\bar{\Pi}_{5}^{\Lambda}(x_{i})=\bar{\Upsilon}_{5}^{B}(x_{i})=0.
\label{barDAs}%
\end{equation}
The normalisation couplings  $f_{B},\ f_{\bot}^{B},\ \lambda_{1,2}^{B}$ and
$\lambda_{\bot}^{\Lambda}$ have dimension of  GeV$^{2}$ and their values  will be
discussed later.  

The conformal expansion and evolution of the moments  for different three-quark  DAs  was discussed in Refs. \cite{ Braun:1999te, Braun:2008ia,Krankl:2011gch}. 
 The explicit analytical expressions for the different terms  in {\it rhs} in
Eqs.(\ref{WWPhi5})-(\ref{WWUps5}) can be found  in Refs.\cite{Braun:2008ia,Wein:2015oqa,Anikin:2015qos}. 
For convenience,  in Appendix \ref{DAmodels} we provide the explicit expressions for the models of basic DAs, which are used in present work.

\subsection{Auxiliary DAs and their properties}
\label{sec:aux}

Let us also consider the {\it auxiliary } set of the
 matrix elements, which is more  convenient  for the actual calculation of the hard kernels. More
details about this calculation will be given in the next section. Here we  only
discuss the formal definition of the auxiliary matrix elements  and their relations with
the basic set of DAs.

The calculation of the decay amplitudes implies that the QCD fields can be
splitted into  sum of hard and collinear fields. The letter describe the
non-perturbative overlap with the baryon-antibaryon state. 
Therefore performing a matching onto collinear
operators one  implies  the effective field theory constructed from the hard
($p_{h}\sim m_{Q}$, $x\sim1/m_{Q}$) and collinear fields. Such theory provides
a definite power counting, which is used for the ranging of the operators.
 The power counting of the collinear fields is associated with  their momentum, if we assume the
collinear sector associated with the momentum $k$ in Eq.(\ref{def:k}) then
($\lambda^{2}\sim\Lambda/m_{Q}$)
\begin{eqnarray}
\nbn q(x_{-})   \sim\mathcal{O}(\lambda^{2}),\ \ 
\nnb q(x_{-})\sim\mathcal{O}(\lambda^{4}),\ \ \\
\ A_{+}(x_{-})   \sim\mathcal{O}(1),\ A_{\bot}(x_{-})\sim\mathcal{O}%
(\lambda^{2}),\ A_{-}(x_{-})\sim\mathcal{O}(\lambda^{4}).
\end{eqnarray}
Performing the multipole expansions of  collinear fields one also obtains
derivatives of the collinear fields, which counts as %
\begin{align}
(n\partial)q(x_{-})  &  \sim\mathcal{O}(1)q(x_{-}),\ \ \partial_{\bot}%
q(x_{-})\sim\mathcal{O}(\lambda)q(x_{-}),\ \ \ (\bar{n}\partial)q(x_{-}%
)\sim\mathcal{O}(\lambda^{4})q(x_{-}),\ \\
(n\partial)A_{+}(x_{-})  &  \sim\mathcal{O}(1)A_{+}(x_{-}), \partial_{\bot}A_{+}(x_{-})\sim\mathcal{O}(\lambda^{2})A_{+}(x_{-}),
 (\bar{n}\partial)A_{+}(x_{-})\sim\mathcal{O}(\lambda^{4})A_{+}(x_{-}).
\end{align}
Remind, that we do not consider the contributions with the quark-gluon
operators, therefore terms like  $A_{\bot},$ $A_{-}$ , $\partial_{\bot}A_{+} $
and $(\bar{n}\partial)A_{+}$ will be excluded from our consideration.  

The auxiliary DAs can be defined in terms of the matrix elements of 3-quark
twist-5 operators with the transverse derivatives. These operators are defined
using the gauge invariant collinear fields, which have a definite scaling
behaviour within the effective field theory framework
\begin{equation}
\chi_{1}(x_{-})=W^{\dag}(x_{-})\nbn q_{1}%
(x_{-}),\ \ \ \ \ W(x_{-})=\text{P}\exp\left\{  ig\int_{-\infty}%
^{0}ds\ A_{+}(x_{-}n/2+sn)\right\}  . \label{def:chi}%
\end{equation}
Obviously%
\begin{equation}
\chi(x_{-})\sim\mathcal{O}(\lambda^{2}),\ \partial_{\bot}\chi(x_{-}%
)\sim\mathcal{O}(\lambda^{4}),\ \ (\bar{n}\partial)\chi(x_{-})\sim
\mathcal{O}(\lambda^{6}).
\end{equation}

The most general 3-quark twist-3 operator is
\begin{equation}
\chi_{1}(x_{-})\chi_{2}(y_{-})\chi_{3}(z_{-})\sim\mathcal{O}(\lambda^{6}),
\end{equation}
where for simplicity  we do not show the colour and flavour structure.

The 3-quark twist-4 operators  include one transverse derivative%
\begin{equation}
\left[  \partial_{\bot}\chi_{1}\right]  \chi_{2}\chi_{3}%
\sim\chi_{1}\left[  \partial_{\bot}\chi_{2}\right]
\chi_{3}\sim\chi_{1}\chi_{2}\left[  \partial_{\bot}%
\chi_{3}\right]  \sim\mathcal{O}(\lambda^{8}),
\label{twist4aux}
\end{equation}
where the derivative is applied inside the brackets only. Notice that we do
not need to write the covariant derivative because the field $\chi(x_{-})$ is
gauge invariant.  The matrix elements of the twist-4 operators in (\ref{twist4aux}) are considered in
Ref.\cite{Kivel:2019wjh}.  The results read
\begin{equation}
~\left\langle 0\right\vert \left[  i\partial_{\bot}^{\alpha}\chi
(x_{-})\right]  (x_{+}%
)C\setbox0=\hbox{$n$}\dimen0=\wd0\setbox1=\hbox{/}\dimen1=\wd1\ifdim\dimen0>\dimen1\rlap{\hbox to \dimen0{\hfil/\hfil}}n\else\rlap{\hbox
to \dimen1{\hfil$n$\hfil}}/\fi\chi(y_{-})\chi_{3}(z_{-})\left\vert
B(k)\right\rangle =k_{+}m_{B}\left[  \gamma_{\bot}^{\alpha}\gamma_{5}%
N_{\bar{n}}\right]  _{3}\text{FT}\left[  \mathcal{V}_{1}^{B}\right]  ,
\label{calV1}%
\end{equation}%
\begin{equation}
\left\langle 0\right\vert ~\chi(x_{-}%
)C\setbox0=\hbox{$n$}\dimen0=\wd0\setbox1=\hbox{/}\dimen1=\wd1\ifdim\dimen0>\dimen1\rlap{\hbox to
\dimen0{\hfil/\hfil}}n\else\rlap{\hbox to \dimen1{\hfil$n$\hfil}}/\fi\left[
i\partial_{\bot}^{\alpha}\chi(y_{-})\right]  \chi_{3}(z_{-})\left\vert
B(k)\right\rangle =k_{+}m_{B}~\left[  \gamma_{\bot}^{\alpha}\gamma_{5}%
N_{\bar{n}}\right]  _{3}\text{FT}\left[  \mathcal{V}_{2}^{B}\right]  .
\end{equation}%
\begin{equation}
\left\langle 0\right\vert \left[  i\partial_{\bot}^{\alpha}\chi(x_{-})\right]
C\setbox0=\hbox{$n$}\dimen0=\wd0\setbox1=\hbox{/}\dimen1=\wd1\ifdim\dimen0>\dimen1\rlap{\hbox to \dimen0{\hfil/\hfil}}n\else\rlap{\hbox to
\dimen1{\hfil$n$\hfil}}/\fi\gamma_{5}\chi(y_{-})\chi_{3}(z_{-})\left\vert
B(k)\right\rangle =k_{+}m_{B}\left[  ~\gamma_{\bot}^{\alpha}N_{\bar{n}%
}\right]  _{3}\text{FT}\left[  \mathcal{A}_{1}^{B}\right]  ,
\end{equation}%
\begin{equation}
\left\langle 0\right\vert \chi(x_{-}%
)C\setbox0=\hbox{$n$}\dimen0=\wd0\setbox1=\hbox{/}\dimen1=\wd1\ifdim\dimen0>\dimen1\rlap{\hbox to
\dimen0{\hfil/\hfil}}n\else\rlap{\hbox to \dimen1{\hfil$n$\hfil}}/\fi\gamma
_{5}\left[  i\partial_{\bot}^{\alpha}\chi(y_{-})\right]  \chi_{3}%
(z_{-})\left\vert B(k)\right\rangle =k_{+}m_{B}\left[  \gamma_{\bot}^{\alpha
}N_{\bar{n}}\right]  _{3}\text{FT}\left[  \mathcal{A}_{2}^{B}\right]  .
\end{equation}%
\begin{align}
\left\langle 0\right\vert \left[  i\partial_{\bot}^{\alpha}\chi(x_{-})\right]
C\setbox0=\hbox{$n$}\dimen0=\wd0\setbox1=\hbox{/}\dimen1=\wd1\ifdim\dimen0>\dimen1\rlap{\hbox to \dimen0{\hfil/\hfil}}n\else\rlap{\hbox to
\dimen1{\hfil$n$\hfil}}/\fi\gamma_{\bot}^{\nu}\chi(y_{-})\chi_{3}%
(z_{-})\left\vert B(k)\right\rangle  &  =~k_{+}m_{B}~g_{\bot}^{\alpha\nu
}\left[  \gamma_{5}N_{\bar{n}}\right]  _{3}\text{FT}\left[  \mathcal{T}%
_{21}^{B}\right] \\
&  +k_{+}m_{B}\left[  i\sigma_{\bot\bot}^{\nu\alpha}\gamma_{5}N_{\bar{n}%
}\right]  _{3}\text{FT}\left[  \mathcal{T}_{41}^{B}\right]  ,
\end{align}%
\begin{align}
\left\langle 0\right\vert \chi(x_{-}%
)C\setbox0=\hbox{$n$}\dimen0=\wd0\setbox1=\hbox{/}\dimen1=\wd1\ifdim\dimen0>\dimen1\rlap{\hbox to
\dimen0{\hfil/\hfil}}n\else\rlap{\hbox to \dimen1{\hfil$n$\hfil}}/\fi\gamma
_{\bot}^{\nu}\left[  i\partial_{\bot}^{\alpha}\chi(y_{-})\right]  \chi
_{3}(z_{-})\left\vert B(k)\right\rangle  &  =k_{+}m_{B}~g_{\bot}^{\alpha\nu
}\left[  \gamma_{5}N_{\bar{n}}\right]  _{3}\text{FT}\left[  \mathcal{T}%
_{22}^{B}\right] \nonumber\\
&  +k_{+}m_{B}\left[  i\sigma_{\bot\bot}^{\nu\alpha}\gamma_{5}N_{\bar{n}%
}\right]  _{3}\text{FT}\left[  \mathcal{T}_{42}^{B}\right]  , \label{calT2i}%
\end{align}
where the color and the flavour structure of the operators are the same as
described before. The DAs in the\textit{\ rhs} of these equations can be
expressed in terms of DAs $\{V_{i},A_{i},T_{i},S_{i},P_{i}\}^{B}$ using the
Lorentz symmetry and QCD equations of motion. This gives\footnote{ If the all function in an equation have the same standard arguments $(x_i)\equiv(x_1,x_2,x_3)$  we use  the simplification $F(x_i)\equiv F$  to simplify the equations. }
\cite{Anikin:2013aka,Kivel:2019wjh}%
\begin{align}
4\mathcal{V}_{i}^{B} &  =x_{3}\left(  V_{2}+A_{2}\right)
^{B}+(-1)^{i}\left[  (x_{1}-x_{2})V_{3}^{B}+\bar{x}_{3}A_{3}^{B}\right]
 +\frac{m_{3}}{m_{B}}\left(  V_{1}^{B}+(-1)^{i}A_{1}^{B}\right)
\nonumber \\ &
-2(-1)^{i}\frac{m_{1}-m_{2}}{m_{B}}T_{1}^{B},
 \label{Vical}
\end{align}%
\begin{align}
4\mathcal{A}_{i}^{B}  &  =-x_{3}\left(  V_{2}+A_{2}\right)
^{B}+(-1)^{i}\left[  (x_{1}-x_{2})A_{3}^{B}+\bar{x}_{3}V_{3}^{B}\right] 
 +\frac{m_{3}}{m_{B}}\left(  A_{1}^{B}+(-1)^{i}V_{1}^{B}\right)
\nonumber \\
& 
-2(-1)^{i}\frac{m_{1}+m_{2}}{m_{B}}T_{1}^{B},
\label{Aical}
\end{align}%
\begin{align}
\left(  \mathcal{T}_{21}+\mathcal{T}_{41}\right)  ^{B} &
=\frac{x_{1}}{2}\left(  T_{3}-T_{7}+S_{1}+P_{1}\right)  ^{B}-\frac
{1}{2}\frac{m_{1}}{m_{B}}\left(  V_{1}^{B}+A_{1}^{B}\right)
,\ \label{rel:TB21pT41}\\
\ \ \ \left(  \mathcal{T}_{22}+\mathcal{T}_{42}\right)  ^{B}  &
=\frac{x_{2}}{2}\left(  T_{3}-T_{7}-S_{1}-P_{1}\right)  ^{B}-\frac
{1}{2}\frac{m_{2}}{m_{B}}\left(  V_{1}^{B}-A_{1}^{B}\right)  ,
\label{rel:TB22p42}%
\end{align}%
\begin{align}
\left(  \mathcal{T}_{21}-\mathcal{T}_{41}\right)  ^{B}  &
=\frac{x_{1}}{2}\left(  T_{3}+T_{7}+S_{1}-P_{1}\right)  ^{B}+\frac
{1}{2}\frac{m_{1}}{m_{B}}\left(  A_{1}^{B}-V_{1}^{B}\right)
,\ \ \ \label{rel:T21mT41}\\
\ \left(  \mathcal{T}_{22}-\mathcal{T}_{42}\right)  ^{B}  &
=\frac{x_{2}}{2}\left(  T_{3}+T_{7}-S_{1}+P_{1}\right)  ^{B}-\frac
{1}{2}\frac{m_{2}}{m_{B}}\left(  V_{1}^{B}+A_{1}^{B}\right)  ,
\label{rel:T22mT42}%
\end{align}
where the quark masses $m_{i}$ correspond to  quarks $q_{1}q_{2}q_{3}$ in
the matrix elements in Eqs.(\ref{qqq1})-(\ref{qqq3}), respectively.

 The additional relation is derived in Appendix \ref{AppA}
\begin{equation}
T_{2}^{B}=\frac{1}{x_{3}}(\mathcal{T}_{21}+\mathcal{T}_{22}+\mathcal{T}%
_{41}+\mathcal{T}_{42})^{B}+\frac{~m_{3}}{m_{B}}T_{1}^{B}. \label{Rel:TB2}%
\end{equation}
Combaining Eq.(\ref{Rel:TB2}) with Eqs.(\ref{rel:TB21pT41}),
(\ref{rel:TB22p42}) and definitions in Eqs. (\ref{Ksi4pm}) and (\ref{Ups4})
one finds for  $\Upsilon_{4}^{B}$ the following equation%
\begin{align}
\Upsilon_{4}(x_{123})  &  =\frac{c_{B}^{-}}{c_{B}^{+}}\frac{1}{x_{1}}\left\{
(-1)_{B}\ x_{3}\Xi_{4+}^{B}(x_{321})+x_{2}\Xi_{4+}^{B}(x_{231})\right\}
\nonumber \\
&  +\frac{1}{x_{1}}\left\{  (-1)_{B}\ x_{3}\Xi_{4-}^{B}(x_{321})+x_{2}\Xi
_{4-}^{B}(x_{231})\right\} 
\nonumber \\
&  +\frac{3c_{B}^{-}}{x_{1}}\left\{  \frac{m_{2}-m_{1}}{m_{B}}A_{1}%
^{B}(x_{321})-\frac{m_{1}+m_{2}}{m_{B}}V_{1}^{B}(x_{321})+\frac{~2m_{3}}%
{m_{B}}T_{1}^{B}(x_{321})\right\}  .
\label{UpsKsi}
\end{align}
Therefore the DA $\Upsilon_{4}^{B}$ is not an independent function in the 3-quark approximation.

The 3-quark twist-5 operators  include  two transverse derivatives%
\begin{equation}
\partial_{\bot}^{2}\chi_{1}\chi_{2}\chi_{3}\sim\mathcal{O}(\lambda^{10}).
\end{equation}
$\ $The longitudinal derivatives $(\bar{n}\partial)\chi$\ can be excluded with 
the help of QCD equation of motion%
\begin{equation}
(i\bar{n}\partial)\chi=(in\partial)^{-1}\partial_{\bot}^{2}\chi.
\end{equation}
Therefore the basis set of the 3-quarks operators can be defined  as
\begin{align}
 \left[  \partial_{\bot}^{\alpha}\partial_{\bot}^{\beta}\chi_{1}\right]  \chi_{2}\chi_{3},
 \quad \chi_{1}\left[
\partial_{\bot}^{\alpha}\partial_{\bot}^{\beta}\chi_{2}\right]
\chi_{3}, 
\quad
\left[  \partial_{\bot}^{\alpha}\chi_{1}\right]
\left[  \partial_{\bot}^{\beta}\chi_{2}\right]  \chi_{3}.
\label{Otw5}%
\end{align}
The operators with $\left[  \partial_{\bot}\chi_{3}\right]  $ can be
 excluded because%
\begin{equation}
\chi_{1}\chi_{2}\left[  \partial_{\bot}\chi_{3}\right]  =\partial_{\bot
}\left[  \chi_{1}\chi_{2}\chi_{3}\right]  -\left[  \partial_{\bot}\chi
_{1}\right]  \chi_{2}\chi_{3}-\chi_{1}\left[  \partial_{\bot}\chi_{2}\right]
\chi_{3}\simeq-\left[  \partial_{\bot}\chi_{1}\right]  \chi_{2}\chi_{3}%
-\chi_{1}\left[  \partial_{\bot}\chi_{2}\right]  \chi_{3},
\end{equation}
where it is used that the matrix element from the operator with the total transverse
derivative vanishes because the  baryon momenta have longitudinal components only. 
 The matrix elements of the three operators in Eq.(\ref{Otw5})
define the auxiliary set of the twist-5 DAs.  It is convenient to write their definitions for the operators with projected Dirac indices $\{12\}$. \ We
introduce the following set of the matrix elements 
\begin{equation}
\left\langle 0\right\vert \left[  \partial_{\bot}^{2}\chi(x_{-})\right]
C\ns  \chi(y_{-})\chi_{3}(z_{-})\left\vert B(k)\right\rangle =-m_{B}%
^{2}k_{+}~\left[  \gamma_{5}N_{\bar{n}}\right]  _{3}~\text{FT}[V_{1xx}^{B}],
\label{V1xx}%
\end{equation}%
\begin{equation}
\left\langle 0\right\vert \chi(x_{-})C\ns  \left[  \partial_{\bot}^{2}%
\chi(y_{-})\right]  \chi_{3}(z_{-})\left\vert B(k)\right\rangle =-m_{B}%
^{2}k_{+}~\left[  \gamma_{5}N_{\bar{n}}\right]  _{3}~\text{FT}\left[
V_{1yy}^{B}\right]  , \label{V1yy}%
\end{equation}%
\begin{equation}
\left\langle 0\right\vert \left[  \partial_{\bot}^{2}\chi(x_{-})\right]
C\ns  \gamma_{5}\chi(y_{-})\chi_{3}(z_{-})\left\vert B(k)\right\rangle
=-m_{B}^{2}k_{+}~\left[  N_{\bar{n}}\right]  _{3}~\text{FT}\left[  A_{1xx}%
^{B}\right]  ,
\end{equation}%
\begin{equation}
\left\langle 0\right\vert \chi(x_{-})C\ns  \gamma_{5}\left[  \partial
_{\bot}^{2}\chi(y_{-})\right]  \chi_{3}(z_{-})\left\vert B(k)\right\rangle
=-m_{B}^{2}k_{+}\left[  N_{\bar{n}}\right]  _{3}~\text{FT}\left[  A_{1yy}%
^{B}\right]  ,
\end{equation}
\begin{align}
\left\langle 0\right\vert \left[  \partial_{\bot}^{\alpha}\chi(x_{-})\right]
C\ns  \left[  \partial_{\bot}^{\beta}\chi(y_{-})\right]  \chi_{3}%
(z_{-})\left\vert B(k)\right\rangle  &  =-\frac{m_{B}^{2}}{2}k_{+}~g_{\bot
}^{\alpha\beta}\left[  \gamma_{5}N_{\bar{n}}\right]  _{3}~\text{FT}\left[
V_{1xy}^{B}\right] \label{V1xy}
\nonumber \\
&  -\frac{m_{B}^{2}}{2}k_{+}\left[  i\sigma_{\bot\bot}^{\alpha\beta}\gamma
_{5}N_{\bar{n}}\right]  _{3}\text{FT}\left[  T_{xy}^{B}\right]  ,
\end{align}%
\begin{align}
\left\langle 0\right\vert \left[  \partial_{\bot}^{\alpha}\chi(x_{-})\right]
C\ns  \gamma_{5}\left[  \partial_{\bot\beta}\chi(y_{-})\right]  \chi
_{3}(z_{-})\left\vert B(k)\right\rangle  &  =-\frac{m_{B}^{2}}{2}k_{+}g_{\bot
}^{\alpha\beta}\left[  N_{\bar{n}}\right]  _{3}~\text{FT}\left[  A_{1xy}%
^{B}\right]  ~\nonumber\\
&  -\frac{m_{B}^{2}}{2}k_{+}\left[  i\sigma_{\bot\bot}^{\alpha\beta}N_{\bar
{n}}\right]  _{3}\text{FT}\left[  G_{xy}^{B}\right] , \label{A1xy}%
\end{align}%
\begin{eqnarray}
&\left\langle 0\right\vert \left[  \partial_{\bot}^{\alpha}\partial_{\bot
}^{\beta}\chi(x_{-})\right]  C\ns  \gamma_{\bot}^{\sigma}\chi(y_{-})
\chi_{3}(z_{-})\left\vert B(k)\right\rangle =\frac{m_{B}^{2}}{4}~k_{+}\left[
\gamma_{\bot}^{\rho}\gamma_{5}N_{\bar{n}}\right]  _{3}
\nonumber \\
&\times\left(  g_{\bot}^{\alpha\beta}g_{\bot}^{\sigma\rho}\text{FT}\left[  T_{0xx}^{B}\right]
+~G_{\bot}^{\alpha\beta\sigma\rho}~\text{FT}\left[  T_{2xx}^{B}\right]
\right)  , \label{T0xx}%
\end{eqnarray}%
\begin{eqnarray}
&\left\langle 0\right\vert \chi(x_{-})C\bar{n}\gamma_{\bot}^{\sigma}\left[
\partial_{\bot}^{\alpha}\partial_{\bot}^{\beta}\chi(y_{-})\right]  \chi
_{3}(z_{-})\left\vert B(k)\right\rangle =\frac{m_{B}^{2}}{4}~k_{+}\left[
\gamma_{\bot}^{\rho}\gamma_{5}N_{\bar{n}}\right]  _{3}~
\nonumber \\
&\left(  g_{\bot
}^{\alpha\beta}g_{\bot}^{\sigma\rho}\text{FT}\left[  T_{0yy}^{B}\right]
+G_{\bot}^{\alpha\beta\sigma\rho}~\text{FT}\left[  T_{2yy}^{B}\right]
\right)  ,
\end{eqnarray}%
\begin{equation}
\left\langle 0\right\vert \left[  \partial_{\bot}^{\alpha}\chi(x_{-})\right]
C\ns  \gamma_{\bot}^{\sigma}\left[  \partial_{\bot}^{\beta}\chi
(y_{-})\right]  \chi_{3}(z_{-})\left\vert B(k)\right\rangle =\frac{m_{B}^{2}%
}{4}k_{+}\left[  \gamma_{\bot}^{\rho}\gamma_{5}N_{\bar{n}}\right]  _{3}%
~T_{xy}^{\alpha\beta\sigma\rho}, \label{Txy}%
\end{equation}
with%
\begin{equation}
T_{xy}^{\alpha\beta\sigma\rho}=~g_{\bot}^{\sigma\rho}g_{\bot}^{\alpha\beta
}~\text{FT}\left[  T_{0xy}^{B}\right]  +G_{\bot}^{\alpha\beta\sigma\rho
}~\text{FT}\left[  T_{2xy}^{B}\right]  +\left(  g_{\bot}^{\alpha\rho}g_{\bot
}^{\beta\sigma}-g_{\bot}^{\alpha\sigma}g_{\bot}^{\beta\rho}\right)
\text{FT}\left[  \tilde{T}_{2xy}^{B}\right]  ,
\end{equation}
where we use the short notaion for the following combination%
\begin{equation}
G_{\bot}^{\alpha\beta\sigma\rho}=g_{\bot}^{\alpha\sigma}g_{\bot}^{\beta\rho
}+g_{\bot}^{\alpha\rho}g_{\bot}^{\beta\sigma}~-g_{\bot}^{\alpha\beta}g_{\bot}^{\sigma\rho}.
\end{equation}

This set of DAs can be splitted into two groups: chiral even and chiral-odd
DAs defined in Eqs.(\ref{V1xx})-(\ref{A1xy}) and Eqs.(\ref{T0xx})-(\ref{Txy}),
respectively.  These groups can be considered independently.

In order to establish  relations of the auxiliary  DAs with the basic
ones, one needs to relate the operator $\mathcal{O}_{123}$ defined in
Eq.(\ref{calO123}) with the operators in Eq.(\ref{Otw5}).  This can be done
with the help of the QCD equations of motion.  Let us write the  quark field  as a sum
\begin{equation}
q(x_{-})=\xi(x_{-})+\eta(x_{-}),
\end{equation}
where
\begin{equation}
\xi(x_{-})=\nbn q(x_{-}),\ \ \eta(x_{-})=\nnb q(x_{-}). 
\label{xidef}%
\end{equation}
The small field $\eta(x_{-})$ can be written as, see {\it e.g.} Ref.\cite{Beneke:2002ph} 
\begin{equation}
\eta(x_{-})=-\frac{\ns}{2}\left(  in\cdot D\right)^{-1}%
i{\Dsl{D}  }_{\bot}\xi(x_{-})=-\frac{\ns  }{2}W(x_{-})\left(in\partial\right)  ^{-1}i\Dsl{\partial} _{\bot}\chi(x_{-})+\ldots,
\label{etaEOM}%
\end{equation}
where dots denote the terms with insertions of the transverse gluon components
$\left[  W^{\dag}(x_{-})i{\Dsl D  }_{\bot}W(x_{-})\right]  $,
which  give the quark-gluon  operators and therefore can be omitted. Hence 
\begin{equation}
\ W^{\dag}(x_{-})\eta(x_{-}%
)\simeq-\frac{\ns  }{2}\left(  in\partial\right)  ^{-1}i\Dsl \partial
_{\bot}\chi(x_{-}). \label{eta=dXi}%
\end{equation}
It is clear that each field $\eta$ gives the one transverse drivative,
therefore the twist-5 operators in the expansion of $\mathcal{O}_{123}$ in
Eq.(\ref{calO123}) have two fields $\eta$%
\begin{align}
\left[  \mathcal{O}_{123}(x_{-},y_{-},z_{-})\right]  _{\text{tw5}}  &
=W^{\dag}(x_{-})\eta_{1}(x_{-})W^{\dag}(y_{-})\eta_{2}(y_{-})W^{\dag}%
(z_{-})\xi_{3}(z_{-})\nonumber\\
&  +W^{\dag}(x_{-})\eta_{1}(x_{-})W^{\dag}(y_{-})\xi_{2}(y_{-})W^{\dag}%
(z_{-})\eta_{3}(z_{-})\nonumber\\
&  +W^{\dag}(x_{-})\xi_{1}(x_{-})W^{\dag}(y_{-})\eta_{2}(y_{-})W^{\dag}%
(z_{-})\eta_{3}(z_{-}). \label{O123exp}%
\end{align}

The main idea is the following: we rewrite the terms in Eq.(\ref{O123exp}) using
Eq.(\ref{eta=dXi}) in terms of the basis operators (\ref{Otw5}) and compute
the matrix element using Eqs.(\ref{V1xx})-(\ref{UpsB5}).  On the other hand,
 using  the definitions (\ref{xidef}) and one can express the $\left[
\mathcal{O}_{123}\right]  _{\text{tw5}}$ in terms of different projections of the three quark operator
defined in Eq.(\ref{calO123}) and use Eq.(\ref{meO123}) in  order to get the
matrix element. Comparing the two results one finds the  relations, which allows one to establish a connection of  the auxiliary DAs with the basis ones.
 It is convenient to perform such calculations for the chiral-even and
chiral-odd operators separately.  Let us first consider the chiral-even
 sector, which includes 8 DAs:%
\begin{equation}
V_{1xx}^{B},V_{1yy}^{B},\ V_{1xy}^{B},\ A_{1xx}^{B},\ A_{1yy}^{B}%
,\ A_{1xy}^{B},T_{xy}^{B},G_{xy}^{B}. 
\label{DAseven}%
\end{equation}

Let us consider as example the following operator%
\begin{equation}
\mathcal{O}_{3}^{V}(x_{-},y_{-},z_{-})=W^{\dag}(x_{-})\eta(x_{-})C\nbs W^{\dag}(y_{-})\eta(y_{-})W^{\dag}(z_{-})\xi_{3}(z_{-}),
\end{equation}
where the Dirac indices $\{12\}$ are contracted with the matrix $C\nbs$.
Using Eq. (\ref{eta=dXi}) one finds 
\begin{eqnarray}
&\mathcal{O}_{3}^{V}(x_{-},y_{-},z_{-})=\left[  -\left(  in\partial\right)
^{-1}\frac{\ns }{2}i\Dsl \partial _{\bot}\chi(x_{-})\right]  C\nbs\left[  -\left(  in\partial\right)  ^{-1}\frac{\ns  }{2}i\Dsl \partial
_{\bot}\chi(y_{-})\right]  \chi_{3}(z_{-})
\\
&  =\left[  \left(  in\partial\right)  ^{-1}\partial_{\bot}^{\alpha}\chi
(x_{-})\right]  C\ns\left[  \left(  in\partial\right)  ^{-1}\partial_{\bot
}^{\alpha}\chi(y_{-})\right]  \chi_{3}(z_{-})
\\
&  +\frac{1}{2}i\varepsilon\lbrack\alpha\beta]\left[  \left(  in\partial
\right)  ^{-1}\partial_{\bot}^{\alpha}\chi(x_{-})\right]  C\ns\gamma_{5}\left[
\left(  in\partial\right)  ^{-1}\partial_{\bot}^{\beta}\chi(y_{-})\right]
\chi_{3}(z_{-}),
\end{eqnarray}
where we use the short notation%
\begin{equation}
i\varepsilon\lbrack\alpha\beta]=\frac{1}{2}i\varepsilon_{\alpha\beta\sigma
\rho}n^{\sigma}\bar{n}^{\rho}.
\end{equation}
The matrix element of this operator can be computed with the help of
Eqs.(\ref{V1xy}) and (\ref{A1xy}). One finds%
\begin{equation}
\left\langle 0\right\vert \mathcal{O}_{3}^{V}(x_{-},y_{-},z_{-})\left\vert
B(k)\right\rangle =-\frac{m_{B}^{2}}{k_{+}}\left[  \gamma_{5}N_{\bar{n}%
}\right]  _{3}~\text{FT}\left[  \frac{1}{x_{1}x_{2}}\left(  V_{1xy}^{B}%
+G_{xy}^{B}\right)  \right]  , \label{O3V-1}%
\end{equation}
where the factor $1/(x_{1}x_{2})$ appears due to the inverse derivatives $\left(
in\partial\right)  ^{-1}$ acting on the first and second fields. Remind, that
the inverse derivatives $\left(  in\partial\right)  ^{-1}$ must be undestood
as nonlocal operator%
\begin{equation}
\left(  in\partial\right)  ^{-1}\phi(x)\equiv\left(  in\partial+i\varepsilon
\right)  ^{-1}\phi(x)=-i\int_{-\infty}^{0}ds\ \phi(x+sn).
\end{equation}

On the other hand, using Eqs.(\ref{xidef}) one finds
\begin{align}
\mathcal{O}_{3}^{V}(x_{-},y_{-},z_{-})  &  =W^{\dag}(x_{-})q(x_{-})\left(
\frac{n\bar{n}}{4}\right)  ^{\top}C\nbs\frac{n\bar{n}}{4}W^{\dag}%
(y_{-})q(y_{-})\ W^{\dag}(z_{-})\left[  \frac{\bar{n}n}{4}q(z_{-})\right]
_{3}\\
&  =\left[  \frac{\bar{n}n}{4}\right]  _{33^{\prime}}\ W^{\dag}(x_{-}%
)q(x_{-})C\nbs W^{\dag}(y_{-})q(y_{-})W^{\dag}(z_{-})q_{3^{\prime}}(z_{-}).
\end{align}
Computing the matrix element of this operator with the help of
Eq.(\ref{<O123>}) one gets
\begin{equation}
\left\langle 0\right\vert \mathcal{O}_{3}^{V}(x_{-},y_{-},z_{-})\left\vert
B(k)\right\rangle =-\frac{m_{B}^{2}}{k_{+}}\left[  \gamma_{5}N_{\bar{n}%
}\right]  _{3}\text{FT}\left[  V^B_{5}\right]  . \label{O3V-2}%
\end{equation}
Comparing Eqs.(\ref{O3V-1}) and (\ref{O3V-2}) \ one obtains%
\begin{equation}
\frac{1}{x_{1}x_{2}}\left(  V_{1xy}^{B}+G_{xy}^{B}\right)  =V_{5}^{B},
\label{V1xy=V5}%
\end{equation}
where, for simplicity, the arguments are not shown: $F(x_{123})\equiv F$.

The similar consideration for the operators $\eta C\gamma_{\bot}\xi\eta_{3}$
and $\xi C\gamma_{\bot}\eta\eta_{3}$\footnote{The Wilson lines are not shown
for simplicity.} yields%
\begin{equation}
V^B_{1xx}+V^B_{1xy}+A^B_{1xx}+A^B_{1xy}+T^B_{xy}+G^B_{xy}=-x_{1}x_{3}\left(  V^B_{4}%
-A^B_{4}\right)  , \label{V1xx=V4mA4}%
\end{equation}%
\begin{equation}
V^B_{1yy}+V^B_{1xy}-A^B_{1yy}-A^B_{1xy}-T^B_{xy}+G^B_{xy}=-x_{2}x_{3}\left(  V^B_{4}%
+A^B_{4}\right)  , \label{V1yy=V4pA4}%
\end{equation}
respectively. \ The analysis for the axial operators $\eta C\bar{n}\gamma
_{5}\eta\xi_{3}$ gives%
\begin{equation}
A^B_{1xy}+T^B_{xy}=x_{1}x_{2}A^B_{5}. \label{A1xy=A5}%
\end{equation}

Two more equations can be obtained using the QCD EOM, which can be rewritten
as
\begin{equation}
i(\bar{n}\partial)\xi(x_{-})\simeq(in\partial)^{-1}\partial_{\bot}^{2}%
\xi(x_{-}), \label{xiEOM}%
\end{equation}
and describes the free collinear quark.  Consider the following operator with
the total derivative%
\begin{align}
i(\bar{n}\partial)\left[  \chi(x_{-})C\nbs\chi(y_{-})\chi_{3}%
(z_{-})\right]    \simeq\left[  (in\partial)^{-1}\partial_{\bot}^{2}%
\chi(x_{-})\right]  C\nbs\chi(y_{-})\chi_{3}(z_{-})
\phantom{
i(\bar{n}\partial)  \chi(x_{-})C\nbs\chi(y_{-})
}
\nonumber \\
 +\chi(x_{-})C\nbs\left[  (in\partial)^{-1}\partial_{\bot}^{2}\chi
(y_{-})\right]  \chi_{3}(z_{-})
  +\chi(x_{-})C\nbs\chi(y_{-})\left[  (in\partial)^{-1}\partial_{\bot}%
^{2}\chi_{3}(z_{-})\right]  .
\label{dOeven}
\end{align}
Taking the matrix elements from the both sides one finds%
\begin{equation}
\frac{x_{1}+x_{3}}{x_{1}x_{3}}V^B_{1xx}+\frac{x_{2}+x_{3}}{x_{2}x_{3}}%
V^B_{1yy}+\frac{2}{x_{3}}V^B_{1xy}=V^B_{1}.
\end{equation}
The same consideration for the axial operator  $\chi C\nbs\gamma_{5}%
\chi\chi_{3}$ gives%
\begin{equation}
\frac{x_{1}+x_{3}}{x_{1}x_{3}}A^B_{1xx}+\frac{x_{2}+x_{3}}{x_{2}x_{3}}%
A^B_{1yy}+\frac{2}{x_{3}}A^B_{1xy}=A^B_{1}.
\end{equation}
Therefore we obtain six linear algebraic relations. The investigation of
other possible 3-quark light-cone operators using the same technique does not
provide any new information. We did not find any method in order to get two
more pure algebraic equations.  However one needs at least two more equations
in order to express the eight auxiliary DAs  (\ref{DAseven}) in  terms of
the basis ones.  In order to derive these equations we introduce  off
light-cone correlation function (CF). Performing the expansion of this CF
around the light-cone we derive two more equations, which allow us to solve
the problem. However these relations are already not algebraic. The
derivation of the corresponding relations is more involved and the details are
given in Appendix \ref{AppA}. The obtained equations  read%
\begin{equation}
2i(kx)\text{FT}\left[  T^B_{xy}-G^B_{xy}\right]  =\text{FT}\left[  \left(
x_{1}+x_{2}\right)  \left(  A^B_{5}-V^B_{5}\right)  -\left(  x_{2}+x_{3}\right)
\left(  V^B_{4}+A^B_{4}\right)  -2\left(  \mathcal{V}^B_{2}+\mathcal{A}^B_{2}\right)
\right]  , \label{TmG}%
\end{equation}%
\begin{equation}
2i(ky)\text{FT}\left[  T^B_{xy}+G^B_{xy}\right]  =\text{FT}\left[  (x_{1}%
+x_{2})\left(  V^B_{5}+A^B_{5}\right)  +\left(  x_{1}+x_{3}\right)  \left(
V^B_{4}-A^B_{4}\right)  +2\left(  \mathcal{V}^B_{1}-\mathcal{A}^B_{1}\right)  \right]
, \label{TpG}%
\end{equation}
where the factors $2(kx)=k_{+}x_{-}$, $2(ky)=k_{+}y_{-}$ and the coordinates
$x$ and $y$ enter in the definition of the FT in Eq.(\ref{FT})\footnote{ We also assume
here that the argument of the third quark field is trivial: $z=0$ }. Remind, that the DAs $\mathcal{V}_{i}$ and $\mathcal{A}_{i}$ on
the \textit{rhs} denote the auxiliary twist-4 DAs from Eqs.(\ref{Vical})-(\ref{Aical}). For instance, for the nucleon, taking the
asymptotic contributions of twist-3 and twist-4 DAs one finds
\begin{equation}
\left.  \left(  \mathcal{A}_{1}-\mathcal{V}_{1}\right)  ^{N}(x_{i})\right\vert
_{as}=-20\ f_{N}~x_{1}^{2}x_{2}x_{3},\ \ \ \left.  \left(  \mathcal{V}%
_{2}+\mathcal{A}_{2}\right)  ^{N}(x_{i})\right\vert _{as}=-20\ f_{N}%
\ x_{1}x_{2}^{2}x_{3}.
\end{equation}
 Solving two equations (\ref{TmG}) and (\ref{TpG}) one finds expressions for
$T_{xy}$ and $G_{xy}$. Then one can easily solve the system of the linear equations.

The properties of Eqs. (\ref{TmG}) and (\ref{TpG})  imply that in case of
$x=0$ or $y=0$ the  one \textit{lhs} vanishes and therefore must vanish the 
corresponding \textit{rhs}. This gives two non-trivial conditions ($x_3=1-x_1-x_2$)%
\begin{equation}
\int_{0}^{\bar{x}_{2}}dx_{1}\left[  \left(  x_{1}+x_{2}\right)  \left(
A^B_{5}-V^B_{5}\right)  -\left(  x_{2}+x_{3}\right)  \left(  V^B_{4}+A^B_{4}\right)
-2\left(  \mathcal{V}^B_{2}+\mathcal{A}^B_{2}\right)  \right]  (x_{1}%
,x_{2},x_3)=0. \label{x=0}%
\end{equation}%
\begin{equation}
\int_{0}^{\bar{x}_{1}}dx_{2}\left[  (x_{1}+x_{2})\left(  V^B_{5}+A^B_{5}\right)
+\left(  x_{1}+x_{3}\right)  \left(  V^B_{4}-A^B_{4}\right)  +2\left(
\mathcal{V}^B_{1}-\mathcal{A}^B_{1}\right)  \right]  (x_{1},x_{2},x_3)=0. \label{y=0}%
\end{equation}
These conditions allows one to perform the integrations by parts on
\textit{rhs} of Eqs.(\ref{TmG}) and (\ref{TpG}),  that reduce these equations to a
simple algebraic form. Notice, that the DAs on the \textit{\ rhs} of
Eqs.(\ref{x=0}) and (\ref{y=0})  depends on the twist-3 and twist-4 moments
only.  We have found  that
Eqs.(\ref{x=0}) and (\ref{y=0}) are satisfied if one uses for description of
the DAs in Eqs.(\ref{x=0}) and (\ref{y=0}) the first moments (or equivalently  the asymptotic contributions) for
twist-3 and twist-4 DAs only.  We believe that this observation is closely
related to the  disregard of  the quark-gluon contributions and the
genuine twist-5 moments  Eq.(\ref{barDAs}). Such simple recipe does not work  for the description of the
twist-5 contributions associated with higher moments of the DAs and this makes
such analysis to be more complicated in comparison with the twist-4 DAs.  In order to 
 illustrate this point we discuss a simple example, which is considered in the Appendix \ref{AppB}.

Therefore in this work we only use the  asymptotic expressions for the
twist-3 and twist-4 DAs for the description of  auxiliary twist-5 DAs.
Let us also mention that obtained  description of the twist-5 DAs ensures
their proper normalisations, which are consistent with the Lorentz structure
for the local matrix elements. This is automatically provided by the algebraic
equations derived above. 

It is convenient to solve the equations for twist-5 auxiliary DAs
using the explicit expressions for the asymptotic DAs.  Then the solutions
are given by the simple analytical expressions.  Assuming below $B\neq
\Lambda$ we write the final result as
\begin{align}
T_{xy}^{B}=x_{1}x_{2}x_{3}(x_{1}-x_{2})(5f_{B}-\lambda_{1}^{B}),\ \ T_{xy}%
^{\Lambda}=-x_{1}x_{2}x_{3}\left\{  \frac{5f_{\Lambda}}{c_{\Lambda}^{+}%
}(3x_{3}-1)+\frac{\lambda_{1}^{\Lambda}}{c_{\Lambda}^{-}}\left(
1+x_{3}\right)  \right\}  ,
\label{TBxy}
\\
G_{xy}^{B}=x_{1}x_{2}x_{3}\left\{  5f_{B}(3x_{3}-1)+(1+x_{3})\lambda_{1}%
^{B}\right\}  ,\ G_{xy}^{\Lambda}=-~x_{1}x_{2}x_{3}(x_{1}-x_{2})\left\{
\frac{5f_{\Lambda}}{c_{\Lambda}^{+}}-\frac{\lambda_{1}^{\Lambda}}{c_{\Lambda
}^{-}}\right\}  ,
\end{align}
\begin{align}
A_{1xy}^{B}  &  =x_{1}x_{2}x_{3}(x_{1}-x_{2})(5f_{B}-\lambda_{1}^{B}),\\
A_{1xy}^{\Lambda}  &  =-~x_{1}x_{2}x_{3}\left\{  ~\frac{5f_{\Lambda}%
}{c_{\Lambda}^{+}}(3x_{3}-1-8x_{1}x_{2})+\frac{\lambda_{1}^{\Lambda}}{c_{-}%
}\left(  1+x_{3}\right)  \right\}  ,
\end{align}%
\begin{align}
V_{1xy}^{B}  &  =x_{1}x_{2}x_{3}(~5f_{B}(3x_{3}-1-8x_{1}x_{2})+\lambda_{1}%
^{B}(1+x_{3})),\\
V_{1xy}^{\Lambda}  &  =-~x_{1}x_{2}x_{3}(x_{1}-x_{2})\left\{  \frac
{5f_{\Lambda}}{c_{\Lambda}^{+}}-\frac{\lambda_{1}^{\Lambda}}{c_{\Lambda}^{-}%
}\right\}  ,
\end{align}%
\begin{align}
V_{1xx}^{B}=2x_{1}^{2}x_{2}x_{3}\left(  5f_{B}(5-4x_{1})-\lambda_{1}%
^{B}\right)  ,\ \ V_{1xx}^{\Lambda}=2x_{1}^{2}x_{2}x_{3}\left(  \frac
{5f_{\Lambda}}{c_{\Lambda}^{+}}-\frac{\lambda_{1}^{\Lambda}}{c_{\Lambda}^{-}%
}\right)  ,
\end{align}%
\begin{align}
V_{1yy}^{B}=2x_{1}x_{2}^{2}x_{3}\left(  5f_{B}(5-4x_{2})-\lambda_{1}%
^{B}\right)  ,\ \ V_{1yy}^{\Lambda}=-2x_{1}x_{2}^{2}x_{3}\left(
\frac{5f_{\Lambda}}{c_{\Lambda}^{+}}-\frac{\lambda_{1}^{\Lambda}}{c_{\Lambda
}^{-}}\right)  ,
\end{align}%
\begin{align}
A_{1xx}^{B}=2x_{1}^{2}x_{2}x_{3}\left(  -5f_{B}+\lambda_{1}^{B}\right)
,\ \ A_{1xx}^{\Lambda}=2x_{1}^{2}x_{2}x_{3}\left(  \frac{5f_{\Lambda}%
}{c_{\Lambda}^{+}}(4x_{1}-5)+\frac{\lambda_{1}^{\Lambda}}{c_{\Lambda}^{-}%
}\right)  ,
\end{align}
\begin{align}
A_{1yy}^{B}=2x_{1}x_{2}^{2}x_{3}\left(  5f_{B}-\lambda_{1}^{B}\right)
,\ \ A_{1yy}^{\Lambda}=2x_{1}x_{2}^{2}x_{3}\left(  \frac{5f_{\Lambda}%
}{c_{\Lambda}^{+}}(4x_{2}-5)+\frac{\lambda_{1}^{\Lambda}}{c_{\Lambda}^{-}%
}\right)  .
\label{A1Byy}
\end{align}

The similar discussion also holds for the chiral-odd set of the auxiliary DAs.
 The details are given in Appendix \ref{AppA}.  The results read%
\begin{equation}
T_{0xx}^{B}=40f_{B}^{~\bot}~x_{1}^{2}x_{2}x_{3}\left(  3-2x_{1}\right)
,\ T_{0xx}^{\Lambda}=8\frac{\lambda_{\bot}^{\Lambda}}{c_{\Lambda}^{-}}\medskip
x_{1}^{2}x_{2}x_{3}, \label{TB0xx}%
\end{equation}%
\begin{equation}
T_{0yy}^{B}=40f_{B}^{~\bot}~x_{1}x_{2}^{2}x_{3}\left(  3-2x_{2}\right)
,\ T_{0yy}^{\Lambda}=-8\frac{\lambda_{\bot}^{\Lambda}}{c_{\Lambda}^{-}}%
x_{1}x_{2}^{2}x_{3},
\end{equation}%
\begin{equation}
T_{0xy}^{B}=-20f_{B}^{~\bot}~x_{1}x_{2}x_{3}\left(  1-x_{3}+4x_{1}%
x_{2}\right)  ,\ ~T_{0xy}^{\Lambda}=-4\frac{\lambda_{\bot}^{\Lambda}%
}{c_{\Lambda}^{-}}x_{1}x_{2}x_{3}(x_{1}-x_{2}),
\end{equation}%
\begin{align}
~T_{2xy}^{B}  &  =-\lambda_{2}^{B}\frac{2}{3}x_{1}x_{2}x_{3}(1+x_{3}%
),~~T_{2xy}^{\Lambda}=\frac{\lambda_{2}^{\Lambda}}{c_{\Lambda}^{+}}\frac{2}%
{3}x_{1}x_{2}x_{3}(x_{1}-x_{2}),\\
 \tilde{T}_{2xy}^{B}(x_{i})  &  =-20f_{B}^{~\bot}~x_{1}x_{2}x_{3}%
(x_{1}-x_{2}),\ \ \tilde{T}_{2xy}^{\Lambda}=-4\frac{\lambda_{\bot}^{\Lambda}%
}{c_{\Lambda}^{-}}x_{1}x_{2}x_{3}(1-x_{3}),
\end{align}%
\begin{align}
T_{2xx}^{B}  &  =\lambda_{2}^{B}\frac{4}{3}x_{1}^{2}x_{2}x_{3},~~T_{2xx}%
^{\Lambda}=-\frac{\lambda_{2}^{\Lambda}}{c_{\Lambda}^{+}}\frac{4}{3}x_{1}%
^{2}x_{2}x_{3},\\
 T_{2yy}^{B}  &  =\lambda_{2}^{B}\frac{4}{3}x_{1}x_{2}^{2}x_{3}%
,\ \ \ T_{2yy}^{\Lambda}=\frac{\lambda_{2}^{\Lambda}}{c_{\Lambda}^{+}}\frac
{4}{3}x_{1}x_{2}^{2}x_{3}. \label{TB2yy}%
\end{align}
For the nucleon set one has to substitute in the  formulas  $f_{B}^{\bot}\rightarrow f_{N}$. The coefficients $c_{\Lambda}^{\pm}$ are defined in
Eq.(\ref{cBpm}). It is useful to note that all auxiliary DAs  are proportional to the factor 
$x_{1}x_{2}x_{3}$ that simplifies an investigation of analytic properties of the
collinear integrals.

In addition to  the identities, which follow from the Lorentz symmetry,  one can also derive symmetry relations that follow from
 the condition that a baryon state has a certain isospin. For the auxiliary matrix elements this can be done in the same way as for the  basis one,  see {\it e.g.} Ref.\cite{Braun:2000kw}.    For instance,   for the nucleon ($I_N=1/2$)   this gives 15 different relations for the auxiliary DAs. We provide these relations in Appendix \ref{App:iso}. These relations  can also be used for the finding of the relation between the auxiliary and basic DAs but different baryons has different isospin, which  complicates the analysis. 
  At the same time, the relations that follow from  the Lorentz symmetry are universal and valid for all baryons.  Nevertheless, the  isospin relations provide a powerful check of the obtained result. We have verified that the obtained expressions for the auxiliary DAs in Eqs.(\ref{TBxy})-(\ref{A1Byy}) and (\ref{TB0xx})-(\ref{TB2yy}) satisfy all 15 isotopic  relations given in  Appendix   \ref{App:iso}.

\section{Calculation of  hard kernels}
\label{calc}

\begin{figure}[ptb]%
\centering
\includegraphics[width=3.50in]{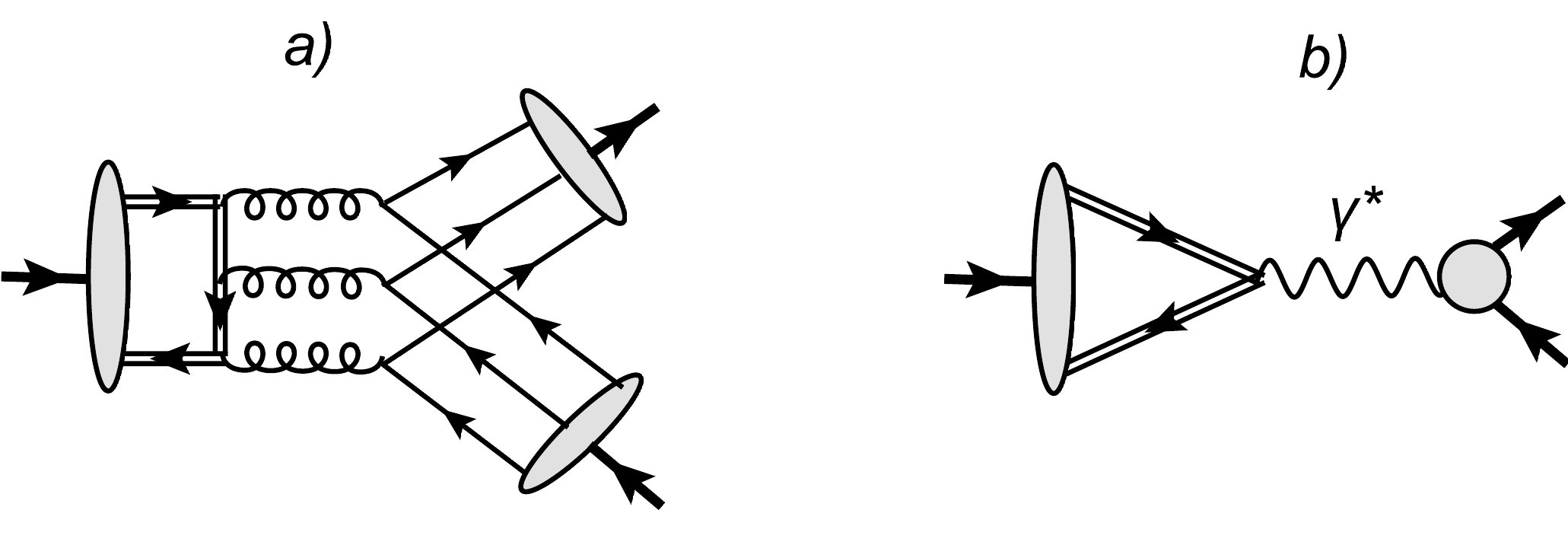}
\caption{ Hadronic $a)$ and electromagnetic $b)$ contributions to the decay amplitudes. }
\label{figure}
\end{figure}

 The decay amplitude  can be written as \cite{Kivel:2019wjh}
\begin{equation}
M_{B}=\bar{N}(k)\left\{  A_{1}^{B}%
~\setbox0=\hbox{$\epsilon$}\dimen0=\wd0\setbox1=\hbox{/}\dimen1=\wd1\ifdim\dimen0>\dimen1\rlap{\hbox to
\dimen0{\hfil/\hfil}}\epsilon\else\rlap{\hbox to
\dimen1{\hfil$\epsilon$\hfil}}/\fi_{\psi}+A_{2}^{B}\left(  \epsilon_{\psi}\right)
_{\mu}(k^{\prime}+k)_{\nu}\frac{i\sigma^{\mu\nu}}{2m_{N}}\right\}
V(k^{\prime})~, \label{def:M}%
\end{equation}
where $\epsilon_{\psi}$ is the charmonium polarisation vector, $\bar{N}(k)$
and $V(k^{\prime})$ denote baryon and antibaryon spinors, respectively. The
scalar amplitudes $A_{i}^{B}$ can be expanded within the effective field theory
framework in small heavy quark velocity $v^{2}$ (NRQCD) and in powers
$\lambda^{2}\sim\Lambda/m_{c}$ (collinear factorisation), where $\Lambda$ is
the typical hadronic scale.  In this work we consider the leading-oder
contribution with respect to small velocity $v$ and compute the next-to-leading power (NLP) correction
to the amplitude $A_{1}^B$, which is provided by the diagrams in Fig.$\ $\ref{figure}$a)$.

The expansion of the amplitudes $A_{i}^B$ in powers of $\lambda$ is associated
with the power counting of the collinear matrix elements, which describe the
overlap with outgoing baryons. The structure of the collinear expansion allows one to write the following expansions%
\begin{equation}
A_{1}^{B}=A_{1}^{\text{lp}}+A_{1}^{\text{nlp}}+\mathcal{O}(\lambda
^{14}),\ \ A_{2}^{B}=A_{2}^{\text{lp}}+\mathcal{O}(\lambda^{14}).
\end{equation}
The leading-power (LP) contribution scales as $A_{1}^{\text{lp}}\sim\mathcal{O}(\lambda^{8})$ because it is described by the two light-cone
matrix elements, each of which is of order $\mathcal{O}(\lambda^{4})$.  The NLP
correction $A_{1}^{\text{nlp}}$ includes collinear operators, which  are
suppressed by the overall factor $\lambda^{4}$ and therefore  $A_{1}^{\text{nlo}%
}$ $\sim\mathcal{O}(\lambda^{12}).$ For the second amplitude one finds
$A_{2}^{\text{lp}}\sim\mathcal{O}(\lambda^{12})\footnote{The additional power
$\lambda^{2}$ appears due to the factor $1/m_{B}$ in the definition in
Eq.(\ref{def:M}).}$.  

The expansion in powers of  $1/m_c$ mixes up  power corrections from the expansions of the scalar amplitudes $A_{i}^B$ and from the baryon spinors.  
The calculation of the amplitude in the factorisation framework  yields the result  in the following form
\begin{equation}
M_{B}=\left\{  \delta A_{1}^{\text{lp}}+\delta A_{1}^{\text{nlp}}\right\}
\bar{N}_{\bar{n}}\Dsl \epsilon _{\psi}^{\bot}V_{n}+\bar{N}_{\bar{n}%
}\boldsymbol{1}V_{n}~\delta A_{2}^{\text{lp}}, \label{Mope}%
\end{equation}
where the coefficients  $\delta A_{i}$ scale as
\begin{equation}
\delta A_{1}^{\text{lp}}\sim\mathcal{O}(\lambda^{8}),\ \delta A_{1}%
^{\text{nlp}}\ \sim\delta A_{2}^{\text{lp}}\sim\mathcal{O}(\lambda^{12}).
\end{equation}
The expressions for  $\delta A_{i} $ are given by the convolution integrals of the hard kernels
with the different baryon DAs. 

In order to perform the interpretation of $\delta A_{i} $ in terms of the scalar amplitudes $A_{1,2}^B$ one needs
to perform the expansion of the {\it rhs} of Eq.(\ref{def:M}). This gives
\begin{align}
M_{B}  &  \simeq\bar{N}_{\bar{n}}\Dsl \epsilon _{\psi}^{\bot}V_{n}\left(
A_{1}^{\text{lp}}+A_{1}^{\text{nlp}}+A_{2}^{\text{lp}}-A_{1}^{\text{lp}}%
\frac{m_{B}^{2}}{k_{+}k_{-}^{\prime}}\right) \nonumber\\
&  +\frac{1}{4m_{B}}\left\{  \left(  \epsilon_{\psi}\cdot n\right)
k_{-}^{\prime}~-k_{+}\left(  \epsilon_{\psi}\cdot\bar{n}\right)  \right\}
~\bar{N}_{\bar{n}}\boldsymbol{1}V_{n}~\left(  A_{2}^{\text{lp}}+\frac
{4m_{B}^{2}}{k_{+}k_{-}^{\prime}}A_{1}^{\text{lp}}\right)  , \label{Mexp}%
\end{align}
where we use that $k_{+}\sim k_{-}^{\prime}\sim m_{c}$. Comparing
Eqs.(\ref{Mope}) and (\ref{Mexp}) one finds
\begin{equation}
A_{1}^{\text{lp}}=\delta A_{1}^{\text{lp}},\ \ A_{2}^{\text{lp}}=\ \delta
A_{2}^{\text{lp}}-\frac{4m_{B}^{2}}{k_{+}k_{-}^{\prime}}A_{1}^{\text{lp}}%
\end{equation}%
\begin{equation}
\ A_{1}^{\text{nlp}}=\delta A_{1}^{\text{nlp}}-\ \delta A_{2}^{\text{lp}%
}+\frac{5m_{B}^{2}}{k_{+}k_{-}^{\prime}}\delta A_{1}^{\text{lp}}%
\end{equation}

The expressions for observables have a compact analytical form if we rewrite them in terms of the linear
combinations $\mathcal{G}_{M}$ and $\mathcal{G}_{E}$, which read
\begin{equation}
\mathcal{G}_{M}^{B}=A_{1}^{B}+A_{2}^{B}\simeq\delta A_{1}^{\text{lp}}\left(
1+\frac{m_{B}^{2}}{M_{\psi}^{2}}\right)  +\delta A_{1}^{\text{nlp}%
},\ \ \ \mathcal{G}_{E}^{B}=A_{1}^{B}+\frac{M_{\psi}^{2}}{4m_{B}^{2}}A_{2}%
^{B}\simeq\frac{M_{\psi}^{2}}{4m_{B}^{2}}\delta A_{2}^{\text{lp}}.
\label{calGME}%
\end{equation}

Notice that the total power correction to $A_{1}^{B}$ also includes the simple
kinematical term $\delta A_{1}^{\text{lo}}\times m_{B}^{2}/M_{\psi}^{2}$.

The LP term $\delta A_{1}^{\text{lo}}$ is calculated long time ago, see {\it e.g.} Refs.\cite{ Brodsky:1981kj, Chernyak:1987nv}.  
The contribution $\delta A_{2}^{\text{lo}}$ is
obtained in Ref.\cite{Kivel:2019wjh}.  The correction  $\delta A_{1}^{\text{nlo}}$ will be considered  for the first time.  
In this calculation we only consider the contributions
of light-cone 3-quark operators and do not  include the various quark-gluon
matrix elements.  

We use the same technique as  in Ref.\cite{Kivel:2019wjh}.  Here we  just provide a schematic
description and  discuss some important points.  The different contributions
in $\delta A_{1}^{\text{nlo}}$ can be divided into two groups, which  are given by the product
of the two twist-4 operators $ \left[  \chi_{1}\chi_{2}\chi_{3}\right]  _{\text{tw4}}\otimes\left[  \bar
{\chi}_{1^{\prime}}\bar{\chi}_{2^{\prime}}\bar{\chi}_{3^{\prime}}\right]
_{\text{tw4}}$ or by the product twist-3 and twist-5 operators $\left[  \chi_{1}\chi_{2}\chi_{3}\right]
_{\text{tw3}}\otimes\left[  \bar{\chi}_{1^{\prime}}\bar{\chi}_{2^{\prime}}%
\bar{\chi}_{3^{\prime}}\right]  _{\text{tw5}}$.  Remind,  $\chi_{1}$ and $\bar{\chi}_{1^{\prime}}$ denote the  collinear quark
fields, see Eq.(\ref{def:chi}). Let us also remind  that $\chi_{1}\equiv\chi_{\alpha_{1}}$ and $\bar{\chi}_{1^{\prime}}\equiv
\bar{\chi}_{\alpha_{1}^{\prime}}$.  Each group can be further divided
into two subgroups with respect to Dirac projections: $[\chi\Gamma_{even}%
\chi]\chi_{3}$ and $[\chi\Gamma_{odd}\chi]\chi_{3}$, where  $\Gamma_{even/odd}$
denote the chiral-even or chiral-odd Dirac matrices.  The explicit definitions of the
various operators and their matrix elements are discussed in Sec.\ref{sec:aux}. 

The contribution of the sum of diagrams  in Fig.$\,$\ref{figure}$\, a)$ to $\delta A_{1}^{\text{nlo}}$ can be written as
\begin{equation}
\delta A_{1}^{\text{nlo}}= f_{\psi}\int Dx_{i}\int Dy_{i}\ \left[
P(x_{i})\right]  _{123}\left[  \bar{P}(y_{i})\right]  _{1^{\prime}2^{\prime
}3^{\prime}}\frac{1}{4}\text{Tr}\left[  (1-\Dsl \omega )\Dsl \epsilon _{\psi}\
 \sum_i D^i_{\left(  123\right)  \left(  1^{\prime}2^{\prime}3^{\prime
}\right)  }\right]  , \label{dA1_PPD}%
\end{equation}
where $D^i_{\left(  123\right)  \left(  1^{\prime}2^{\prime}3^{\prime}\right)}$ 
denotes the analytical expression  of the diagram $i$ in momentum space.
It can be written as
\begin{equation}
\ D_{\left(  123\right)  \left(  1^{\prime}2^{\prime}3^{\prime}\right)  }%
^{i}=\ D_{Q}^{(i)\alpha\beta\gamma}(k_{i},k_{i}^{\prime},k,k^{\prime
})\ (ig_{s})^{3}\left[  \gamma^{\alpha}\right]  _{11^{\prime}} [
\gamma^{\beta}]  _{22^{\prime}}\left[  \gamma^{\gamma}\right]
_{33^{\prime}}.
\end{equation}
Here the matrices $\gamma^{\alpha}\otimes\gamma^{\beta}\otimes\gamma^{\gamma}$
corresponds to the light quark-antiquark vertices, $D_{Q}^{(i)\alpha
\beta\gamma}$ denotes the remnant part of the diagram, the trace in
Eq.(\ref{dA1_PPD}) is taken with respect of Dirac indices of the heavy quark fermion line. For
simplicity,  the colour structure of  diagrams is not shown. The coupling $f_{\psi}$
describes the matrix element in NRQCD and can be related with the radial wave
function at the origin%
\begin{equation}
f_{\psi}=\sqrt{2M_{\psi}}\sqrt{\frac{3}{2\pi}}~R_{10}(0). \label{fpsi}%
\end{equation}
The definition of the  NRQCD matrix element is the same as  in Ref.\cite{Kivel:2019wjh}.  The expression $(1-\Dsl{ \omega} )\Dsl {\varepsilon}_\psi$ in the trace in Eq.(\ref{dA1_PPD}) represent the Dirac part of the projector on the NRQCD matrix element,  $\omega=(1,0,0,0)$ and $\varepsilon_\psi^\mu$ denote the velocity and the polarisation vector of  charmonium,  respectively. 

The operators $P(x_{i})$ and $\bar{P}(y_{i})$ describe the projections on the matrix elements of the light-cone 3-quark operators for
baryon and antibaryon, respectively. For the twist-3 case these  operators
depend on the nucleon spinors, Dirac matrices and twist-3 DAs
and can be
obtained just from the definition in Eq.(\ref{Metw3}). For instance, for the
matrix element $\left\langle 0\right\vert \mathcal{O}_{123}\left\vert
B(k)\right\rangle _{\text{tw3}}$ one easily obtains
\begin{align}
P_{\text{tw3}}(x_{i})  &  =~\frac{1}{8}k_{+}\left[  \nbs C\right]
_{12}~\left[  \gamma_{5}N_{\bar{n}}\right]  _{3}V_{1}^{B}(x_{i})+\frac{1}%
{8}k_{+}\left[  \nbs \gamma_{5}C\right]  _{12}~\left[  N_{\bar{n}}\right]
_{3}\ A_{1}^{B}(x_{i})\\
&  +\frac{1}{8}k_{+}\left[ \nbs\gamma_{\bot}^{\alpha}C\right]
_{12}~\left[  \gamma_{\bot}^{\alpha}\gamma_{5}N_{\bar{n}}\right]  _{3}%
T_{1}^{B}(x_{i}).
\end{align}
For the higher twist matrix elements these projections also include the
derivatives with respect to  quark and antiquark momenta $k_{i} $
and $k_{i}^{\prime}$, which are assumed to be off-shell $k_{i}^{2}\neq0$,
$k_{i}^{\prime}\neq0$. Only after the differentiations one can neglect the
small momentum components and set $k_{i}\simeq x_{i}k_{+}\bar{n}/2$ and $k_{i}^{\prime
}\simeq y_{i}k_{-}^{\prime}n/2$.  The derivation of the twist-4 projections
 have been considered in Ref.\cite{Kivel:2019wjh} and here we follow the same
 technique. The analytical results for the different projectors are given in Appendix \ref{AppC}.
Substituting the various projectors in Eq.(\ref{dA1_PPD}) one obtains the  expression,  which allows one
to calculate  the hard kernels. The  analytical calculations were
performed using the package FeynCalc \cite{Kublbeck:1992mt,Shtabovenko:2020gxv}.

The obtained result can be written as
\begin{equation}
\delta A_{1}^{\text{nlo}}=A_{0}^{B}\frac{m_{B}^{2}}{m_{c}^{2}}\left\{
J_{44}^{B}+J_{35}^{B}\right\}  , \label{dA1nlo=A0}%
\end{equation}
where the normalisation coefficient $A_{0}$ read
\begin{equation}
A_{0}^{B}=\frac{f_{\psi}}{m_{c}^{2}}\frac{f_{B}^{2}}{m_{c}^{4}}(\pi\alpha
_{s})^{3}\frac{10}{81}.
\end{equation}
The dimensionless integrals $J_{ij}^{B}$ have\ subscripts \textquotedblleft%
$44$\textquotedblright\ and \textquotedblleft$35$\textquotedblright\ , which
are related with the structure of the collinear matrix elements: twist-4$\times
$twist-4 and twist-3$\times$twist-5.  Each  integral in Eq.(\ref{dA1nlo=A0}) is given by the sum of  simpler
integrals $J[X,Y]$ where $X$ and $Y$ denote the DAs, which are included in the
integrand%
\begin{equation}
J[X,Y]=\frac{1}{f_{B}^{2}}\int Dx_{i}\ X(x_{i}) \int Dy_{i}Y(y_{i})\ K_{XY}(x_{i},y_{i}), \label{def:J}%
\end{equation}
where $K_{XY}(x_{i},y_{i})$ is the universal hard kernel. The calculation
of the traces in Eq.(\ref{dA1_PPD}) gives the sum of the collinear integrals, which
can be  simplified using the symmetry properties of the DAs and the
hard kernels with respect to interchange $x_{1}\leftrightarrow x_{2}$ and
$x_{i}\leftrightarrow y_{i}$. Below we provide the analytical expressions for the integrals in Eq.(\ref{dA1nlo=A0}), where such
simplifications are taken into account. Let us write each term in
Eq.(\ref{dA1nlo=A0}) as
\begin{equation}
J_{ij}^{B}=\left.  J_{ij}^{B}\right\vert _{\text{even}}+\left.  J_{ij}%
^{B}\right\vert _{\text{odd}}.
\end{equation}
The subscript \textquotedblleft
even/odd\textquotedblright\ corresponds to the chiral-even and chiral-odd
subsets of the DAs.  Then%
\begin{equation}
\left.  J_{44}^{B}\right\vert _{\text{even}}=J[\mathcal{V}^B_{1},\mathcal{V}^B
_{1}]+J[\mathcal{V}^B_{2},\mathcal{V}^B_{1}]+J[\mathcal{A}^B_{1},\mathcal{A}^B
_{1}]+J[\mathcal{A}^B_{2},\mathcal{A}^B_{1}]+J[\mathcal{A}^B_{1},\mathcal{V}^B
_{1}]+J[\mathcal{A}^B_{2},\mathcal{V}^B_{1}].
\label{J44Beven}
\end{equation}%
\begin{equation}
\left.  J_{44}^{B}\right\vert _{\text{odd}}=J[\mathcal{T}^B_{21}+\mathcal{T}^B
_{41},\mathcal{T}^B_{21}+\mathcal{T}^B_{41}]+J[\mathcal{T}^B_{21}-\mathcal{T}^B_{41},\mathcal{T}^B_{21}-\mathcal{T}^B_{41}]
+J[\mathcal{T}^B_{22}+\mathcal{T}^B_{42},\mathcal{T}^B_{21}+\mathcal{T}^B_{41}].
\label{J44Bodd}
\end{equation}%
\begin{eqnarray}
&&\left.  J_{35}^{B}\right\vert _{\text{even}}= J[V^B_{5},V^B_{1}]+J[V^B_{5},A^B_{1}]+J[A^B_{5},A^B_{1}]+J[A^B_{5},V^B_{1}]
\\ &&
\phantom{J[V^B_{5},V^B_{1}]}
+J_{Z}[V^B_{Z},V^B_{1}]+J_{Z}[A^B_{Z},V^B_{1}]+J_{Z}[A^B_{Z},A^B_{1}]+J_{Z}[V^B_{Z},A^B_{1}].
\label{J35Beven}
\end{eqnarray}
\begin{equation}
\left.  J_{35}\right\vert _{\text{odd}}=J_Z[T^B_Z,T^B_1]. 
\label{J35Bodd}
\end{equation}
Each integral on the {\it rhs} of Eqs.(\ref{J44Beven})-(\ref{J35Bodd})  is IR-finite, {\it i.e.} does not have endpoint
singularities.   The integrals $J_Z[X,Y]$ are given by the sum of singular integrals and read%
\begin{eqnarray}
J_{Z}[V^B_{Z},V^B_{1}]&=&J[V^B_{1xx},V^B_{1}]+J[V^B_{1yy},V^B_{1}]+J[V^B_{1xy},V^B_{1}
]+(1-\delta_{B\Lambda})J[V^B_{1},V^B_{1}], \label{JVZV1}%
\\
J_{Z}[V^B_{Z},A^B_{1}]&=&J[V^B_{1xx},A^B_{1}]+J[V^B_{1yy},A^B_{1}]+J[V^B_{1xy},A^B_{1}
]+(1-\delta_{B\Lambda})J[V^B_{1},A^B_{1}], \label{JVZA1}%
\\
J_{Z}[A^B_{Z},V^B_{1}]&=&J[A^B_{1xx},V^B_{1}]+J[A^B_{1yy},V^B_{1}]+J[A^B_{1xy},V^B_{1}
]+\delta_{B\Lambda}J[A^B_{1},V^B_{1}], \label{JAZV1}%
\\
J_{Z}[A_{Z},A_{1}]&=&J[A^B_{1xx},A^B_{1}]+J[A^B_{1yy},V^B_{1}]+J[A^B_{1xy},V^B_{1}%
]+\delta_{B\Lambda}J[A^B_{1},A^B_{1}], \label{JAZA1}%
\\
J_Z[T^B_Z,T^B_1]&=&  J[T^B_{0xx},T^B_{1}]+J[T^B_{0yy},T^B_{1}]+J[T^B_{0xy},T^B_{1}]+J[\tilde{T}^B_{2xy},T^B_{1}]
\nonumber  \\
&& \phantom{J[T^B_{0xx},T^B_{1}]+J[T^B_{0yy},T^B_{1}]+J[T^B_{0xy},T^B_{1}}
+(1-\delta_{B\Lambda})J[T^B_{1},T^B_{1}].
\label{JTZT1}
\end{eqnarray}
where $\delta_{B\Lambda}=1$ if $B=\Lambda$ and $0$ otherwise. Notice that
expressions in Eqs.(\ref{JVZV1}) - (\ref{JTZT1}) include the integrals, which have  twist-3 DAs only. These are contributions, which
appears from twist-5 matrix elements with the total derivatives, see
discussion in Appendix \ref{AppC}. The Kronecker  symbol $\delta_{B\Lambda}$ appears in
Eqs.(\ref{JVZV1}) - (\ref{JAZA1}) because we only use asymptotic twist-3 and twist-4 DAs  for the evaluation of  the twist-5 DAs.  The integrals with the twist-3 DAs in Eqs.(\ref{JVZV1}) - (\ref{JTZT1})  are  singular and  they are included
 in the given expressions because they are required for the
cancellation of the endpoint singularities in the  integrals $J_{Z}[X,Y]$.

The analytical expressions for the hard kernels $K_{XY}(x_{i},y_{i})$, as defined  in Eq.(\ref{def:J}), are given in the
Appendix \ref{app:d}.  It has been  analytically verified  that these kernels  correctly describe  the  relations between the amplitudes in the $SU(3)$ limit. 
 The corresponding  convolution integrals can be relatively
easily  calculated.  Substituting the explicit expressions for DAs with
arbitrary moments and summing the singular integrands gives analytical
expressions with the well defined four-dimensional integrals. These integrals can be
easily computed numerically using the standard {\it mathematica} packages for numerical integrations.

\section{Phenomenology}
\label{phen}

A systematic numerical analysis of existing data is still very challenging due to
various theoretical and experimental uncertainties. Therefore, the main goal of the following discussion is to understand how well the existing data can be described using plausible assumptions about various nonperturbative parameters.

 In this analysis  we  take into account the electromagnetic amplitudes, which are 
shown by the diagram in Fig.\ref{figure}$\, (b)$. With these contributions the decay amplitudes read
\begin{equation}
M[J/\psi\rightarrow B\bar{B}]=\bar{N}\Dsl \epsilon _{\psi}V~\left(  A_{1}%
^{B}+F_{1}^{B}\right)  +\frac{1}{2m_{N}}\epsilon_{\psi}^{\mu}(k+k^{\prime
})^{\nu}~\bar{N}i\sigma^{\mu\nu}V~\left(  A_{2}^{B}+F_{2}^{B}\right)  ,
\end{equation}
where $F_{1,2}^{B}$ are the time-like e.m. form factors. 
Expression for the width reads%
\begin{equation}
\Gamma\lbrack J/\psi\rightarrow B\bar{B}]=\frac{M_{\psi}\beta}{12\pi
}|\mathcal{\tilde{G}}_{M}^{B}|^{2}\left(  1+\frac{2m_{B}^{2}}{M_{\psi}^{2}%
}\gamma_{B}^{2}\right)  ,\ \ \label{width}%
\end{equation}
where $\beta=\sqrt{1-4m_{B}^{2}/M_{\psi}^{2}}$,
\begin{equation}
~\mathcal{\tilde{G}}_{M}^{B}=\mathcal{G}_{M}^{B}+\zeta G_{M}^{B}%
,~~\ \mathcal{\tilde{G}}_{E}^{B}=\mathcal{G}_{E}^{B}+\zeta G_{E}%
^{B},\ \ \gamma_{B}\equiv\left\vert \mathcal{\tilde{G}}_{E}^{B}\right\vert
/|\mathcal{\tilde{G}}_{M}^{B}|.
\label{zeta}
\end{equation}
The amplitudes $\mathcal{G}_{M}$ and $\mathcal{G}_{E}$ are defined in
Eq.(\ref{calGME}), the time-like Sachs form factors $G^B_{E,M}$ read%
\begin{equation}
G_{M}^{B}=F_{1}^{B}+F_{2}^{B},\ \ G_{E}^{B}=F_{1}^{B}+\frac{M_{\psi}^{2}%
}{4m_{B}^{2}}F_{2}^{B}.
\end{equation}
The coupling $\zeta$ in Eq.(\ref{zeta}) is defined as
\begin{equation}
\zeta=  4\pi\alpha_{em}\  e_{c}\frac{f_{\psi}}{M_{\psi}^{2}}.
\end{equation}

Therefore, the observables  depend on the  time-like form factors of  baryons,
which are still not well known. For the given value of the $q^{2}=M_{\psi}^{2}$
 their values are still depend on the large non-perturbative corrections and
can not be sufficiently well estimated within the QCD factorisation framework. This
introduces additional uncertainty into the phenomenological analysis.

The  hadronic amplitudes calculated  to this accuracy within the factorisation
framework are real.  The e.m. FFs have real and imaginary parts, therefore to
this accuracy one finds
\begin{equation}
\left\vert ~\mathcal{\tilde{G}}_{M}^{B}\right\vert ^{2}/\left\vert
\mathcal{G}_{M}^{B}\right\vert ^{2}\simeq1+2\zeta\left\vert G_{M}%
^{B}\right\vert \cos\varphi_{M}/\left\vert \mathcal{G}_{M}^{B}\right\vert
+\zeta^{2}\left\vert G_{M}^{B}\right\vert ^{2}/\left\vert \mathcal{G}_{M}%
^{B}\right\vert ^{2},
\end{equation}%
\begin{equation}
\left\vert ~\mathcal{\tilde{G}}_{E}^{B}\right\vert ^{2}/\left\vert
\mathcal{G}_{E}^{B}\right\vert ^{2}\simeq1+2\zeta\left\vert G_{E}%
^{B}\right\vert \cos\varphi_{E}/\left\vert \mathcal{G}_{E}^{B}\right\vert
+\zeta^{2}\left\vert G_{E}^{B}\right\vert ^{2}/\left\vert \mathcal{G}_{E}%
^{B}\right\vert ^{2},
\end{equation}
where $\varphi_{M,E}$ are phases of the e.m. form factors.  Let us 
rewrite the ratio as%
\begin{equation}
\gamma_{B}=\gamma_{B}^{g}\mathcal{R}_{em}^{B},
\end{equation}
with%
\begin{equation}
\gamma_{B}^{g}=\frac{\left\vert \mathcal{G}_{E}^{B}\right\vert }{\left\vert
\mathcal{G}_{M}^{B}\right\vert },\ \ \ \mathcal{R}_{em}^{B}=\sqrt
{\frac{1+2\zeta\left\vert G_{E}^{B}\right\vert /\left\vert \mathcal{G}_{E}%
^{B}\right\vert \cos\varphi_{E}+\zeta^{2}\left\vert G_{E}^{B}\right\vert
^{2}/\left\vert \mathcal{G}_{E}^{B}\right\vert ^{2}}{1+2\zeta\left\vert
G_{M}^{B}\right\vert /\left\vert \mathcal{G}_{M}^{B}\right\vert \cos
\varphi_{M}+\zeta^{2}\left\vert G_{M}^{B}\right\vert ^{2}/\left\vert
\mathcal{G}_{M}^{B}\right\vert ^{2}}},
\end{equation}
where the factor $\mathcal{R}_{em}^{B}$ defines how  the measured value $\gamma_{B}$  depends on the interference
with e.m. amplitude. 

At present, the time-like baryon FFs are little known. The best experimental data for the proton can be found in Ref.\cite{BESIII:2019hdp}. For
other baryons one can only find the data for the effective FF  obtained
from the analysis of the e.m. cross section under assumption $G_{\text{eff}}^{B}=|G_{M}^{B}|=|G_{E}^{B}|$, see  Refs.$\,$\cite{BaBar:2007fsu,BESIII:2017hyw,BESIII:2020ktn,BESIII:2020uqk,BESIII:2021dfy,BESIII:2021aer,BESIII:2021rkn}. Therefore an
estimation of $\mathcal{R}_{em}^{B}$ can not be done without assumptions. In
the following we assume that $0\ll\cos\varphi_{M,E}<1$\footnote{The small
value of $\cos\varphi_{M,E}\simeq0$ implies that $G_{M,E}$ dominates by imaginary
part. We assume that such a scenario is unlikely.}. The value of $|G_{M}^{p}|$
we take from data in Ref.\cite{BESIII:2019hdp}, the maximal value $|G_{E}^{p}|$ is taken to be $|G_{E}^{p}|\leq 3.29\times 10^{-2}$ as it follows  from analysis of data in Ref.\cite{BESIII:2019hdp}. In other cases we assume that $|G_{M}^{B}|\approx$
$G_{\text{eff}}^{B}$.  The  value of   $\cos\varphi_{M}$ can be estimated
from  the data for the width (\ref{width}), where $\gamma_{B}$ is fixed by the experimental value.  
An estimate of  $G_{E}^{B}$ is the most challenging point.
We assume  that $\left\vert G_{E}^{B}\right\vert \leq$1.5 $G_{\text{eff}}^{B}$, 
where the numerical coefficient $1.5$ is our  conservative estimate. Numerical value of $\zeta
^{2}\left\vert G_{E}\right\vert ^{2}/\left\vert \mathcal{G}_{E}\right\vert^{2}$ is quite small and can be neglected, which yields
\begin{equation}
\mathcal{R}_{em}^{B}\approx\sqrt{\frac{1+2\zeta\left\vert G_{E}^{B}\right\vert
/\left\vert \mathcal{G}_{E}^{B}\right\vert \cos\varphi_{E}}{1+2\zeta\left\vert
G_{M}^{B}\right\vert /\left\vert \mathcal{G}_{M}^{B}\right\vert \cos
\varphi_{M}+\zeta^{2}\left\vert G_{M}^{B}\right\vert ^{2}/\left\vert
\mathcal{G}_{M}^{B}\right\vert ^{2}}}.
\end{equation}
Therefore,  assuming the estimate
\begin{equation}
0<\left\vert G_{E}^{B}\right\vert \cos\varphi_{E}\leq\ 1.5 G_{\text{eff}}^{B},
\end{equation}
we get  an interval, which is considered as an estimate of the possible numerical effect provided 
by $\mathcal{R}_{em}^{B}$.


Let us now discuss  the non-perturbative input, which is used for calculation
of the hadronic amplitudes. Recall that we consider only  three-quark DAs.  It is convenient  to describe the non-perturbative input
 in terms of basic DAs as discussed in the Sec.\ref{T5DA}. Analytical expressions for the corresponding models  are  described in Appendix \ref{DAmodels}.  The  values of  DAs parameters are summarised in Table \ref{DAparameters}.
\begin{table}[th]
\caption{The parameters, which define the twist-3 and twist-4 models of the
baryon DAs (upper and bottom tables, respectively). All values are given in units of $10^{-2}$ at
the scale $\mu^{2}=1.5~$GeV$^{2}$. \newline}%
\label{DAparameters}%
\centering%
\begin{tabular}
[c]{|l|l|l|l|l|l|l|l|l|l|}\hline
$B$ & $f_{B},$ GeV$^{2}$ & $\phi_{10}$ & $\phi_{11}$ & $\phi_{20}$ &
$\phi_{21}$ & $\phi_{22}$ & $f_{\bot}^{B},$ GeV$^{2}$ & $\pi_{10}^{B}$ &
$\pi_{11}^{B}$\\\hline
$N$ & $0.54$ & $5.1$ & $5.15$ & $7.82$ & $-2.84$ & $17.95$ &
$-$ & $-$ & $-$\\\hline
$\Lambda$ & $0.60$ & $10.9$ & $4.42$ & $0$ & $0$ & $0$ & $-$ &
$3.83$ & $-$\\\hline
$\Sigma$ & $0.51$ & $1.90$ & $4.18$ & $0$ & $0$ & $0$ &
$0.53$ & $-$ & $-1.90$\\\hline
$\Xi$ & $0.62$ & $6.10$ & $-0.25$ & $0$ & $0$ & $0$ &
$0.61$ & $-$ & $7.10$\\\hline
\end{tabular}
\\[4mm]%
\begin{tabular}
[c]{|c|c|c|c|c|c|c|c|c|}\hline
$B$ & $\lambda_{1}^{B}$, GeV$^{2}$ & $\eta_{10}^{B}$ & $\eta_{11}^{B}$ &
$\lambda_{\bot}^{B}$, GeV$^{2}$ & $\zeta_{10}^{B}$ & $\zeta_{11}^{B}$ &
$\lambda_{2}$, GeV$^{2}$ & $\xi_{10}$\\\hline
$N$ & $-2.81$ & $-4.00$ & $16.34$ & $-$ & $-$ & $-$ &
$5.29$ & -27.24\\\hline
$\Lambda$ & $-3.93$ & $-4.00$ & $18.86$ & $-4.87$ &
$-4.00$ & $-$ & $9.21$ & $-$\\\hline
$\Sigma$ & $-4.29$ & $-4.62$ & $21.99$ & $-$ & $-$ & $20.95$ &
$7.93$ & $-37.56$\\\hline
$\Xi$ & $-4.64$ & $-2.05$ & $13.62$ & $-$ & $-$ & $15.71$ &
$9.27$ & $-27.24$\\\hline
\end{tabular}
\end{table}

The parameters of nucleon chiral-even DAs correspond to the model ABO-I from
Ref.\cite{Anikin:2013aka}.  This set of parameters have beed obtained from the fit of the space-like  electromagnetic FF data using the light-cone sum rules.   The  moments $\lambda_{2}$ and $\xi_{10}$   are
evaluated  in Ref.\cite{Braun:2000kw} using the QCD sum rule approach.  Some twist-3
moments for different baryons are also obtained using QCD sum rules in Ref.\cite{Chernyak:1987nv}.
Recently,  the different twist-3 and twist-4 moments have  also been computed on the lattice, see
Ref.\cite{Bali:2019ecy}. We use these results in order to fix parameter values in DA models.
 In particular, the values for the  moments $\phi_{10}^{B}, \phi_{11}^{B}$, $\lambda_{1}^{B}$,$ \lambda_{\bot
}^{B}$, $\lambda_{2}^{B}, \pi_{10}^{B}$ and $\pi_{11}^{B}$ ($B\neq N$) are
fixed according to data in  Ref.\cite{Bali:2019ecy}. 

For the nucleon twist-3 normalisation coupling we take the  value $f_{N}(1$\, GeV$)=5.5\times10^{-3}\ $GeV$^{2}$, which is in agreement with the sum rule estimate $f_{N}(1$\, GeV$)=(5.3\pm0.5)\times10^{-3} $GeV$^{2}$ obtained in Ref.\cite{Chernyak:1987nv}. 
The other  twist-3 normalisation couplings
$f_{B}$  are fixed from the fit of data. The values obtained in Ref.\cite{Bali:2019ecy}  were used as the first estimate and then changed if this improves the description of the data.  Therefore the resulting  value $f_{\Lambda}$  in Tab.\ref{DAparameters}  is  larger than the estimates in Refs.\cite{Bali:2019ecy, Chernyak:1987nv}, which read $0.50$ and $0.48$, respectively.\footnote{For a convenience,  these values are given in the same notation as in the Tab.\ref{DAparameters} . }  We also find a relatively small difference for  values of  $f_{\Sigma}$ ($0.54$ \cite{Bali:2019ecy} and $0.48$ \cite{Chernyak:1987nv}) and $f_{\bot}^{\Xi}$  ($0.64$ \cite{Bali:2019ecy} and $0.50$ \cite{Chernyak:1987nv}).  The values for  $f_{\bot}^{\Sigma}$  and $f_{\Xi}$  are the same as in  Ref.\cite{Bali:2019ecy}.

 The values of the twist-4 moments $\eta_{10}^{B},\eta_{11}^{B}$,
$\zeta_{10}^{B}, \zeta_{11}^{B}$ and $\xi_{10}^{B}$ are fixed from the data, assuming that
their values are not significantly  different  from the nucleon ones because of
approximate $SU(3)$ symmetry. 

In our calculation we fix relatively low renormalisation scale $\mu^{2}=1.5\,$GeV$^{2}$.  
This  gives for the strong coupling $\alpha_{s}(\mu
^{2})=0.35$.   The twist-4 DAs include terms with the quark masses, see Eqs.(\ref{Vical})-(\ref{rel:T22mT42}).  For these terms we set $m_{u}\approx
m_{d}\approx0$,  $m_{s}(1.5$GeV$^{2})=105\,$MeV. The value of $J/\psi$ total width $\Gamma_{tot}=93\,$keV  is taken from Ref.\cite{Zyla:2020zbs} .

For the value $R_{10}(0)$ in Eq.(\ref{fpsi}) \ we use the estimate obtained
for the Buchmuller-Tye potential \cite{Eichten:1995ch}
\begin{equation}
\left\vert R_{10}(0)\right\vert ^{2}\simeq0.81\text{GeV}^{3}, \label{R10BT}%
\end{equation}
which implies for the value of charm quark mass $m_{c}=1.48\,$GeV.

One of the elements of the current analysis is the estimation of the power-law correction
to  $\mathcal{G}_{M}^{B}$, see Eq.(\ref{calGME}).  In Table \ref{nlp} we show the numerical results obtained for these corrections. 
The total effect of the power corrections is quite moderate due to the partial cancellation in the total sum, see Eq.(\ref{calGME}). 
Therefore  the total  numerical effect from the next-to-leading power correction  varies from
$1.7\%$ for $\Xi$  up to $16\%$ for nucleon. 
\begin{table}[th]
\caption{ Numerical value of the power corrections for different baryons, see Eq.(\ref{calGME}).}
\label{nlp}%
\centering%
\begin{tabular}
[c]{|l|l|l|l|l|}\hline
& $N$ & $\Lambda$ & $\Sigma$ & $\Xi$\\\hline
$m_{B}^{2}/M_{\psi}^{2}\times100\%$ & $9.2$ & $13.0$ & $14.8$ & $18.1$\\\hline
$\delta A_{1}^{\text{nlo}}/\delta A_{1}^{\text{lo}}\times100\%$ & $-25.4$ &
$-16.4$ & $-23.1$ & $-19.8$\\\hline
\end{tabular}
\end{table}
  
The numerical results for branching fractions and  ratios $\gamma_{B}$ are
presented in Table \ref{results}.  The obtained values of $\gamma_B$ include the error bars, which describe the  estimate of the  uncertainty  occurring from  to the interference with the electromagnetic FF's as described by the coefficient $R_{em}$. 
\begin{table}[th]
\caption{ The  numerical  results for  branching ratios and for the ratio $\gamma_B$ in comparison with 
experimental data. The values of the electromagnetic FF's $|G^B_M|$ and the branching rations  are given in units of $10^{-2}$ and $10^{-3}$, respectively.   For more explanations see discussion  in the text. \newline\phantom{xxx} }%
\label{results}
\begin{center}%
\begin{tabular}
[c]{|c|c|c|c|c|c|c|}\hline
\phantom{$\int^C_{C_C}$} & $|G^{B}_{M}|$ & $\cos\phi_{M}$ &
$Br[J/\psi\to B\bar B]$ & $Br$[exp] & $\gamma_{B}$ 
 & $\gamma_{B}$[exp]
 \\ \hline
$p$ & $3.47$ & $0.70$ & $2.10$ & $2.12(3)$ & $0.82^{+0.10^{\phantom{X}} }_{-0.12_{\phantom{X}} }$ & 
$0.83(2)$\cite{BESIII:2012ion} 
\\  \hline
$n$ & $2.10$ & $0.80$ & $1.94$ & $2.09(2)$ & $0.82^{+0.10^{\phantom{X}} }_{-0.10_{\phantom{X}} }$ & 
$0.95(6)$\cite{BESIII:2012ion} 
\\ \hline
$\Lambda$ & $2.29$ & $0.80$ & $1.94$ & $1.89(9)$ & $0.85^{+0.15^{\phantom{X}} }_{-0.07_{\phantom{X}} }$ &
$0.83(4)$ \cite{BESIII:2017kqw} 
\\ \hline
$\Sigma^{0}$ 
& \multirow{2}{1.8em}{$2.00 $} 
& \multirow{2}{2em}{$0.80$} 
&\multirow{2}{2em}{$1.10 $} 
& $1.17(3)$ 
& \multirow{2}{4.5em}{ $1.85^{+0.10}_{-0.24}$ } 
& $2.11(5)$   \cite{BESIII:2017kqw} 
\\ 
$\Sigma^{+}$ &  &  &  & $1.50(3)$ &  &$2.27(5)$ \cite{BESIII:2020fqg}
\\ \hline
$\Xi^{+}$ 
&  \multirow{2}{2em}{$1.60$}  
& \multirow{2}{2em}{$0.50$} 
&  \multirow{2}{2.1em}{$1.38$ } 
& $0.97(8)$ 
& \multirow{2}{4.8em}{  $0.65^{+0.10}_{-0.03}$ }  
&$0.61(5)$ \cite{BESIII:2016ssr}
\\ 
$\Xi^{0}$ &  &  &  & $1.16(44)$ &  &
$0.66(3)$  \cite{BESIII:2016nix}
\\  \hline
\end{tabular}
\end{center}
\end{table}
The given choice of the different parameters for  baryon DAs allows one to describe the data for the $\gamma_B$ within the accuracy $20\%$, which is sufficiently reasonable taking into account various higher order effects.  The largest differences are obtained for neutron and for the  $\Sigma$ decay channels.  In considered description, the neutron channel differs from the proton one  only in the contribution of  electromagnetic form factors.
  However this interference can not provide a sufficient numerical impact.  On the other hand  the neutron  data  in Ref.\cite{BESIII:2012ion} have very large systematic error,  which gives   $\gamma_n=0.95(6)\pm 0.27_{syst}$.  Therefore  the obtained value of $\gamma_n$  can be accepted  as  quite reasonable result.  For  $\Sigma$ decay channel, it is possible to obtain the larger value of  $\gamma_\Sigma$ but  corresponding parameter set gives a worse description of the branching ratio because of relatively large and negative power correction.     
  
  Comparing the current analysis with that in Ref.\cite{Kivel:2021uzl} we conclude that  the NLP corrections  improve the phenomenological description and make it possible to better constrain  the unknown parameters of  baryon DAs.  The negative effects of the power contribution reduces the absolute value of  the amplitude $\mathcal{G}_M$ and increases the  value of the ratio $\gamma_B$ that improves  the phenomenological description. However, this effect cannot be very large, since this can significantly worsen the description of the branching fractions.

\section{Conclusions}
\label{conc}
In conclusion,  in this work we continue to develop the analysis proposed in Ref.\cite{Kivel:2021uzl}. The  decay amplitudes  $J/\psi$ into baryon-antibaryon pairs are considered within the NRQCD and  collinear factorisation approach, power suppressed  corrections  and electromagnetic contributions  are taken into account. This allows us to better appreciate some of the theoretical uncertainties.

 The next-to-leading power correction, which is described by  twist-4 and twist-5  three-quark baryon DAs is calculated  and discussed  for the first time.  It is shown that this contribution is well defined  and  this allows one to study their  effects  in a more systematic way.  From the phenomenological point of view the NLP correction allows to improve the description of the  data and provide more rigorous constrains on the baryons  DAs parameters.   

In the given analysis  we also take into account  the interference with the  electromagnetic amplitude, which depends on the time-like baryon form factors. Unfortunately, these values are  little known due to the lack of experimental information. Therefore, an evaluation of their numerical impact can only be made under certain assumptions. 

The existing experimental data  for decays $J/\psi\to B\bar B$  provide an interesting  information about  two observables: branching fractions and angular distribution coefficients $\alpha_B$. The value of $\alpha_B$  is determined by the  ratio of the  two independent decay amplitudes, which is especially  interesting because it does not contain  many theoretical uncertainties that are significant for the branching fractions.  This also provides  an attractive opportunity to study  the baryon distribution amplitudes. 
 
 One of the main goal of the given phenomenological analysis is to verify that  the current data can be described within the factorisation framework.  In order to constrain the non-perturbative baryon DAs we use  results  obtained in the framework of  QCD sum rules \cite{Anikin:2013aka ,Chernyak:1987nv} and from the lattice calculations  \cite{Bali:2019ecy}.  Some of unknown moments have been estimated using flavour $SU(3)$-symmetry and from data fit.  It is found that under reasonable assumptions about the non-perturbative parameters, the data can be described with an  accuracy of $5\% - 20\% $.
 
  It is shown that the power corrections arising from higher twist baryonic DAs are moderate.  This conclusion is, of course,  sensitive to the models  of  baryonic  DAs.  The maximal numerical effect is about $ 16 \%$  is obtained for nucleon only.  It is interesting to note that the model ABOI proposed in 
  Ref.\cite{Anikin:2013aka} for describing nucleon electromagnetic form factors also  works well  for describing charmonium decay, if we use  the value for the normalisation couplings $f_N$ obtained from  the QCD sum rules \cite{Chernyak:1987nv}.   The values of the branching fractions are very sensitive to the normalisation scale,  a reliable description is obtained  only for the  relatively  low  scale  $\mu^2=1.5\ $GeV$^2$.  
   
 Note also that a large value of $\gamma\Sigma\sim 2$ leads to a large numerical effect, which is crucially important for the correct estimate of the decay width. Unlike other baryons, in the case of $\Sigma$  the contribution of the power suppressed term  in the expression  for the width, see the equation (\ref{width} ), is of the order of one, because  it is  enhanced  by a factor of four due to large  $\gamma_\Sigma$.

  The  angular coefficient  $\gamma_B$  is quite sensitive to the  interference with the electromagnetic decay amplitudes. Under the described assumptions, the corresponding numerical effect is estimated in the range of $ 10 \% - 25 \%$. This definitely reduces the accuracy  in the constraining  of  DA parameters.  Nevertheless, the obtained results allow us to conclude that the calculated hadronic amplitudes make a dominant contribution  and therefore the QCD factorisation provides  a suitable basis for understanding dynamics of  these decays.

\section{Acknowledgements}

This work is supported by the Deutsche Forschungsgemeinschaft (DFG, German
Research Foundation) Project-ID 445769443.

\section*{\Huge Appendix}
\appendix
\numberwithin{equation}{section}
\setcounter{equation}{0}

\section{ Analytical expressions for the models of  baryon DAs   }
\lab{DAmodels}
In this Appendix we provide the explicit expressions for the models of the  basic light-cone baryon DAs, which define the QCD nonperturbative input in
 the compact form.  These DAs are used in order to compute the  auxiliary DAs, which are convenient for the calculation of the 
 decay amplitudes.  Recall, that in this work we follow the definitions from Ref.\cite{Anikin:2013aka}.  
 
 We start from the description  of the  twist-3  DAs. For the nucleon DA  introduced in Eq.(\ref {phi3})  we use  the following expression
\begin{align}
\varphi^N _{3}(x_i)=120x_{1}x_{2}x_{3}\left( 1+\phi^N _{10}\mathcal{P}%
_{10}(x_i)+\phi^N _{11}\mathcal{P}_{11}(x_i)
+ \phi ^N_{20}\mathcal{P}%
_{20}(x_i)+\phi^N _{21}\mathcal{P}_{21}(x_{i})+\phi^N _{22}\mathcal{P}%
_{22}(x_i)\right).  \label{phi3N}
\end{align}%
For other baryons $B\neq N$ the DAs are defined in Eqs.(\ref{phi3Bpm}) and (\ref{pi3B}). Corresponding models read  
\begin{eqnarray}
\varphi_{3+}^{B}(x_{i})&=&120x_{1}x_{2}%
x_{3}~\left(  1+\phi_{11}^{B}\mathcal{P}_{11}(x_{i})\right),\, 
\varphi_{3-}^{B}(x_{i})=120x_{1}x_{2}x_{3}\, \phi_{10}^{B}\mathcal{P}_{10}(x_{i}) .
\\
\Pi_{3}^{B}(x_{i})&=&120x_{1}x_{2}x_{3}\left( (1-\delta_{\Lambda B}) \{1+\pi_{11}^{B}P_{11}(x_{i})\} + \delta_{\Lambda B}\,  \pi^{B}_{10}\mathcal{P}_{10}(x_{i})  \right).
\end{eqnarray}
The orthogonal polynomials $\mathcal{P}_{21}(x_{i})$ read
\begin{equation}
\mathcal{P}_{10}(x_{i})=21(x_{1}-x_{3}),~~\mathcal{P}%
_{11}(x_{i})=7(x_{1}-2x_{2}+x_{3}),  \label{P1i}
\end{equation}%
\begin{align}
& \mathcal{P}_{20}(x_{i})=\frac{63}{10}\left[
3(x_{1}-x_{3})^{2}-3x_{2}(x_{1}+x_{3})+2x_{2}^{2}\right] ,~ \\
& \mathcal{P}_{21}(x_{i})=\frac{63}{2}(x_{1}-3x_{2}+x_{3})(x_{1}-x_{3}), \\
& \mathcal{P}_{22}(x_{i})=\frac{9}{5}\left[
x_{1}^{2}+9x_{2}(x_{1}+x_{3})-12x_{1}x_{3}-6x_{2}^{2}+x_{3}^{2}\right] .
\label{P2i}
\end{align}
The moments $\phi _{ik}^{B}$ are multiplicatively renormalisable and the corresponding anomalous dimensions can be found  in Ref.\cite{Anikin:2013aka}.

The set of the models for the twist-4 DAs are defined as following.  The nucleon DAs  from Eqs.(\ref {Phi4}) and(\ref {Psi4}) have the 
following twist decomposition
\begin{equation}
\Phi _{4}(x_{123})=f_{N}\Phi _{4}^{(3)}(x_{123})+\lambda _{1}\bar{\Phi}%
_{4}(x_{123}),  \label{Phi4N}
\end{equation}%
\begin{equation}
\Psi _{4}(x_{123})=f_{N}\Psi _{4}^{(3)}(x_{123})-\lambda _{1}\bar{\Psi}%
_{4}(x_{123}),  \label{Psi4N}
\end{equation}%
The kinematical twist-3  contributions read
\begin{align}
\Phi _{4}^{(3)}(x_{123})=& 40x_{1}x_{2}\left( 1-2x_{3}\right)
-20x_{1}x_{2}\sum_{k=0,1}\phi^N _{1k}\left( 3-\frac{\partial }{\partial x_{3}}%
\right) x_{3}\mathcal{P}_{1k}(x_{123})  \notag \\
& -12x_{1}x_{2}\sum_{k=0,1,2}\phi^N _{2k}\left( 4-\frac{\partial }{\partial
x_{3}}\right) x_{3}\mathcal{P}_{2k}(x_{123}).  \label{Phi4WW}
\end{align}%
\begin{equation}
\Psi _{4}^{(3)}(x_{123})=40x_{1}x_{3}\left( 1-2x_{2}\right)
-20x_{1}x_{3}\sum_{k=0,1}\phi^N _{1k}\left( 3-\frac{\partial }{\partial x_{2}}%
\right) x_{2}\mathcal{P}_{1k}(x_{213})
\end{equation}%
\begin{equation}
-12x_{1}x_{3}\sum_{k=0,1,2}\phi^N _{2k}\left( 4-\frac{\partial }{\partial x_{2}%
}\right) x_{2}\mathcal{P}_{2k}(x_{213}).
\end{equation}%
Notice that the differentiations must be computed with the unmodified
expressions of the polynomials $\mathcal{P}_{nk}(x_{i})$ in Eqs.(\ref{P1i})-(\ref{P2i}) and only after that one can apply the condition $%
x_{1}+x_{2}+x_{3}=1$.

The  genuine twist-4 DAs in Eqs.(\ref{Phi4N}), (\ref{Psi4N}) and (\ref{Ksi4}) reads
\begin{equation}
\bar{\Phi}_{4}(x_{123})=24x_{1}x_{2}\left( 1+\eta^N _{10}\mathcal{R}%
_{10}(x_{312})-\eta^N _{11}\mathcal{R}_{11}(x_{312})\right) ,  \label{PhiN4bar}
\end{equation}%
\begin{equation}
\bar{\Psi}_{4}(x_{123})=24x_{1}x_{3}\left( 1+\eta^N _{10}\mathcal{R}%
_{10}(x_{231})+\eta^N _{11}\mathcal{R}_{11}(x_{231})\right), \label{PsiN4bar}
\end{equation}%
\begin{equation}
 {\Xi}_{4}(x_{i})=24x_{2}x_{3}\left(  1+\frac{9}{4}\xi^N_{10}\mathcal{R}_{10}(x_{132})\right)  .\
\end{equation}
where the orthogonal polynomials $\mathcal{R}_{1i}$ read
\begin{equation}
\mathcal{R}_{10}(x_{1},x_{2},x_{3})=4\left( x_{1}+x_{2}-3/2x_{3}\right) ,~ 
\mathcal{R}_{11}(x_{1},x_{2},x_{3})=\frac{20}{3}\left(
x_{1}-x_{2}+x_{3}/2\right) .
\end{equation}

For other baryons we use definitions from Eqs.(\ref{WWPhi4pm})-(\ref{WWKsi4}).  Recall, that  DA $\Upsilon_4$ from Eq.(\ref{WWUps4}) is not independent and  can be obtained  from  Eq.(\ref{UpsKsi} ).  The kinematical twist-3  contributions in Eqs.(\ref{WWPhi4pm})-(\ref{WWPi4L})  read%
\begin{equation}
\Phi _{4+}^{B(3)}(x_{123})=40x_{1}x_{2}\left( 1-2x_{3}\right)
-20x_{1}x_{2}\phi^B _{11}\left( 3-\frac{\partial }{\partial x_{3}}\right) x_{3}%
\mathcal{P}_{11}(x_{123}),
\end{equation}%
\begin{equation}
\Phi _{4-}^{B(3)}(x_{123})=-20x_{1}x_{2}\ \phi^B _{10}\left( 3-\frac{\partial 
}{\partial x_{3}}\right) x_{3}\mathcal{P}_{10}(x_{123}).
\end{equation}%
\begin{equation}
\Pi _{4}^{B(3)}(x_{123})=40x_{1}x_{2}\left( 1-2x_{3}\right) -20x_{1}x_{2}\pi
_{11}^{B}\left( 3-\frac{\partial }{\partial x_{3}}\right) x_{3}\mathcal{P}%
_{11}(x_{123}).
\end{equation}%
\begin{equation}
\Pi _{4}^{\Lambda (3)}(x_{123})=-20x_{1}x_{2}\ \pi _{10}^{\Lambda }\left( 3-%
\frac{\partial }{\partial x_{3}}\right) x_{3}\mathcal{P}_{10}(x_{123}).
\end{equation}
The genuine twist-4 contributions read
\begin{eqnarray}
\bar{\Phi}^B_{4+}(x_{123})&=&24x_{1}x_{2}\left( -\eta^B_{11}\right) \mathcal{R}_{11}(x_{312}) ,
\\
\bar{\Phi}^B_{4-}(x_{123})&=&24x_{1}x_{2}\left( 1+\eta^B_{10} \mathcal{R}_{10}(x_{312}) \right).
\end{eqnarray}%
\begin{eqnarray}
\bar{\Xi}^B_{4+}(x_{123})&=&24x_{2} x_{3}\left( 1+\frac98 \zeta^B_{10}( \mathcal{R}_{10}(x_{132})+ \mathcal{R}_{10}(x_{123}) )\right),
\\
\bar{\Xi}^B_{4-}(x_{123})&=&27 x_{2}x_{3} (4^{-\delta_{\Lambda B} }) \zeta^B_{10} (\mathcal{R}_{10}(x_{132})-\mathcal{R}_{10}(x_{123})).
\end{eqnarray}%
\begin{eqnarray}
\bar{\Pi}^B_{4}(x_{123})&=&24x_{1}x_{2}\left( -\zeta^B_{11}\right) \mathcal{R}_{11}(x_{312}) ,
\\
\bar{\Pi}^\Lambda_{4}(x_{123})&=&24x_{1}x_{2}\left( 1+\zeta^\Lambda_{10} \mathcal{R}_{10}(x_{312}) \right).
\end{eqnarray}%

The basic twist-5 DAs are used in order to obtain the auxiliary DAs, which are discussed in Sec.\ref{sec:aux}.  To obtain  them,  only  the  asymptotic expressions for   DAs of twist-3 and twist-4  are used, i.e. all higher moments except $f_B,\, f^B_\perp$, $\lambda^B_{1,2}$ and 
$\lambda^B_{\bot}$ are neglected in the above expressions.  For the twist-5 basic  DAs we used the following input 
\begin{align}
\Phi^{B(3)}_{5+}(x_{i})= 40x_{3}(1-x_{2}-2x_{1}-2x_{1}x_{2}), \quad 
\Phi^{B(3)}_{5-}(x_{i})=0. 
\\
\Phi^{B(4)}_{5+}(x_{i})= 0,\quad \Phi^{B(4)}_{5-}(x_{i})= 8x_{3}(1-x_{2}).
\end{align}
\begin{eqnarray}
\Xi^{B(4)}_{5+}(x_{i})&=&4x_1(1+x_1), \quad  \Xi^{B(4)}_{5-}(x_{i})= -12 x_{1}(x_{2}-x_{3}),
\\
\Upsilon_{5}^{B(4)}(x_i) &=&(1-\delta_{B\Lambda}) 4x_{1}\left(  1+x_{1}\right)-\delta_{B\Lambda}12 x_{1}(x_{2}-x_{3}). 
\end{eqnarray}
\begin{eqnarray}
 \Pi_{5}^{B(3)}(x_{i}) &=& (1-\delta_{B\Lambda}) 40x_{3}(1-2x_{1}-x_{2}-2x_{1}x_{2}),
 \\
 \Pi_{5}^{B(4)}(x_{i})&=&\delta_{B\Lambda} 8x_{3}(1-x_{2}).
\end{eqnarray}
Numerical values of all  moments are given  in the Tab.\ref{DAparameters}.  All these moments are multiplicatively renomalisable  and the corresponding anomalous dimensions can be found in Refs.\cite{Braun:2008ia, Anikin:2013aka}.

\section{ Relations between the  twist-5 DAs }
\lab{AppA}

Here we discuss  the technical details that clarify the derivation of
relationships between the various DAs introduced in Sec.\ref{BDA}.

Let us start from  discussion of the chiral even DAs. There are two ways to
get additional equations, which relate the auxiliary and the basic DAs. One way is
to consider the light-cone operator with one derivative  and to repeat the
same analysis as in Sec.\ref{BDA}.  The second method is to use the off light-cone
correlation function (CF). Below we use this one. For simplicity we  consider  the nucleon
case. We also do not write explicitly the colour indices and the Wilson lines.

Consider the following CF ($x^2\neq 0,\ y^2\neq 0$)
\begin{align}
  \left\langle 0\right\vert u(x)C\gamma^{\sigma}u(y)d_{3}(0)\left\vert
B(k)\right\rangle =-k^{\sigma}\left[  \gamma_{5}N\right]  _{3}\text{FT}\left[
V^B_{1}\right] 
\phantom{
\left\langle 0\right\vert u(x)C\gamma^{\sigma}u(y)d_{3}(0)\left\vert
k\right\rangle =}
\nonumber\\
    -\frac{m_{N}}{2}\left[  \gamma^{\sigma}\gamma_{5}N\right]  _{3}%
\text{FT}\left[  V^B_{3}\right]  -m_{N}~ik^{\sigma}\left[  \gamma_{\alpha}%
\gamma_{5}N\right]  _{3}\left(  x^{\alpha}\text{FT}\left[  \mathcal{V}^B%
_{1}\right]  +y^{\alpha}\text{FT}\left[  \mathcal{V}^B_{2}\right]  \right)
\nonumber\\
-\frac{m_{N}^{2}}{4}k^{\sigma}~\left[  \gamma_{5}N\right]  _{3}\left\{
x^{2}\text{FT}\left[  V^B_{1xx}\right]  +y^{2}\text{FT}\left[  V^B_{1yy}\right]
+2(xy)\text{FT}\left[  V^B_{1xy}\right]  \right\}
\nonumber\\
-\frac{m_{N}^{2}}{2}k^{\sigma}x_{\alpha}y_{\beta}\left[  i\sigma^{\alpha\beta
}\gamma_{5}N\right]  _{3}\ \text{FT}\left[  T^B_{xy}\right]  -\frac{m_{N}^{2}%
}{2}~\left[  \gamma_{5}N\right]  _{3}\left\{  x^{\sigma}\text{FT}\left[
V^B_{x4}\right]  +y^{\sigma}\text{FT}\left[  V^B_{y4}\right]  \right\}
\nonumber\\
-\frac{m_{N}^{2}}{4}~\left[  i\sigma^{\alpha\sigma}\gamma_{5}N\right]
_{3}~\left\{  x_{\alpha}\text{FT}\left[  V^B_{x5}\right]  +y_{\alpha}%
\text{FT}\left[  V^B_{y5}\right]  \right\}  . \label{VDAs}%
\end{align}
Performing the expansion of the operator in \textit{lhs} of Eq.(\ref{VDAs})
around the light-cone direction up to twist-5 terms and expanding the
\textit{rhs } of Eq.(\ref{VDAs}) one obtains the set of relations. 

The twist-5 operators resulting  from the expansion  can
be divided into three groups. The first group includes  the operators with
two transverse derivatives, which arise from the expansion of the arguments of
the fields
\begin{align}
u(x)  &  \rightarrow \frac12 x_{\bot}x_{\bot}\partial_{\bot}\partial_{\bot}%
u(x_{-}),\text{ }u(y)\rightarrow  \frac12 y_{\bot}y_{\bot}\partial_{\bot}\partial
_{\bot}u(y_{-}),\ \label{2perp}\\
\ u(x)u(y)  &  \rightarrow x_{\bot}\left[  \partial_{\bot}u(x_{-})\right]
y_{\bot}\left[  \partial_{\bot}u(y_{-})\right]  .
\end{align}
 There second group includes the terms with one transverse derivative and one
small field $\eta$
\begin{equation}
u(x)\rightarrow x_{\bot}\partial_{\bot}\eta(x_{-}),\ \ u(x)u(y)\rightarrow
x_{\bot}\left[  \partial_{\bot}u(x_{-})\right]  \eta(y_{-}),...\ .
\label{1perp}%
\end{equation}
 The third group includes the operators with two small fields $\eta$%
\begin{equation}
u(x)u(y)\rightarrow\eta(x_{-})\eta(y_{-}),...\ .
\end{equation}
All operators with fields $\eta$ can be converted into the operators with two
transverse derivatives using Eq.(\ref{etaEOM}) and  corresponding matrix
elements can be obtained using Eqs. (\ref{V1xx})-(\ref{V1xy}).

The matrix elements of the operators from the first group  simply
reproduce the definitions in Eqs. (\ref{V1xx}), (\ref{V1yy}) and
(\ref{V1xy}).  

The required  relations can be obtained from the matrix elements of the
operators from the second group (\ref{1perp}).  Consider the  terms
with $x_{\bot}$.  There are three  twist-5 light-cone operators, which appear in this case%
\begin{equation}
\left\{  O_{21},O_{22},O_{23} \right\}  =\left\{  \left[  \partial_{\bot}^{\alpha}\xi\right]
C\gamma_{\bot}^{\sigma}\eta\xi_{3},\, \left[  \partial_{\bot}^{\alpha}\xi\right]
C\gamma_{\bot}^{\sigma}\eta\xi_{3},\, \left[  \partial_{\bot}^{\alpha}\xi\right]
C\gamma^{\sigma}\xi\eta_{3}\right\}  ,
\end{equation}
where we assume $qqq_{3}\equiv q(x_{-})q(y_{-})q_{3}(0)$.  The matrix
elements of these operators can be obtaned using QCD EOM (\ref{etaEOM}) and
definitions Eqs. (\ref{V1xx}), (\ref{V1yy}) and (\ref{V1xy}). On the other
side the same matrix elements  can be obtained from  the expansion of {\it rhs} in Eq.(\ref{VDAs}).
 Comparing these results one obtains%
\begin{equation}
\left\langle 0\right\vert O_{23}^{\sigma}\left\vert B(k)\right\rangle
:-2i(py)\text{FT}\left[  T_{xy}\right]  =\text{FT}\left[  \frac{1}{x_{3}%
}\left(  V^B_{1xx}+V^B_{1xy}+T^B_{xy}\right)  -2\mathcal{V}^B_{1}-V^B_{x5}\right]  ,
\end{equation}%
\begin{equation}
\left\langle 0\right\vert O_{21}^{\sigma}+O_{22}^{\sigma}\left\vert
B(k)\right\rangle :V^B_{x5}=\frac{1}{x_{2}}\left(  A^B_{1xy}+T^B_{xy}\right)  -\frac
{1}{x_{1}}A^B_{1xx}.
\end{equation}
Excluding  $V^B_{x5}$ one obtains
\begin{equation}
-2i(ky)\text{FT}\left[  T^B_{xy}\right]  =\text{FT}\left[  \frac{1}{x_{3}%
}\left(  V^B_{1xx}+V^B_{1xy}+T^B_{xy}\right) +\frac{1}{x_{1}%
}A^B_{1xx}-\frac{1}{x_{2}}\left(  A^B_{1xy}+T^B_{xy}\right)  -2\mathcal{V^B}_{1} \right]  ,
\label{Txy:py}%
\end{equation}
where we assume $2(ky)=k_{+}y_{-}$. 

The similar  analysis for the operators  with $y_{\bot}$ 
\begin{equation}
\left\{  O_{21}, O_{22}, O_{23}\right\}  =\left\{  
\xi C\gamma_{\bot}^{\sigma}\left[\partial_{\bot}^{\alpha}\eta\right]  \xi_{3},\,
\xi C\gamma_{\bot}^{\sigma}\left[  \partial_{\bot}^{\alpha}\eta\right]  \xi_{3},\,
\xi C\gamma^{\sigma}\left[  \partial_{\bot}^{\alpha}\xi\right]  \eta_{3}\right\} ,
\end{equation}
allows one to obtain one more relation%
\begin{equation}
2i(kx)\text{FT}\left[  T^B_{xy}\right]  =\text{FT}\left[  \frac{1}{x_{3}}\left(
V^B_{1yy}+V^B_{1xy}-T^B_{xy}\right)  -\frac{1}{x_{2}}A^B_{1yy}+\frac{1}{x_{1}}\left(
A^B_{1xy}+T^B_{xy}\right)  -2\mathcal{V}^B_{2}\right]  . \label{Txy:px}%
\end{equation}

Consider now the axial CF%
\begin{align}
\left\langle 0\right\vert u(x)C\gamma^{\sigma}\gamma_{5}u(y)d_{\gamma
}(0)\left\vert B(k)\right\rangle =-k^{\sigma}\left[  N\right]  _{3}%
\text{FT}\left[  A^B_{1}\right] 
\phantom{
\left\langle 0\right\vert u(x)C\gamma^{\sigma}\gamma_{5}u(y)d_{\gamma
}(0)\left\vert B(k)\right\rangle =
}
 \nonumber \\
-\frac{m_{N}}{2}\left[  \gamma^{\sigma}N\right]  _{3}\text{FT}\left[
A^B_{3}\right]  -ik^{\sigma}m_{N}~\left[  \gamma^{\alpha}N\right]  _{3}\left(
x_{\alpha}\text{FT}\left[  \mathcal{A}^B_{1}\right]  +y_{\alpha}\text{FT}\left[
\mathcal{A}^B_{2}\right]  \right)
 \nonumber  \\ 
  -\frac{m_{N}^{2}}{4}k^{\sigma}\left[  N\right]  _{3}\left\{  x^{2}%
\text{FT}\left[  A^B_{1xx}\right]  +y^{2}\text{FT}\left[  A^B_{1yy}\right]
+2(xy)\text{FT}\left[  A^B_{1xy}\right]  \right\} 
 \nonumber  \\ 
  -\frac{m_{N}^{2}}{2}k^{\sigma}\left[  i\sigma^{\alpha\beta}N\right]
_{3}x^{\alpha}y^{\beta}\text{FT}\left[  G^B_{xy}\right]  -i\frac{m_{N}^{2}}%
{2}~\left[  N\right]  _{3}\left\{  x^{\sigma}\text{FT}\left[  A^B_{4x}\right]
+y^{\sigma}\text{FT}\left[  A^B_{4y}\right]  \right\}
  \nonumber  \\
-i\frac{m_{N}^{2}}{4}~\left[  i\sigma^{\alpha\sigma}N\right]  _{3}~\left\{
x^{\alpha}\text{FT}\left[  A^B_{5x}\right]  +y^{\alpha}\text{FT}\left[
A^B_{5y}\right]  \right\}  .
\end{align}
The analogous consideration of  terms with $x_\bot$ and $y_\bot$   yields the two relations
\begin{align}
2i(ky)\text{FT}\left[  G^B_{xy}\right]  =\text{FT}\left[  -\frac{1}{x_{3}%
}\left(  A^B_{1xx}+A^B_{1xy}+G^B_{xy}\right)  -2\mathcal{A}^B_{1}+\frac{1}{x_{2}%
}\left(  V^B_{1xy}+G^B_{xy}\right)  -\frac{1}{x_{1}}V^B_{1xx}\right]  ,
\end{align}
\begin{align}
2i(kx)\text{FT}\left[ G^B_{xy}\right]  =\text{FT}\left[  \frac{1}{x_{3}%
}\left(  A^B_{1yy}+A^B_{1xy}-G^B_{xy}\right)  +2\mathcal{A}^B_{2}+\frac{1}{x_{1}%
}\left(  V^B_{1xy}+G^B_{xy}\right)  -\frac{1}{x_{2}}V^B_{1yy}\right]  .
\end{align}
It is convenient to combine these two relations with ones in Eqs.(\ref{Txy:py}%
) and (\ref{Txy:px})  and simplify the obtained expressions  using  the relations (\ref{V1xy=V5})--(\ref{A1xy=A5}).
 This gives the relations  given in  Eqs.(\ref{TmG}) and (\ref{TpG}).

Consider now the auxiliary chiral-odd DAs. The set of these functions is
defined in Eqs.(\ref{T0xx})-(\ref{Txy}) and reads%
\begin{equation}
T^B_{0xx},T^B_{0yy},T^B_{2xx},T^B_{2yy},T^B_{0xy},T^B_{2xy},\tilde{T}^B_{2xy}. \label{Tlist}%
\end{equation}
 The algebraic relations can be obtained in the same way as described for the
chiral-even DAs in Sec.\ref{sec:aux}. Therefore we skip the details and only provide
a schematical description.  The consideration of the scalar operators $\eta
C\xi\eta_{3}$ and $\xi C\eta\eta_{3}$ yields%
\begin{equation}
T^B_{0xx}+2T^B_{2xx}+T^B_{0xy}+2T^B_{2xy}-\tilde{T}^B_{2xy}=2x_{1}x_{3}\left(
S^B_{2}-T^B_{4}\right)  , \label{T0xx=S2mT4}%
\end{equation}%
\begin{equation}
T^B_{0yy}+2T^B_{2yy}+T^B_{0xy}+2T^B_{2xy}+\tilde{T}^B_{2xy}=2x_{2}x_{3}\left(
-S^B_{2}-T^B_{4}\right)  , \label{T0yy=S2pT4}%
\end{equation}
respectively.  From the analysis of the pseuvdoscalar operators $\eta
C\gamma_{5}\xi\eta_{3}$ and $\xi C\gamma_{5}\eta\eta_{3}$ one finds%
\begin{equation}
T^B_{0xx}-2T^B_{2xx}+T^B_{0xy}-2T^B_{2xy}-\tilde{T}^B_{2xy}=-2x_{1}x_{3}\left(
P^B_{2}+T^B_{8}\right)  , \label{T0xx=P2pT8}%
\end{equation}%
\begin{equation}
T^B_{0yy}-2T^B_{2yy}+T^B_{0xy}-2T^B_{2xy}+\tilde{T}_{2xy}=2x_{2}x_{3}\left(
P^B_{2}-T^B_{8}\right)  . \label{T0yy=P2mT8}%
\end{equation}
One more relation follows from the consideration of the tensor operator $\eta
C\bar{n}\gamma_{\bot}\eta\xi_{3}$:%
\begin{equation}
T^B_{2xy}=-x_{1}x_{2}T^B_{5}. \label{T2xy=T5}%
\end{equation}
Using the same consideration as in Eq.(\ref{dOeven}) we consider  the matrix
element of the operator $(\bar{n}\partial)\left[  \xi Cn\gamma_{\bot}^{\nu}%
\xi\xi_{3}\right] $.  This yields%
\begin{equation}
-2x_{1}x_{2}x_{3}T^B_{1}=x_{2}\bar{x}_{2}T^B_{0xx}+x_{1}\bar{x}_{1}T^B_{0yy}%
+2x_{1}x_{2}T^B_{0xy}. \label{T1EOM}%
\end{equation}
All other operators give the linear combinations of these six relations. 

In order to get one more relation we use the method of CF, which is described above. Consider the
following CF
\begin{align}
\left\langle 0\right\vert u(x)C\sigma^{\mu\nu}u(y)d_{3}(0)\left\vert
B(k)\right\rangle =-~i(k^{\nu}g^{\alpha\mu}-k^{\mu}g^{\alpha\nu})\left[
\gamma^{\alpha}\gamma_{5}N\right]  _{3}\text{FT}\left[  T^B_{1}\right]
\phantom{
\langle 0\vert u(x)C\sigma^{\mu\nu} u(y) 
}
\nonumber \\
-~2m_{N}\left[  \sigma^{\mu\nu}\gamma_{5}N\right]  _{3}\text{FT}\left[
\mathcal{T}^B_{3}\right]  -m_{N}\left(  k^{\nu}g^{\alpha\mu}-k^{\mu}g^{\alpha
\nu}\right)  \left[  \gamma_{5}N\right]  _{3}\left(  x^{\alpha}\text{FT}%
\left[  \mathcal{T}^B_{2x}\right]  +y^{\alpha}\text{FT}\left[  \mathcal{T}^B%
_{2y}\right]  \right)
\nonumber \\
~-m_{N}\left(  k^{\nu}g^{\alpha\mu}-k^{\mu}g^{\alpha\nu}\right)  \left[
i\sigma^{\alpha\beta}\gamma_{5}N\right]  _{3}\left(  x^{\beta}\text{FT}\left[
\mathcal{T}^B_{4x}\right]  +y^{\beta}\text{FT}\left[  \mathcal{T}^B_{4y}\right]
\right)
\nonumber \\
~-m_{N}^{2}\left[  \gamma^{\alpha}\gamma_{5}N\right]  _{3}i(p^{\nu}%
g^{\alpha\mu}-p^{\mu}g^{\alpha\nu})\left(  x^{2}\text{FT}\left[
T^B_{1xx}\right]  +y^{2}\text{FT}\left[  T^B_{1yy}\right]  +2(xy)\text{FT}\left[
T^B_{1xy}\right]  \right)
\nonumber \\
-m_{N}^{2}\left[  \gamma^{\alpha}\gamma_{5}N\right]  _{3}i(p^{\nu}g^{\beta\mu
}-p^{\mu}g^{\beta\nu})\left(  x^{\alpha}x^{\beta}\text{FT}\left[
T^B_{2xx}\right]  +y^{\alpha}y^{\beta}\text{FT}\left[  T^B_{2yy}\right]  \right)
\nonumber \\
-m_{N}^{2}\left[  \gamma^{\alpha}\gamma_{5}N\right]  _{3}i(p^{\nu}g^{\beta\mu
}-p^{\mu}g^{\nu\beta})\left\{  x^{\alpha}y^{\beta}+x^{\beta}y^{\alpha
}\right\}  \text{FT}\left[  T^B_{2xy}\right]
\nonumber \\
-m_{N}^{2}\left[  \gamma^{\alpha}\gamma_{5}N\right]  _{3}i(p^{\nu}g^{\beta\mu
}-p^{\mu}g^{\nu\beta})\left\{  x^{\alpha}y^{\beta}-x^{\beta}y^{\alpha
}\right\}  \text{FT}\left[  \tilde{T}^B_{2xy}\right]
\nonumber \\
-m_{N}^{2}\left[  \gamma^{\alpha}\gamma_{5}N\right]  _{3}\left(  g^{\alpha\nu
}g^{\mu\beta}-g^{\alpha\mu}g^{\beta\nu}\right)  \left(  x^{\beta}%
\text{FT}\left[  T^B_{5x}\right]  +y^{\beta}\text{FT}\left[  T^B_{5y}\right]
\right)  ~
\nonumber \\
-m_{N}^{2}\left[  \gamma^{\alpha}\sigma^{\mu\nu}\gamma_{5}N\right]
_{3}\left(  x^{\alpha}\text{FT}\left[  T^B_{8x}\right]  +y^{\alpha}%
\text{FT}\left[  T^B_{8y}\right]  \right)  ,
\end{align}
where the \textit{rhs} includes all terms up to twist-5 accuracy.  Expanding
the operator in the  \textit{lhs} we consider the following linear in $x_{\bot}$
and $y_{\bot}$ twist-5 terms%
\begin{align}
\left.  u(x)C\sigma^{\mu\nu}u(y)d_{3}(0)\right\vert _{\text{tw5}}%
=x_{\bot\alpha}\mathcal{O}_{x\bot}^{\alpha\mu\nu}+y_{\bot\alpha}%
\mathcal{O}_{y\bot}^{\alpha\mu\nu}
\phantom{
\left.  u(x)C\sigma^{\mu\nu}u(y)d_{3}(0)\right\vert _{\text{tw5}}
}
\nonumber \\
+\frac{i}{2}\left(  \bar{n}^{\mu}g_{\bot
}^{\nu\sigma}-\bar{n}^{\nu}g_{\bot}^{\mu\sigma}\right)  \left(  x_{\bot\alpha
}\mathcal{O}_{x}^{\alpha\sigma}+y_{\bot\alpha}\mathcal{O}_{y}^{\alpha\sigma
}\right)  +\ldots,
\end{align}
where%
\begin{equation}
\mathcal{O}_{x\bot}^{\alpha\mu\nu}=\left[  \partial_{\bot}^{\alpha}\eta
(x_{-})\right]  C\sigma_{\bot\bot}^{\mu\nu}\xi(y_{-})\xi_{3}(0)+\left[
\partial_{\bot}^{\alpha}\xi(x_{-})\right]  C\sigma_{\bot\bot}^{\mu\nu}%
\eta(y_{-})\xi_{3}(0),
\end{equation}%
\begin{equation}
\mathcal{O}_{y\bot}^{\alpha\mu\nu}=\eta(x_{-})C\sigma_{\bot\bot}^{\mu\nu
}\left[  \partial_{\bot}^{\alpha}\xi(y_{-})\right]  \xi_{3}(0)+\xi
(x_{-})C\sigma_{\bot\bot}^{\mu\nu}\left[  \partial_{\bot}^{\alpha}\eta
(y_{-})\right]  \xi_{3}(0),
\end{equation}%
\begin{equation}
\mathcal{O}_{x}^{\alpha\sigma}=\left[  \partial_{\bot}^{\alpha}\xi
(x_{-})\right]  C\Dsl n  \gamma_{\bot}^{\sigma}\xi(y_{-})\eta_{3}(0),
\end{equation}%
\begin{equation}
\mathcal{O}_{y}^{\alpha\sigma}=\xi(x_{-})C\Dsl n  \gamma_{\bot}^{\sigma
}\left[  \partial_{\bot}^{\alpha}\xi(y_{-})\right]  \eta_{3}(0).
\end{equation}
The consideration of the corresponding matrix elements yields%
\begin{equation}
\left\langle 0\right\vert \mathcal{O}_{x\bot}^{\alpha\mu\nu}\left\vert
B(k)\right\rangle :T^B_{5x}+T^B_{8x}=\frac{1}{8}\frac{1}{x_{1}}\left(  T^B_{0xx}%
-2T^B_{2xx}\right)  +\frac{1}{8}\frac{1}{x_{2}}\left(  T^B_{0xy}-2T^B_{2xy}%
+\tilde{T}^B_{2xy}\right)  ,
\end{equation}%
\begin{equation}
\left\langle 0\right\vert \mathcal{O}_{y\bot}^{\alpha\mu\nu}\left\vert
B(k)\right\rangle :T^B_{5y}+T^B_{8y}=\frac{1}{8}\frac{1}{x_{2}}\left(  T^B_{0yy}%
-2T^B_{2yy}\right)  +\frac{1}{8}\frac{1}{x_{1}}\left(  T^B_{0xy}-2T^B_{2xy}%
-\tilde{T}^B_{2xy}\right)  .
\end{equation}
The sum of these two relations gives%
\begin{align}
T^B_{5y}+T^B_{8y}+T^B_{5x}+T^B_{8x}  &  =\frac{1}{8}\frac{1}{x_{2}}\left(
T^B_{0yy}-2T^B_{2yy}+T^B_{0xy}-2T^B_{2xy}+\tilde{T}^B_{2xy}\right)
\nonumber  \\
&  +\frac{1}{8}\frac{1}{x_{1}}\left(  T^B_{0xx}-2T^B_{2xx}+T^B_{0xy}-2T^B_{2xy}%
-\tilde{T}^B_{2xy}\right)  .
\end{align}
Using Eqs.(\ref{T0xx=P2pT8}) and (\ref{T0yy=P2mT8}) one finds
\begin{equation}
T^B_{5y}+T^B_{8y}+T^B_{5x}+T^B_{8x}=-\frac{1}{2}x_{3}T^B_{8}. \label{T5x=T8}%
\end{equation}

The consideration of the matrix elements $\ \left\langle 0\right\vert
\mathcal{O}_{x}^{\alpha\sigma}\left\vert B(k)\right\rangle $ and $\ \left\langle
0\right\vert \mathcal{O}_{y}^{\alpha\sigma}\left\vert B(k)\right\rangle $ yields
\begin{align}
i(kx)\text{FT}\left[  T^B_{2xx}\right]  +i(ky)\text{FT}\left[  T^B_{2xy}-\tilde
{T}^B_{2xy}\right]  =\text{FT}\left[  -x_{1}\left(  S^B_{2}-T^B_{4}\right)
-4\left(  T^B_{5x}+T^B_{8x}\right)  +2\mathcal{T}^B_{2x}\right] ,
\\
i(ky)\text{FT}\left[  T^B_{2yy}\right]  +i(kx)\text{FT}\left[  T^B_{2xy}+\tilde
{T}^B_{2xy}\right]  =\text{FT}\left[  x_{2}\left(  S^B_{2}+T^B_{4}\right)  -4\left(
T^B_{5y}+T^B_{8y}\right)  +2\mathcal{T}^B_{2y}\right]  ,
\end{align}
respectively. The sum of these two equations gives
\begin{align}
ip(x-y)\text{FT}\left[  \tilde{T}^B_{2xy}\right]   &  =-i(kx)\text{FT}\left[
T^B_{2xx}+T^B_{2xy}\right]  -i(ky)\text{FT}\left[  T^B_{2yy}+T^B_{2xy}\right]
\nonumber\\
&  \text{FT}\left[  2\left(  \mathcal{T}^B_{2x}+\mathcal{T}^B_{2y}\right)
+2x_{3}T^B_{8}\right]  +\text{FT}\left[  \left(  x_{1}+x_{2}\right)
T^B_{4}+(x_{2}-x_{1})S^B_{2}\right]  , \label{Tt2xy=}%
\end{align}
where we used Eq.(\ref{T5x=T8}).  The \textit{rhs} of this equation can be
rewritten in terms of basic DAs. Combaining Eqs.(\ref{T0xx=S2mT4}) and
(\ref{T0xx=P2pT8}) one finds%
\begin{equation}
T^B_{2xx}+T^B_{2xy}=\frac{1}{2}x_{1}x_{3}\left(  S^B_{2}-T^B_{4}+P^B_{2}+T^B_{8}\right)  .
\end{equation}
Similar combination Eqs.(\ref{T0yy=S2pT4}) and (\ref{T0yy=P2mT8}) gives%
\begin{equation}
T^B_{2yy}+T^B_{2xy}=\frac{1}{2}x_{2}x_{3}\left(  -S^B_{2}-T^B_{4}-P^B_{2}+T^B_{8}\right)
.
\end{equation}
Substituting this into Eq.(\ref{Tt2xy=}) one obtains the relation, which allows one
to find the DA $\tilde{T}_{2xy}$. The expressions for other DAs in the
list (\ref{Tlist}) can be derived from the linear relations (\ref{T0xx=S2mT4}%
)-(\ref{T1EOM}). The corresponding  results are given in Eqs.(\ref{TB0xx})-(\ref{TB2yy}). 

Let us also notice that consistency of Eq.(\ref{Tt2xy=}) implies (setting
$x=y=0$)%
\begin{equation}
\int_{0}^{1}dx_{1}\int_{0}^{1-x_{1}}dx_{2}\left\{  2\left(  \mathcal{T}^B%
_{2x}+\mathcal{T}^B_{2y}\right)  +2x_{3}T^B_{8}+\left(  x_{1}+x_{2}\right)
T^B_{4}+(x_{2}-x_{1})S^B_{2}\right\}  (x_{123})=0,
\end{equation}
where we assume $x_3=1-x_1-x_2$. This provides a good check of the obtained expressions.

Finally, let us also mention that the additional relation for the twist-4
chiral-odd DAs  given in Eq.(\ref{Rel:TB2}), can be obtained  from
the  consideration  of the matrix element of the operator $\xi(x_{-})C\sigma^{\mu\nu}\xi
(y_{-})\eta_{3}(0)$.

\section{ On the determination of quark and quark-gluon contributions for the next-to-leading power correction}
\label{AppB}
In this section, we discuss a simple example that demonstrates a difficulty in the  determination 
of the pure quark contribution for the subleading  power corrections.
 We assume the same kinematical notations as for the baryon with momentum $k$.
 Let us consider the light-cone matrix element for the vector meson state
 ($J^{PC}=1^{--}$), which can be parametrised as
\begin{equation}
\left\langle 0\right\vert \bar{\psi}(x_{-})\gamma^{\sigma}\psi(0)\left\vert V(k) \right\rangle =f_{V}m_{V}\ e_{+}\frac{\bar{n}^{\sigma}}{2}\text{FT}\left[
\phi\right]  +f_{V}m_{V}e_{\sigma}^{\bot}\text{FT}\left[  g_{v}\right]
-f_{V}m_{V}^{2}\frac{n^{\sigma}}{2}\frac{e_{+}}{k_{+}^{2}}\text{FT}\left[
g_{3}\right]  , \label{lcmeV}%
\end{equation}
where $e_{\mu}\equiv e^\lambda_{\mu}(k)$ denotes the polarisation vector, FT
denotes the appropriate Fourier transformation%
\begin{equation}
\text{FT}\left[  \phi \right]  =\int_{0}^{1}due^{-i(1-u)(x_{-}k_{+})/2}%
\phi(u).
\end{equation}
All defined DAs are normalised to one
\begin{equation}
\int_{0}^{1}du~\phi(u)\equiv\left\langle \phi\right\rangle =\left\langle
g_{v,3}\right\rangle =1.
\end{equation}
The twist-3 DA $g_{v}$ and the twist-4 DA $g_{3}$ in the  {\it rhs} have been studied  in
Refs.\cite{Ball:1998ff, Ball:1998sk}. The obtained results read%
\begin{equation}
g_{v}(u)=\frac{1}{2}\int_{0}^{u}dv\frac{\phi(v)}{1-v}+\frac{1}{2}\int_{u}%
^{1}dv\frac{\phi(v)}{v}+\bar{g}_{v}(u), \label{gvtot}%
\end{equation}%
\begin{equation}
g_{3}(x)=\phi(x)+\bar{g}_{3}(x), \label{g3tot}%
\end{equation}
where $\bar{g}_{v}$ and $\bar{g}_{3}$ denote the contributions of quark-gluon
matrix elements. \ \ 

The local limit in Eq.(\ref{lcmeV}) gives%
\begin{equation}
\left\langle 0\right\vert \bar{\psi}(0)\gamma^{\sigma}\psi(0)\left\vert V(k)\right\rangle =f_{V}m_{V}e_{+}\frac{1}{2}\bar{n}^{\sigma}\left\langle
\phi\right\rangle +f_{V}~m_{V}e_{\sigma}^{\bot}\left\langle g_{v}\right\rangle
-f_{V}m_{V}\frac{1}{2}n^{\sigma}m_{V}^{2}\frac{e_{+}}{k_{+}^{2}}\left\langle
g_{3}\right\rangle
\end{equation}%
\begin{equation}
=f_{V}m_{V}\left\{  e_{+}\frac{1}{2}\bar{n}^{\sigma}+e_{\sigma}^{\bot}%
-\frac{1}{2}n^{\sigma}m_{V}^{2}\frac{e_{+}}{k_{+}^{2}}\right\}  =f_{V}%
m_{V}e^{\sigma}, \label{me_local}%
\end{equation}
where it is used that
\begin{equation}
e\cdot k=0\Rightarrow e_{-}=-e_{+}k_{-}/k_{+}=-e_{+}k_{-}/k_{+}=-e_{+}%
\frac{m_{V}^{2}}{k_{+}^{2}}.
\end{equation}

The off light-cone CF can be written as \cite{Ball:1998ff}
\begin{align}
\left\langle 0\right\vert \bar{\psi}(x)\gamma^{\sigma}\psi(0)\left\vert
V(k) \right\rangle =f_{V}m_{V}\frac{(ex)}{(kx)}k^{\sigma}\left(  \text{FT}\left[
\phi\right]  +\frac{m^{2}_Vx^{2}}{16}\text{FT}\left[ \mathbf{ A}\right]  \right)
\phantom{
\left\langle 0\right\vert \bar{\psi}(x)\gamma^{\sigma}\psi(0)\left\vert
k\right\rangle 
}
\nonumber \\
+f_{V}m_{V}\left(  e_{\sigma}-k^{\sigma}\frac{(ex)}{(kx)}\right)
\text{FT}\left[  g_{v}\right]  -f_{V}m_{V}\frac{1}{2}x^{\sigma}m_{V}^{2}%
\frac{(ex)}{(kx)^{2}}\text{FT}\left[  g_{3}+\phi-2g_{v}\right]  . 
\label{CF}%
\end{align}
The derivation of the expression for the DA $\mathbf{ A}$ is given in Ref.\cite{Ball:1998ff},  the result reads%
\begin{equation}
\mathbf{ A}(u)=32\int_{0}^{u}dv\int_{0}^{v}d\omega(g_{v}-\phi)(\omega)+\bar{\mathbf{ A}}(u),
\label{Atot}%
\end{equation}
where $\bar{\mathbf{ A}}$ again denotes the quark-gluon contributions. Such notation for the contributions of quark-gluon matrix elements is also accepted  below. 
All results in Eqs.(\ref{gvtot}),(\ref{g3tot}) and (\ref{Atot}) are obtained
with the help of differentiation of the CF, see details in Refs.\cite{Ball:1998ff,Ball:1998sk}.

Let us now consider the  twist-4 DAs with help of the technique
discussed in this work. The twist-4 operators, which appear in the expansion of
the off light-cone operator $\bar{\psi}(x)\gamma^{\sigma}\psi(0)$ read
\begin{eqnarray}
\left[  \bar{\psi}(x)\gamma^{\sigma}\psi(0)\right]  _{\text{tw4}}=\frac{1}%
{4}x_{\bot}^{2}\left[  \partial_{\bot}^{2}\bar{\xi}(x_{-})\right]
\gamma^{\sigma}\xi(0)+\frac{1}{2}x_{+}\left[  (\bar{n}\partial)\bar{\xi}%
(x_{-})\right]  \gamma^{\sigma}\xi(0)\nonumber\\
~\ \ \ +x_{\bot}^{\alpha}\left[  \partial_{\bot}^{\alpha}\bar{\xi}%
(x_{-})\right]  \gamma_{\bot}^{\sigma}\eta(0)+x_{\bot}^{\alpha}\left[
\partial_{\bot}^{\alpha}\bar{\eta}(x_{-})\right]  \gamma_{\bot}^{\sigma}%
\xi(0)+\bar{\eta}(x_{-})\gamma^{\sigma}\eta(0)+\mathcal{O}_{4}^{\sigma
}[\mathcal{A}], \label{Otw4}
\end{eqnarray}
where, remind, the collinear fields  $\xi$ and $\eta$ are defined in Eq.(\ref{xidef}). The operator $\mathcal{O}_{4}^{\sigma}[\mathcal{A}]$ denotes the contribution of the quark-gluon operators. 

 The matrix element of the twist-4 operator with two derivatives can be
obtained from the expansion of the CF
\begin{equation}
\left\langle 0\right\vert \left[  \partial_{\bot}^{2}\bar{\xi}(x_{-})\right]
\ns  \xi(0)\left\vert V(k) \right\rangle =f_{V}m_{V}^{3}\frac{e_{+}~}%
{4}~\text{FT}\left[  \mathbf{A}+\bar{G}_{2}\right]. \label{d2O}%
\end{equation} 
Consider now the operators with the longitudinal derivative $i(\bar{n}\partial)$ in (\ref{Otw4}). Using QCD EOM
\begin{equation}
\left\langle 0\right\vert \left[  i(\bar{n}\partial)\bar{\xi}(x_{-})\right]
\ns  \xi(0)\left\vert V(k)\right\rangle =\left\langle 0\right\vert \left[
(i\bar{n}\partial)^{-1}\partial_{\bot}^{2}\bar{\xi}(x_{-})\right]  \ns
\xi(y_{-})\left\vert V(k) \right\rangle +\ldots
\end{equation}%
\begin{equation}
=f_{V}m_{V}^{3}\frac{e_{+}}{k_{+}}~\frac{1}{4}\text{FT}\left[  \frac{1}%
{\bar{u}}\mathbf{A}(u)\right]  +\ldots\ ,
\end{equation}
where dots denote the quark-gluon contributions. Similarly%
\begin{equation}
\left\langle 0\right\vert \left[  \bar{\xi}(x_{-})\right]  \ns  \left[
i(\bar{n}\partial)\xi(0)\right]  \left\vert V(k) \right\rangle =f_{V}~\frac
{m_{V}^{3}}{4}\frac{e_{+}}{k_{+}}~\text{FT}\left[  \frac{1}{u}\mathbf{A}(u)\right]
+\ldots\ .
\end{equation}
The sum of these terms gives the operator m.e. of the operator $i(\bar
{n}\partial)\left[  \bar{\xi}(x_{-})\ns  \xi(0)\right]  $ with the total
derivative, which can be computed from Eq.(\ref{CF}).  This  gives the
following relation%
\begin{equation}
\mathbf{A}(u)=4u\bar{u}\phi(u)+\bar{G}_{1}(u), \label{A=4uubphi}%
\end{equation}
where $\bar{G}_{1}(u)$ again schematically denotes  the contributions of 
quark-gluon DAs.

Consider now the m.e.%
\begin{eqnarray}
\left\langle 0\right\vert \bar{\eta}(x_{-})\nbs\eta(0)\left\vert
V(k) \right\rangle = 
\left\langle 0\right\vert \left[  (in\partial)^{-1}\partial_{\bot}^{\alpha
}\bar{\xi}(x_{-})\right]  \ns  \left[  (in\partial)_{\bot}^{-1}%
\partial_{\bot}^{\alpha}\xi(y_{-})\right]  \left\vert k\right\rangle +\dots
\nonumber \\
=-\left\langle 0\right\vert \left[  (in\partial)^{-1}\partial_{\bot}^{2}%
\bar{\xi}(x_{-})\right]  \ns  \left[  (in\partial)^{-1}\xi(y_{-})\right]
\left\vert k\right\rangle +\dots
\nonumber \\
=-f_{V}m_{V}\frac{m_{V}^{2}}{4}\frac{e_{+}}{k_{+}^{2}}~\text{FT}\left[
\frac{1}{u\bar{u}}\mathbf{A}(u)\right] +\dots,
\end{eqnarray}
where we  show  the two-particle quark-antiquark operator only. On the other hand%
\begin{equation}
\left\langle 0\right\vert \bar{\eta}(x_{-})\nbs\eta(y_{-})\left\vert
V(k) \right\rangle  =\left\langle
0\right\vert \bar{\psi}(x_{-})\nbs\psi(y_{-})\left\vert k\right\rangle
=-f_{V}m_{V}^{3}\frac{e_{+}}{k_{+}^{2}}\text{FT}\left[  g_{3}\right]  .
\end{equation}
Therefore this gives
\begin{equation}
g_{3}=\frac{1}{4}\frac{\mathbf{A}(u)}{u\bar{u}}+\bar{G}_{3}(u),
\end{equation}
where  $\bar{G}_{3}$ denotes  a contribution from quark-gluon DAs.  Using
(\ref{A=4uubphi}) one obtains%
\begin{equation}
g_{3}(u)=\phi(u)+\frac{\bar{G}_{1}(u)}{4u\bar{u}}+\bar{G}_{3}(u).
\end{equation}
Neglecting  the quark-gluon terms $\bar{G}_{1}\sim\bar{G}_{3}\rightarrow0$
 one finds  the result
\begin{equation}
g_{3}(u)\simeq\phi(u),
\end{equation}
that agrees with Eq.(\ref{g3tot}) assumming  $\bar{g}_{3}\rightarrow0.$  At the same
time the results  for  $\mathbf A$ in Eqs.(\ref{Atot}) and (\ref{A=4uubphi}) do
not agree  if one simply  neglects the quark-gluon  contributions
\begin{equation}
4u\bar{u}\phi(u)\neq\left.  32\int_{0}^{u}dv\int_{0}^{v}d\omega(g_{v}%
-\phi)(\omega)\right\vert _{\bar{g}_{3}=\bar{g}_v=0}. \label{problem}%
\end{equation}
This result can be explained by the fact that the neglected quark-gluon
contributions still implicitly depend on the twist-2 moments. For the local
matrix elements this can be seen from the QCD equation of motions, see e.g.
Ref.\cite{Ball:1998ff}. Therefore the simple recipe prescribing just to neglect the
quark--gluon contributions, as for the twist-3 operators,  is not
applicable for the twist-4 calculations. Notice however that 
Eq.(\ref{problem}) is  valid for one particular case:  for the asymptotic shape  $\phi^{as}(u)=6u\bar{u}$.
The corresponding local operator can not be related with any  quark-gluon one. 

This observation  shows that  consistent twist-4 calculation, which  only account  the two-particle quark operator
 requires some prescription, which makes it possible to unambiguously and systematically distinguish between the twist-4 quark and the quark-gluon contribution.

In fact, the Lorentz symmetry of the CF (\ref{CF})  implies that DA $\mathbf{A}$ must satisfy to a certain condition.
 Let us briefly discuss this point.  Consider the following matrix element
\begin{equation}
\left\langle 0\right\vert \bar{\psi}(0)\nbs\psi(0)\left\vert V(k) \right\rangle
=-~f_{V}m_{V}^{3}~\frac{e_{+}}{k_{+}^{2}}, \label{tw4norm}%
\end{equation}
where we used Eq.(\ref{lcmeV}).  Now the operator in {\it lhs} can be rewritten as
\begin{align}
\bar{\psi}(0)\nbs\psi(0)  &  =\bar{\eta}(0)\nbs\eta(0)={i^{2}}%
\bar{\xi}\overleftarrow{\Dsl D  }_{\bot}(i\overleftarrow{D}n%
)^{-1}\ns  (in\overrightarrow{D})^{-1}%
\overrightarrow{\Dsl D  }_{\bot}\xi\\
&  \simeq\left[  (in\partial)^{-1}i{\partial}_{\alpha\bot}  \bar{\xi}\,\right]
\gamma_{\bot}^{\alpha}\ns  \gamma_{\bot}^{\beta}\left[
(in\partial)^{-1}i{\partial}_{\bot\beta}\xi\right],
\end{align}
where we again skip the quark-gluon operators.  The
matrix element of this operator gives%
\begin{equation}
\left\langle 0\right\vert
\left[  (in\partial)^{-1}i{\partial}_{\alpha\bot}  \bar{\xi}\,\right]
\gamma_{\bot}^{\alpha}\ns  \gamma_{\bot}^{\beta}\left[
(in\partial)^{-1}i{\partial}_{\bot\beta}\xi\right]
 \left\vert
V(k) \right\rangle \simeq-\frac{e_{+}}{k_{+}^{2}}~f_{V}m_{V}^{3}\int_{0}^{1}%
du~\frac{\mathbf{A}_{\bar{q}q}(u)}{4u\bar{u}} ,
\end{equation}
where $\mathbf{A}_{\bar{q}q}$ denotes the pure quark part of the DA $\mathbf{A}$%
\begin{equation}
\mathbf{A}=\mathbf{A}_{\bar{q}q}+\bar{\mathbf{A}}.
\end{equation}
Therefore one finds%
\begin{equation}
\left\langle 0\right\vert \bar{\psi}(0)\nbs\psi(0)\left\vert V(k) \right\rangle
=-\frac{e_{+}}{k_{+}^{2}}~f_{V}m_{V}^{3}\int_{0}^{1}du~\frac{\mathbf{A}_{\bar{q}q}%
(u)}{4u\bar{u}}+\bar{G}_{4} ,
\end{equation}
where $\bar G_{4}$ schematically denotes the quark-gluon contributions.
Comparing with Eq.(\ref{tw4norm}) one finds the following relation%
\begin{equation}
\int_{0}^{1}du~\frac{  \mathbf{A}_{ \bar q q } (u) }{4u\bar{u} }+G_{4}=1,
\end{equation}
 One can assume that  consistently defined  quark contribution $\mathbf{A}_{\bar{q}q}$ must satisfy
\begin{equation}
\int_{0}^{1}du~\frac{\mathbf{A}_{ \bar{q}q}(u) }{4u\bar{u}}=1, \label{A2norm}%
\end{equation}
that ensures the correct Lorentz symmetry properties in the local limit, even
if the quark-gluon contributions are not taken into account, see Eq(\ref{me_local}).
 For instance, neglecting  the quark-gluon contributions  in Eq.(\ref{A=4uubphi}) one finds the expression%
\begin{equation}
\mathbf{A}_{\bar{q}q}(u)=4u\bar{u}\phi(u),
\end{equation}
which satisfy to Eq.(\ref{A2norm}). 

In conclusion, the given consideration demonstrates that in general case the
definition of the pure quark contribution for the twist-4 DAs is not universal
due to the implicit contributions from the matrix element of the quark-gluon
operators. In the given example the universal part  can only be defined in
terms of the asymptotic part of the twist-2 DA $\phi$ because the
corresponding local operator do not appear from  the quark-gluon ones. 

The
similar arguments  also applicable  for the three-quark operators. Therefore in
 our calculations we only consider the asymptotic contributions of the
twist-3 and twist-4 3-quark matrix elements in order to define the corresponding universal
three-quark components of twist-5 DAs.

\section{ Isospin relations between the auxiliary twist-5 DAs }
\lab{App:iso}
The standard analysis  of the light-cone matrix elements with two transverse derivatives allows one to obtain many  relations between the auxiliary DAs defined in the Sec.\ref{sec:aux}. Here we consider such relations for a nucleon state. For simplicity, in this section we omit the baryon subscript ``B=N''  assuming for all DAs  $F^N\equiv F$.  Let us define  the operator basis as following
\bea
\epsilon^{ijk}\left[  \partial_{\bot}^{\alpha} u^i_{1}(x_{-}) \right] \left[  \partial_{\bot}^{\beta} u^j_{2}(y_{-})\right]  \ d^k_{3}(z_{-}),  \label{def:O1} \\
 \epsilon^{ijk}\left[  \partial_{\bot}^{\alpha} \partial_{\bot}^{\beta}u^i_{1}(x_{-}) \right]   u^j_{2}(y_{-})  \ d^k_{3}(z_{-}), \label{def:O2}\\
 \epsilon^{ijk} u^i_{1}(x_{-})\left[   \partial_{\bot}^{\alpha} \partial_{\bot}^{\beta} u^j_{2}(y_{-})\right]  \ d^k_{3}(z_{-}). 
 \label{def:O3}
\eea
 In these formulas we assume that each quark field $q={u,d}$ is projected onto the large component $\xi$ as defined in Eq.(\ref{xidef}). Recall, that we use the short notation for  the Dirac indices $q_{\alpha_i}\equiv q_i$. 
From the technical point of view the derivation of such relations is  the same as in Ref.\cite{Braun:2000kw} .  This method gives  the following three relations for the light-cone matrix elements  of defined  operators 
\begin{eqnarray}
\left\langle 0\right\vert 
 \left[  \partial_{\bot}^{\alpha}u_{1}(x_{-})\right]
\left[  \partial_{\bot}^{\beta}u_{2}(y_{-})\right]  \ d_{3}(z_{-})
\left\vert k\right\rangle
 +\left\langle 0\right\vert \left[  \partial_{\bot}^{\alpha
}u_{1}(x_{-})\right]  \ u_{3}(z_{-})\ \left[  \partial_{\bot}^{\beta}%
d_{2}(y_{-})\right]  \left\vert k\right\rangle \nonumber\\
+\left\langle 0\right\vert u_{3}(z_{-})\ \left[  \partial_{\bot}^{\beta}%
u_{2}(y_{-})\right]  \ \left[  \partial_{\bot}^{\alpha}d_{1}(x_{-})\right]
\left\vert k\right\rangle =0.
\end{eqnarray}
\begin{eqnarray}
\left\langle 0\right\vert \left[  \partial_{\bot}^{\alpha}\partial_{\bot
}^{\beta}u_{1}(x_{-})\right]  u_{2}(y_{-})\ d_{3}(z_{-})\left\vert
k\right\rangle +\left\langle 0\right\vert \left[  \partial_{\bot}^{\alpha
}\partial_{\bot}^{\beta}u_{1}(x_{-})\right]  \ u_{3}(z_{-})\ \left[
d_{2}(y_{-})\right]  \left\vert k\right\rangle \nonumber\\
+\left\langle 0\right\vert u_{3}(z_{-})\ u_{2}(y_{-})\ \left[  \partial_{\bot
}^{\alpha}\partial_{\bot}^{\beta}d_{1}(x_{-})\right]  \left\vert
k\right\rangle =0,
\end{eqnarray}%
\begin{eqnarray}
\left\langle 0\right\vert u_{1}(y_{-})\left[  \partial_{\bot}^{\alpha}%
\partial_{\bot}^{\beta}u_{2}(x_{-})\right]  \ d_{3}(z_{-})\left\vert
k\right\rangle +\left\langle 0\right\vert u_{1}(x_{-})u_{3}(z_{-})\ \left[
\partial_{\bot}^{\alpha}\partial_{\bot}^{\beta}d_{2}(y_{-})\right]  \left\vert
k\right\rangle \nonumber\\
+\left\langle 0\right\vert u_{3}(z_{-})\ \left[  \partial_{\bot}^{\alpha
}\partial_{\bot}^{\beta}u_{2}(y_{-})\right]  d_{1}(x_{-})\left\vert
k\right\rangle =0.
\end{eqnarray}
Substituting in these equations  the  definitions of the auxiliary matrix elements from Sec.\ref{sec:aux} and using the Fierz identities  one derives various relations between the auxiliary DAs. It is convenient to  divide the operators $O_i^{\alpha\beta}$ in Eqs.(\ref{def:O1})-(\ref{def:O3})  into three independent groups: symmetric ($O_i^{\alpha\beta}=g_\bot^{\alpha\beta}O_i$), symmetric traceless $g_\bot^{\alpha\beta}O_i^{\alpha\beta}$ and  and antisymmetric $O_1^{\beta\alpha}=-O_1^{\alpha\beta}$. Each of these groups can be considered independently.  

 For simplicity, in what follow we use  the following convenient notations
\bea
F(x_1,x_2, x_3)\equiv F,\,  F(x_2,x_1, x_3)=\hat P_{12} F, \,  F(x_3,x_2, x_1)=\hat P_{13} F \, \text{ and so on}. 
\eea

    For the DAs, which describe the  symmetric traceless matrix elements the set of relations read
\bea
T_{2xy}\ -\hat P_{23}\left(  T_{2xy}+T_{2xx}\right)  -\hat P_{13}\left(  T_{2xy}+T_{2yy}\right)  \ =0.
\\
T_{2xx}+\hat P_{23}T_{2xx}+\hat P_{13}\left(  T_{2xx}+T_{2yy}+2T_{2xy}\right)
\ =0.
\\
T_{2yy}+\hat P_{13}T_{2yy}+\hat P_{23}\left( T_{2xx}+T_{2yy}+2T_{2xy}\right)
=0.
\eea
 For the DAs, which describe the  antisymmetric matrix elements the set of relations read
  \bea
T_{xy}+G_{xy}-\hat P_{23}\left( T_{xy}+G_{xy}\right)  +\hat P_{13}\,  \tilde {T}_{2xy}  =0,
\\
T_{xy}-G_{xy}+\hat P_{23}\, \tilde{T}_{2xy}  -\hat P_{13}\left( T_{xy}-G_{xy} \right)  =0,
\\
\tilde{T}_{2xy}+\hat P_{23}\left(  T_{xy}-G_{xy}\right)  +\hat P_{13}\left(T_{xy}+G_{xy}\right)  =0.
\eea

 The relations for the symmetric  matrix elements  are described by the matrix elements of the operators $\left[  \partial_\bot^{2}u\right]  ud$,  $u\left[  \partial_\bot^{2}u\right]  d$ and $\left[  \partial_\bot^{\alpha}u\right]  \left[  \partial_\bot^{\alpha} u\right]  d$.
 The analysis of the  operator $\left[  \partial_\bot^{2}u\right]  ud$ gives  
  \bea
V_{1xx}+A_{1xx}+\hat P_{23}\left(  V_{1xx}+A_{1xx}\right)  -\hat P_{13}\left(T_{0xx}+T_{0yy}+2T_{0xy}\right)  =0,
\\
V_{1xx}-A_{1xx}-\hat P_{23}T_{0xx}+\hat P_{13}\left(  \ V_{1xx}+V_{1yy}+2V_{1xy}-A_{1xx}-A_{1yy}-2A_{1xy}\right)  =0,
\\
\hat P_{23}\left(  V_{1xx}-A_{1xx}\right)  + \hat P_{13}\left(   V_{1xx}+V_{1yy}+2V_{1xy}+A_{1xx}+A_{1yy}+2A_{1xy}\right)- T_{0xx}=0 .
\eea
  The analysis of the  operator $u\left[  \partial_\bot^{2}u\right]  d$ gives
\bea
V_{1yy}+A_{1yy}+\hat P_{23}\left(  V_{1xx}+V_{1yy}+2V_{1xy}+A_{1xx}+A_{1yy}+2A_{1xy}\right)  -\hat P_{13}T_{0yy}=0.
\\
V_{1yy}-A_{1yy}+\hat P_{13}\left(  V_{1yy}-A_{1yy}\right)  -\hat P_{23}\left( T_{0xx}+T_{0yy}+2T_{0xy}\right)  =0.
\\
\hat P_{13}\left(  V_{1yy}+A_{1yy}\right)  +\hat P_{23}\left(  V_{1xx}+V_{1yy}+2V_{1xy}-A_{1xx}-A_{1yy}-2A_{1xy}\right)- T_{0yy}=0 .
\eea
 The analysis of the  operator $\left[  \partial_\bot u\right] \left[  \partial_\bot u\right] d$ gives 
 \bea
V_{1xy}+A_{1xy}-\hat P_{23}  \left(  V_{1xy}+V_{1xx}  +
A_{1xy}+A_{1xx}\right) +\hat P_{13}\left(  T_{0xy}+T_{0yy}\right)
=0,
\\
V_{1xy}-A_{1xy}+\hat P_{23}\left(  T_{0xy}+T_{0xx}\right)  -\hat P_{13}\left(
V_{1xy}+V_{1yy}-A_{1xy}-A_{1yy}\right)  =0,
\\
T_{0xy}+\hat P_{23}\left( V_{1xy}+V_{1xx}-A_{1xy}-A_{1xx}\right)  +\hat P_{13}\left( V_{1xy}+V_{1yy}+A_{1xy}+A_{1yy}\right)  =0.
\eea

From these equations it follows that  the relations  between  the chiral-odd and chiral-even auxiliary DAs have the following form
\bea
T_{0xx}&=& P_{13}\left(  \ V_{1xx}+V_{1yy}+2V_{1xy}+A_{1xx}+A_{1yy}%
+2A_{1xy}\right)  +P_{23}\left(  V_{1xx}-A_{1xx}\right)  ,
\\
T_{0yy}&=&P_{13}\left(  V_{1yy}+A_{1yy}\right)  +P_{23}\left(  V_{1xx}%
+V_{1yy}+2V_{1xy}-A_{1xx}-A_{1yy}-2A_{1xy}\right)  ,
\\
T_{0xy}&=&-P_{13}\left(  V_{1xy}+V_{1yy}+A_{1xy}+A_{1yy}\right)  -P_{23}\left(
V_{1xy}+V_{1xx}-A_{1xy}-A_{1xx}\right)  ,
\\
\tilde{T}_{2xy}&=&-P_{13}\left(  T_{xy}+G_{xy}\right)  -P_{23}\left(
T_{xy}-G_{xy}\right) .
\eea
These equations are very helpful for the analytical check of the $SU(3)$ limit for the calculated amplitudes.

\section{Expressions for the projections onto higher twist matrix elements}

\label{AppC}

Here we give  expressions for the  projections $P$ and $\bar P$ that  enter  
Eq.(\ref{dA1_PPD}).  In the following we assume the  matrix element
with the baryon in the initial state as in the definitions in Sec.\ref{T5DA}. 
This is convenient  for a  comparison of  definitions of  baryon DAs, which exist in the literature. 
 This also implies that one calculates the amplitude of the time reversal process $B+\bar
{B}\rightarrow J/\psi$. Since  the corresponding amplitude  is real  this also gives the correct result.   The antibaryon projectors can be
obtained from the baryon ones  using the charge conjugation.

\begin{figure}[ptb]%
\centering
\includegraphics[width=1.50in]{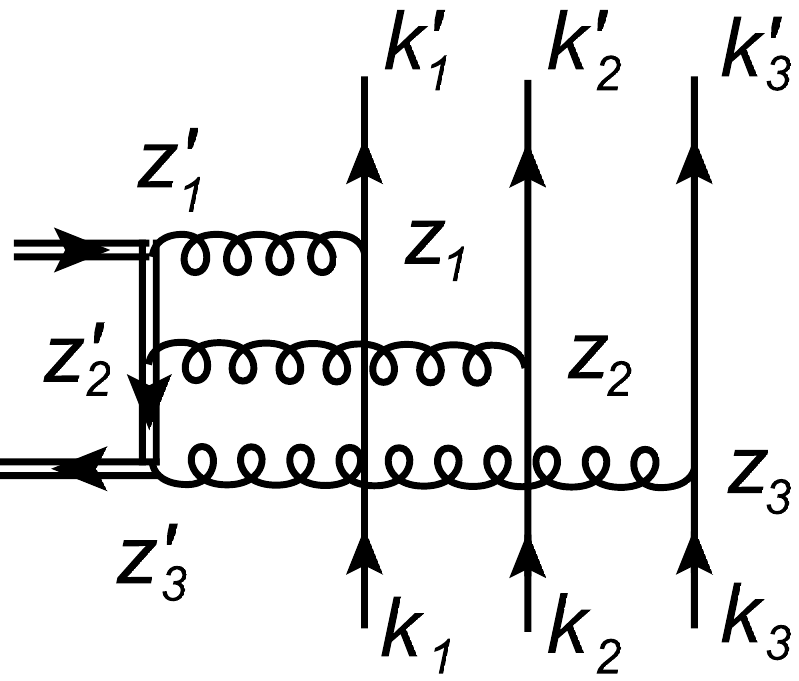}
\caption{ The diagram, which illustrates the choice of  external momenta for the light quarks. The timeline is going to left. }
\label{figure2}
\end{figure}

For partonic diagrams we assume the configuration of the external momenta
as shown in Fig.\ref{figure2}. 
After differentiations with respect to momenta $k_{i}$
and $k'_{i}$ we set%
\begin{equation}
k_{i}=x_{i}k_{+}\frac{\bar{n}}{2},\ k_{i}^{\prime}=y_{i}k_{-}^{\prime}\frac
{n}{2}. \label{def:ki}%
\end{equation}
The derived projectors depend on the auxiliary DAs, which are defined in
Sec.\ref{sec:aux}.

 The twist-4 projectors have already been derived in  Ref.\cite{Kivel:2019wjh}. The
chiral-even projection reads%
\begin{align}
\left.  \left[  P_{\text{tw4}}(x_{i})\right]  _{123}\right\vert _{\text{even}}
=\frac{m_{B}}{4}\left\{ 
 \mathcal{V}_{1}^{B}(x_{i})\left[  C_{\mathcal{V}1}\right]  _{123}
 +\mathcal{A}_{1}^{B}(x_{i})\left[  C_{\mathcal{A}1}\right] _{123}
 \right.
\nonumber \\
\left.
+\mathcal{V}_{2}^{B}(x_{i})\left[  C_{\mathcal{V}2}\right]_{123}
+\mathcal{A}_{2}^{B}(x_{i})\left[  C_{\mathcal{A}2}\right]_{123}
\right\}  ,
\end{align}
where ($i=1,2$)%
\begin{equation}
\left[  C_{\mathcal{V}i}\right]  _{123}=\left[  C_{i\bot}^{\alpha}C^{\top
}\right]  _{21}\left[  \gamma_{\bot}^{\alpha}\gamma_{5}N_{\bar{n}}\right]
_{3}+\left[  C_{i}^{\Vert}C^{\top}\right]  _{21}\left[  \Dsl k  ^{\prime
}\gamma_{5}N_{\bar{n}}\right]  _{3},
\end{equation}
\begin{equation}
\left[  C_{\mathcal{A}i}\right]  _{123}=\left[  \gamma_{5}C_{i\bot}^{\alpha
}C^{\top}\right]  _{21}\left[  \gamma_{\bot}^{\alpha}N_{\bar{n}}\right]
_{3}+\left[  \gamma_{5}C_{i}^{\Vert}C^{\top}\right]  _{21}\left[  \Dsl k
^{\prime}N_{\bar{n}}\right]  _{3},
\end{equation}%
\begin{equation}
C_{i\bot}^{\alpha}=\left[  \Dsl k  \right]  \frac{\partial}{\partial k_{i\bot
}^{\alpha}}-\frac{\delta_{i1}}{2(kk^{\prime})}\frac{1}{x_{1}}\left[
\Dsl k\gamma_{\bot}^{\alpha}\Dsl k  ^{\prime}\right]  -\frac{\delta_{i2}%
}{2(kk^{\prime})}\frac{1}{x_{2}}\left[  \Dsl k  ^{\prime}\gamma_{\bot}%
^{\alpha}\Dsl k\right]  ,\ ~\ \ C_{i}^{\Vert}=\frac{1}{(kk^{\prime})}\frac{1}%
{x_{3}}\left[  \Dsl k  \right]  .
\end{equation}
The chiral-odd projection reads%
\begin{align}
\left.  \left[  P_{\text{tw4}}(x_{i})\right]  _{123}\right\vert _{\text{odd}}
&  =\frac{m_{B}}{4}\left\{  T_{2}^{B}(x_{i})\left[  C_{T}\right]
_{123}+\mathcal{T}_{21}^{B}(x_{i})\left[  C_{\mathcal{T}21}\right]
_{123}+\mathcal{T}_{22}^{B}(x_{i})\left[  C_{\mathcal{T}22}\right]
_{123}\right.
\nonumber \\
&  \left.  +\mathcal{T}_{41}^{B}(x_{i})\left[  C_{\mathcal{T}41}\right]
_{123}+\mathcal{T}_{42}^{B}(x_{i})\left[  C_{\mathcal{T}42}\right]
_{123}\right\}  ,
\end{align}
where%
\begin{equation}
C_{T}=\frac{1}{2(kk^{\prime})}\left[  \Dsl k  \gamma_{\bot\alpha}C\right]
_{12}\left[  \Dsl k  ^{\prime}\gamma_{\bot}^{\alpha}\gamma_{5}N_{\bar{n}%
}\right]  _{3},
\end{equation}%
\begin{align}
C_{\mathcal{T}2i}  &  =\left[  C_{i}^{\Vert}C\right]  _{12}\left[  \gamma
_{5}N_{\bar{n}}\right]  _{3},\ \ C_{\mathcal{T}4i}=\left[  C_{i}^{\Vert}%
\gamma_{5}C\right]  _{12}\ \left[  N_{\bar{n}}\right]  _{3},\\
C_{i}^{\Vert}  &  =\left[  \Dsl k  \gamma_{\bot\alpha}\right]  ~\frac
{\partial}{\partial k_{i\bot}^{\alpha}}+\frac{\delta_{i1}}{(kk^{\prime})}%
\frac{1}{x_{1}}\left[  \Dsl k  ^{\prime}\Dsl k  \right]-\frac
{\delta_{i2}}{(kk^{\prime})}\frac{1}{x_{2}}\left[  \Dsl k  \Dsl k  ^{\prime
}\right] .
\end{align}
 We assume that derivatives in these formulas are applied to the expressions on the
right side, i.e. the derivatives are not applied to the momenta from the left
side and therefore corresponding momenta can be simplified neglecting the power
suppressed components.  For a convenience, we also show the Dirac matrices in
the square brackets: $[\Dsl k  ]$, $\left[  \Dsl k  \gamma_{\bot}^{\alpha
}\right]  $ etc.

The expressions for the twist-5 projectors are more complicated and include
some special contribution, which is completely defined in terms of twist-3 DAs.

 The chiral-even projector is defined as
\begin{align}
\left.  \left[  P_{\text{tw5}}(x_{i})\right]  _{123}\right\vert _{\text{even}%
}    =\frac{m_{B}^{2}}{8}\left\{  \ x_{1}x_{2}\left(  V_{5}^{B}%
(x_{i})\ \left[  C_{V5}\right]  _{123}+A_{5}^{B}(x_{i})\left[  C_{A5}\right]
_{123}\right)  \right. 
\nonumber \\
  +V_{1xy}^{B}(x_{i})\left[  C_{V1xy}\right]  _{123}+A_{1xy}^{B}%
(x_{i})\left[  C_{A1xy}\right]  _{123}%
\nonumber \\
+V_{1xx}^{B}(x_{i})\left[  C_{V1xx}\right]  _{123}+A_{1xx}^{B}(x_{i})\left[
C_{A1xx}\right]  _{123}+V_{1yy}^{B}(x_{i})\left[  C_{V1yy}\right]
_{123}+A_{1yy}^{B}(x_{i})\left[  C_{A1yy}\right]  _{123}%
\nonumber \\
\left.  +V_{1}^{B}(x_{i})\left[  C_{V1}\right]  _{123}+A_{1}^{B}(x_{i})\left[
C_{A1}\right]  _{123}\right\}  ,
\label{Ptw5even}
\end{align}%
where
\begin{equation}
\left[  C_{V5}\right]  _{123}=\left[  C_{5}^{\alpha\beta}\gamma_{5}C\right]
_{12}\left[  i\sigma_{\bot\bot}^{\alpha\beta}N_{\bar{n}}\right]  _{3}+\left[
C_{5}^{\alpha}\gamma_{5}C\right]  _{12}\left[  \Dsl k  ^{\prime}\gamma_{\bot
}^{\alpha}N_{\bar{n}}\right]  _{3},
\end{equation}%
\begin{equation}
\left[  C_{A5}\right]  _{123}=\left[  C_{5}^{\alpha\beta}C\right]
_{12}\left[  i\sigma_{\bot\bot}^{\alpha\beta}\gamma_{5}N_{\bar{n}}\right]
_{3}+\left[  C_{5}^{\alpha}C\right]  _{12}\left[  \Dsl k  ^{\prime}%
\gamma_{\bot}^{\alpha}\gamma_{5}N_{\bar{n}}\right]  _{3},
\end{equation}
with%
\begin{align}
C_{5}^{\alpha\beta}  &  =-\left[  \Dsl k  \right]  \ \frac{\partial}{\partial
k_{1\bot}^{\alpha}}\frac{\partial}{\partial k_{2\bot}^{\beta}}+\frac
{1}{2(kk^{\prime})}\left(  \frac{1}{x_{2}}\left[  \Dsl k  \gamma_{\bot}%
^{\beta}\Dsl k  ^{\prime}\right]  \frac{\partial}{\partial k_{1\bot}^{\alpha}%
}+\frac{1}{x_{1}}\left[  \Dsl k  ^{\prime}\gamma_{\bot}^{\alpha}\Dsl k
\right]  \frac{\partial}{\partial k_{2\bot}^{\beta}}\right)
\nonumber  \\
&  +\frac{1}{2(kk^{\prime})}\frac{1}{x_{1}x_{2}}\left[  \gamma_{\bot}^{\alpha
}\gamma_{\bot}^{\beta}\Dsl k  ^{\prime}\right]  ,
\end{align}%
\begin{equation}
C_{5}^{\alpha}=\frac{1}{x_{3}}\frac{1}{2(kk^{\prime})}\left\{  \left[  \Dsl k
\right]  \left(  \frac{\partial}{\partial k_{2\bot}^{\alpha}}-\frac{\partial
}{\partial k_{1\bot}^{\alpha}}\right)  +\frac{1}{2(kk^{\prime})}\left(
\frac{1}{x_{1}}\left[\Dsl  k^{\prime}\gamma_{\bot}^{\alpha}\Dsl k  \right]
-\frac{1}{x_{2}}~\left[  \Dsl k  \gamma_{\bot}^{\alpha}\Dsl k^{\prime}\right]
\right)  \right\}  .
\end{equation}
The terms in the second line of Eq.(\ref{Ptw5even}) read%
\begin{equation}
\left[  C_{V1xy}\right]  _{123}=\left[  C_{1xy}C\right]  _{12}\left[
\gamma_{5}N_{\bar{n}}\right]  _{3}+\left[  C_{1xy}^{\beta}C\right]
_{12}\left[\Dsl  k^{\prime}\gamma_{\bot}^{\beta}\gamma_{5}N_{\bar{n}}\right]
_{3},
\end{equation}%
\begin{equation}
\left[  C_{A1xy}\right]  _{123}=\left[  C_{1xy}\gamma_{5}C\right]
_{12}\left[  N_{\bar{n}}\right]  _{3}+\left[  C_{1xy}^{\beta}\gamma
_{5}C\right]  _{12}\left[ \Dsl k^{\prime}\gamma_{\bot}^{\beta}N_{\bar{n}}\right]
_{3},
\end{equation}%
\begin{equation}
C_{1xy}=-~\left[  \Dsl k  \gamma_{\bot\beta}\gamma_{\bot\alpha}\right]
\frac{\partial}{\partial k_{1\bot}^{\alpha}}\frac{\partial}{\partial k_{2\bot
}^{\beta}},
\end{equation}%
\begin{equation}
C_{1xy}^{\beta}=-\frac{1}{2}\frac{1}{(kk^{\prime})}\frac{1}{x_{3}}\left(
\left[  \Dsl k  \gamma_{\bot\beta}\gamma_{\bot\alpha}\right]  \frac{\partial
}{\partial k_{1\bot}^{\alpha}}+\left[  \Dsl k  \gamma_{\bot\alpha}\gamma
_{\bot\beta}\right]  \frac{\partial}{\partial k_{2\bot}^{\alpha}}\right)  .
\end{equation}
The terms from the third line of Eq.(\ref{Ptw5even})%
\begin{equation}
\left[  C_{V1xx}\right]  _{123}=\left[  C_{1xx}C\right]  _{12}~~\left[
\gamma_{5}N\right]  _{3}+\left[  C_{1xx}^{\alpha}C\right]  _{12}\left[
\Dsl k^{\prime}\gamma_{\bot}^{\alpha}\gamma_{5}N_{\bar{n}}\right]  _{3},
\end{equation}%
\begin{equation}
\left[  C_{A1xx}\right]  _{123}=~\left[  C_{1xx}\gamma_{5}C\right]
_{12}~~\left[  N\right]  _{3}+\left[  C_{1xx}^{\alpha}\gamma_{5}C\right]
_{12}\left[ \Dsl  k^{\prime}\gamma_{\bot}^{\alpha}N_{\bar{n}}\right]  _{3},
\end{equation}
with%
\begin{equation}
C_{1xx}=-\frac{1}{2}\left[  \Dsl k  \right]  \frac{\partial}{\partial
k_{1\bot}^{\alpha}}\frac{\partial}{\partial k_{1\bot\alpha}}+\frac
{1}{(kk^{\prime})}\frac{1}{x_{1}}\left(  \left[  \Dsl k  \right]  k_{\alpha
}^{\prime}\frac{\partial}{\partial k_{1}^{\alpha}}+\frac{1}{2}\left[
\Dsl k^{\prime}\gamma_{\bot\alpha}\Dsl k\right]  \frac{\partial}{\partial k_{1\bot
}^{\alpha}}\right)  ,
\end{equation}%
\begin{equation}
C_{1xx}^{\alpha}=\frac{1}{2(kk^{\prime})}\frac{1}{x_{3}}\left(  -\left[
\Dsl k  \right]  \frac{\partial}{\partial k_{1\bot}^{\alpha}}+\frac
{1}{2(kk^{\prime})}\frac{1}{x_{1}}~\left[\Dsl  k^{\prime}\gamma_{\bot}^{\alpha
}\Dsl k\right]  \right)  .
\end{equation}
\begin{equation}
\left[  C_{V1yy}\right]  _{123}=\left[  C_{1yy}^{(1)}C\right]  _{12}\left[
\gamma_{5}N\right]  _{3}+\left[  C_{1yy}^{\alpha}C\right]  _{12}\left[
\Dsl k^{\prime}\gamma_{\bot}^{\alpha}\gamma_{5}N\right]  _{3}\ ,
\end{equation}%
\begin{equation}
\left[  C_{A1yy}\right]  _{123}=\left[  C_{1yy}^{(1)}\gamma_{5}C\right]
_{12}\left[  N\right]  _{3}+\left[  C_{1yy}^{\alpha}\gamma_{5}C\right]
_{12}\left[\Dsl  k^{\prime}\gamma_{\bot}^{\alpha}N\right]  _{3}\ ,
\end{equation}
with%
\begin{equation}
C_{1yy}=-\frac{1}{2}\left[  \Dsl k  \right]  \frac{\partial}{\partial
k_{2\bot}^{\alpha}}\frac{\partial}{\partial k_{2\bot\alpha}}~+\frac{1}{\left(
kk^{\prime}\right)  }\frac{1}{x_{2}}\left(  \left[  \Dsl k  \right]
k_{\alpha}^{\prime}\frac{\partial}{\partial k_{2}^{\alpha}}+\frac{1}{2}\left[
\Dsl k\gamma_{\bot\alpha}\Dsl k^{\prime}\right]  \frac{\partial}{\partial k_{2\bot
}^{\alpha}}\right)  ,
\end{equation}%
\begin{equation}
C_{1yy}^{\alpha}=\frac{1}{2(kk^{\prime})}\frac{1}{x_{3}}\left(  -\left[
\Dsl k\right]  \frac{\partial}{\partial k_{2\bot}^{\alpha}}+\frac{1}{2(kk^{\prime
})}\frac{1}{x_{2}}\left[ \Dsl k\gamma_{\bot}^{\alpha}\Dsl k^{\prime}\right]  \right)  .
\end{equation}
The  terms from the last line of Eq.(\ref{Ptw5even})%
\begin{equation}
\left[  C_{V1}\right]  _{123}=\frac{1}{(kk^{\prime})}~\left[  \Dsl k
C\right]  _{12}\left[  \gamma_{5}N_{\bar{n}}\right]  _{3}\ k_{\alpha}^{\prime
}\frac{\partial}{\partial P^{\alpha}},\ \ \ \left[  C_{A1}\right]
_{123}=\frac{1}{(kk^{\prime})}~\left[  \Dsl k  \gamma_{5}C\right]
_{12}\left[  N_{\bar{n}}\right]  _{3}\ k_{\alpha}^{\prime}\frac{\partial
}{\partial P^{\alpha}}.
\end{equation}
Notice that terms $C_{V1}$ and $C_{A1}$ include the derivatives with respect
to initial momentum $P$, see a more detailed explanation below.

The twist-5 chiral-odd projector is defined as
\begin{align}
\left.  \left[  P_{\text{tw5}}(x_{i})\right]  _{123}\right\vert _{\text{odd}}
&  =\frac{m_{B}^{2}}{16}\left(  \ \left[  C_{0xx}\right]  _{123}T_{0xx}%
^{B}(x_{i})+\left[  C_{2xx}\right]  _{123}T_{0xx}^{B}(x_{i})\right.
\nonumber \\
&  +\left[  C_{0yy}\right]  _{123}T_{0yy}^{B}(x_{i})+\left[  C_{2yy}\right]
_{123}T_{0yy}^{B}(x_{i})
\nonumber\\
&  +\ \left[  C_{0xy}\right]  _{123}T_{0xy}^{B}(x_{i})+\left[  C_{2xy}\right]
_{123}T_{2xy}^{B}(x_{i})+\left[  \tilde{C}_{2xy}\right]  _{123}\tilde{T}%
_{2xy}^{B}(x_{i})
\nonumber\\
&  \left.  +\left[  C_{T}\right]  _{123}T_{1}\right)
\end{align}
where the coefficients read%
\begin{equation}
\left[  C_{T}\right]  _{123}=\frac{2}{(kk^{\prime})}\left[ \Dsl k\gamma_{\bot
}^{\mu}C\right]  _{12}\left[  \gamma_{\bot}^{\mu}\gamma_{5}N\right]
_{3}\ k_{\alpha}^{\prime}\frac{\partial}{\partial P^{\alpha}}.
\end{equation}
\begin{align}
C_{0xx}  &  =-\frac{1}{2}\left[  \Dsl k  \gamma_{\bot}^{\alpha}C\right]
_{12}~\left[  \gamma_{\bot}^{\alpha}\gamma_{5}N_{\bar{n}}\right]  _{3}%
\frac{\partial}{\partial k_{1\bot}^{\alpha}}\frac{\partial}{\partial
k_{1\bot\alpha}}+\frac{1}{(kk^{\prime})}\frac{1}{x_{1}}\left[  \Dsl k
\gamma_{\bot}^{\alpha}C\right]  _{12}~\left[  \gamma_{\bot}^{\alpha}\gamma
_{5}N_{\bar{n}}\right]  _{3}\ k_{\alpha}^{\prime}\frac{\partial}{\partial
k_{1}^{\alpha}}
\nonumber\\
&  +\frac{1}{2(kk^{\prime})}\left(  \frac{1}{x_{1}}\left[  \Dsl k^{\prime}%
\gamma_{\bot}^{\alpha}\Dsl k  \gamma_{\bot}^{\sigma}C\right]  _{12}~\left[
\gamma_{\bot}^{\sigma}\gamma_{5}N_{\bar{n}}\right]  _{3}-\frac{1}{x_{3}%
}\left[  \Dsl k  \gamma_{\bot}^{\sigma}C\right]  _{12}~\left[ \Dsl k^{\prime
}\gamma_{\bot}^{\alpha}\gamma_{\bot}^{\sigma}\gamma_{5}N_{\bar{n}}\right]
_{3}\right)  \frac{\partial}{\partial k_{1\bot}^{\alpha}}
\nonumber\\
&  +\frac{1}{4(kk^{\prime})^{2}}\frac{1}{x_{1}x_{3}}\left[  \Dsl k  ^{\prime
}\gamma_{\bot}^{\alpha}\Dsl k  \gamma_{\bot}^{\beta}C\right]  _{12}~\left[
\Dsl k  ^{\prime}\gamma_{\bot}^{\alpha}\gamma_{\bot}^{\beta}\gamma_{5}%
N_{\bar{n}}\right]  _{3}\ .
\end{align}%
\begin{align}
C_{0yy}=-\frac{1}{2}\left[  \Dsl k  \gamma_{\bot}^{\alpha}C\right]
_{12}~\left[  \gamma_{\bot}^{\alpha}\gamma_{5}N_{\bar{n}}\right]  _{3}%
\frac{\partial}{\partial k_{2\bot}^{\alpha}}\frac{\partial}{\partial
k_{2\bot\alpha}}+\frac{1}{(kk^{\prime})}\frac{1}{x_{2}}\left[  \Dsl k
\gamma_{\bot}^{\alpha}C\right]  _{12}~\left[  \gamma_{\bot}^{\alpha}\gamma
_{5}N_{\bar{n}}\right]  _{3}k_{\alpha}^{\prime}\frac{\partial}{\partial
k_{2}^{\alpha}}%
\nonumber \\
+\frac{1}{2(kk^{\prime})}\left(  \frac{1}{x_{2}}\left[  \Dsl k  \gamma_{\bot
}^{\sigma}\gamma_{\bot}^{\alpha}k^{\prime}C\right]  _{12}~\left[  \gamma
_{\bot}^{\sigma}\gamma_{5}N_{\bar{n}}\right]  _{3}-\frac{1}{x_{3}}\left[
\Dsl k  \gamma_{\bot}^{\sigma}C\right]  _{12}~\left[  \Dsl k  ^{\prime}%
\gamma_{\bot}^{\alpha}\gamma_{\bot}^{\sigma}\gamma_{5}N_{\bar{n}}\right]
_{3}\right)  \frac{\partial}{\partial k_{2\bot}^{\alpha}}%
\nonumber \\
+\frac{1}{4(kk^{\prime})^{2}}\frac{1}{x_{2}x_{3}}\left[  \Dsl k  \gamma_{\bot
}^{\sigma}\gamma_{\bot}^{\alpha}\Dsl k^{\prime}C\right]  _{12}~\left[\Dsl  k^{\prime
}\gamma_{\bot}^{\alpha}\gamma_{\bot}^{\sigma}\gamma_{5}N_{\bar{n}}\right]
_{3}\ .
\end{align}
Below we use the short notation%
\begin{equation}
G_{\bot}^{\alpha\beta\sigma\rho}=g_{\bot}^{\alpha\sigma}g_{\bot}^{\beta\rho
}+g_{\bot}^{\alpha\rho}g_{\bot}^{\beta\sigma}-g_{\bot}^{\sigma\rho}g_{\bot
}^{\alpha\beta}.
\end{equation}%
\begin{align}
C_{2yy}=-\frac{1}{2}G_{\bot}^{\alpha\beta\sigma\rho}\left[  \Dsl k
\gamma_{\bot}^{\sigma}C\right]  _{12}~\left[  \gamma_{\bot}^{\rho}\gamma
_{5}N_{\bar{n}}\right]  _{3}\frac{\partial}{\partial k_{2\bot}^{\alpha}}%
\frac{\partial}{\partial k_{2\bot}^{\beta}}%
\hspace{15em} 
\nonumber \\
+G_{\bot}^{\alpha\beta\sigma\rho}\frac{1}{2(kk^{\prime})}\left(  \frac
{1}{x_{2}}\left[  \Dsl k  \gamma_{\bot}^{\sigma}\gamma_{\bot}^{\beta}%
 \Dsl k^{\prime}C\right]  _{12}~\left[  \gamma_{\bot}^{\rho}\gamma_{5}N_{\bar{n}%
}\right]  _{3}-\frac{1}{x_{3}}\left[  \Dsl k  \gamma_{\bot}^{\sigma}C\right]
_{12}~\left[  \Dsl k  ^{\prime}\gamma_{\bot}^{\beta}\gamma_{\bot}^{\rho}%
\gamma_{5}N_{\bar{n}}\right]  _{3}\right)  \frac{\partial}{\partial k_{2\bot
}^{\alpha}}%
\nonumber \\
+G_{\bot}^{\alpha\beta\sigma\rho}\frac{1}{4(kk^{\prime})^{2}}\frac{1}%
{x_{2}x_{3}}\left[  \Dsl k  \gamma_{\bot}^{\sigma}\gamma_{\bot}^{\alpha
} \Dsl k^{\prime}C\right]  _{12}~\left[ \Dsl k^{\prime}\gamma_{\bot}^{\beta}\gamma
_{\bot}^{\rho}\gamma_{5}N_{\bar{n}}\right]  _{3}.
\end{align}%
\begin{align}
C_{2xx}=-\frac{1}{2}G_{\bot}^{\alpha\beta\sigma\rho}\left[  \Dsl k
\gamma_{\bot}^{\sigma}C\right]  _{12}~\left[  \gamma_{\bot}^{\rho}\gamma
_{5}N_{\bar{n}}\right]  _{3}\frac{\partial}{\partial k_{1\bot}^{\alpha}}%
\frac{\partial}{\partial k_{1\bot}^{\beta}}%
\hspace{15em} 
\nonumber \\
+G_{\bot}^{\alpha\beta\sigma\rho}\frac{1}{2(kk^{\prime})}\left(  \frac
{1}{x_{1}}\left[ \Dsl k^{\prime}\gamma_{\bot}^{\beta}\Dsl k  \gamma_{\bot}%
^{\sigma}C\right]  _{12}~\left[  \gamma_{\bot}^{\rho}\gamma_{5}N_{\bar{n}%
}\right]  _{3}-\frac{1}{x_{3}}\left[  \Dsl k  \gamma_{\bot}^{\sigma}C\right]
_{12}~\left[\Dsl  k^{\prime}\gamma_{\bot}^{\beta}\gamma_{\bot}^{\rho}\gamma
_{5}N_{\bar{n}}\right]  _{3}\right)  \frac{\partial}{\partial k_{1\bot
}^{\alpha}}%
\nonumber \\
+G_{\bot}^{\alpha\beta\sigma\rho}\frac{1}{4(kk^{\prime})^{2}}\frac{1}%
{x_{1}x_{3}}\left[  \Dsl k  ^{\prime}\gamma_{\bot}^{\alpha}\Dsl k
\gamma_{\bot}^{\sigma}C\right]  _{12}~\left[  \Dsl k  ^{\prime}\gamma_{\bot
}^{\beta}\gamma_{\bot}^{\rho}\gamma_{5}N_{\bar{n}}\right]  _{3}.
\end{align}%
\begin{align}
C_{0xy}=-\left[  \Dsl k  \gamma_{\bot}^{\sigma}C\right]  _{12}~\left[
\gamma_{\bot}^{\sigma}\gamma_{5}N_{\bar{n}}\right]  _{3}\frac{\partial
}{\partial k_{1\bot}^{\alpha}}\frac{\partial}{\partial k_{2\bot\alpha}}%
\hspace{15em} 
\nonumber \\
  +\frac{1}{2(kk^{\prime})}\left(  \frac{1}{x_{2}}\left[  \Dsl k
\gamma_{\bot}^{\sigma}\gamma_{\bot}^{\alpha} \Dsl k^{\prime}C\right]  _{12}~\left[
\gamma_{\bot}^{\sigma}\gamma_{5}N_{\bar{n}}\right]  _{3}-\frac{1}{x_{3}%
}\left[  \Dsl k  \gamma_{\bot}^{\sigma}C\right]  _{12}~\left[\Dsl  k^{\prime
}\gamma_{\bot}^{\beta}\gamma_{\bot}^{\sigma}\gamma_{5}N_{\bar{n}}\right]
_{3}\right)  \frac{\partial}{\partial k_{1\bot}^{\alpha}}
\nonumber \\
 +\frac{1}{2(kk^{\prime})}\left(  \frac{1}{x_{1}}\left[\Dsl  k^{\prime}%
\gamma_{\bot}^{\alpha}\Dsl k  \gamma_{\bot}^{\sigma}C\right]  _{12}~\left[
\gamma_{\bot}^{\sigma}\gamma_{5}N_{\bar{n}}\right]  _{3}-\frac{1}{x_{3}%
}\left[  \Dsl k  \gamma_{\bot}^{\sigma}C\right]  _{12}~\left[\Dsl  k^{\prime
}\gamma_{\bot}^{\alpha}\gamma_{\bot}^{\sigma}\gamma_{5}N_{\bar{n}}\right]
_{3}\right)  \frac{\partial}{\partial k_{2\bot}^{\alpha}}%
\nonumber \\
+\frac{1}{4(kk^{\prime})^{2}}\left(  \frac{1}{x_{1}x_{3}}\left[ \Dsl k^{\prime
}\gamma_{\bot}^{\alpha}\Dsl k  \gamma_{\bot}^{\sigma}C\right]  _{12}~+\frac
{1}{x_{2}x_{3}}~\left[  \Dsl k  \gamma_{\bot}^{\sigma}\gamma_{\bot}^{\alpha
}\Dsl k^{\prime}C\right]  _{12}~\right)  \left[\Dsl  k^{\prime}\gamma_{\bot}^{\alpha
}\gamma_{\bot}^{\sigma}\gamma_{5}N_{\bar{n}}\right]  _{3}.
\end{align}%
\begin{equation}
C_{2xy}=\tilde{C}_{\alpha\beta\sigma\rho}\ G_{\bot}^{\alpha\beta\sigma\rho
},\ \ \tilde{C}_{2xy}=\tilde{C}_{\alpha\beta\sigma\rho}\left(  g_{\bot
}^{\alpha\rho}g_{\bot}^{\beta\sigma}-g_{\bot}^{\alpha\sigma}g_{\bot}%
^{\beta\rho}\right)  ,
\end{equation}
where%
\begin{align}
\tilde{C}_{\alpha\beta\sigma\rho}=-\left[  \Dsl k  \gamma_{\bot}^{\sigma
}C\right]  _{12}~\left[  \gamma_{\bot}^{\rho}\gamma_{5}N_{\bar{n}}\right]
_{3}\frac{\partial}{\partial k_{1\bot}^{\alpha}}\frac{\partial}{\partial
k_{2\bot\alpha}}%
\hspace{15em} 
\nonumber \\
  +\frac{1}{2(kk^{\prime})}\left(  \frac{1}{x_{2}}\left[  \Dsl k
\gamma_{\bot}^{\sigma}\gamma_{\bot}^{\beta}\Dsl k^{\prime}C\right]  _{12}~\left[
\gamma_{\bot}^{\rho}\gamma_{5}N_{\bar{n}}\right]  _{3}-\frac{1}{x_{3}}\left[
\Dsl k  \gamma_{\bot}^{\sigma}C\right]  _{12}~\left[\Dsl  k^{\prime}\gamma_{\bot
}^{\beta}\gamma_{\bot}^{\rho}\gamma_{5}N_{\bar{n}}\right]  _{3}\right)
\frac{\partial}{\partial k_{1\bot}^{\alpha}}
\nonumber \\
  +\frac{1}{2(kk^{\prime})}\left(  \frac{1}{x_{1}}\left[\Dsl  k^{\prime}%
\gamma_{\bot}^{\alpha}\Dsl k  \gamma_{\bot}^{\sigma}C\right]  _{12}~\left[
\gamma_{\bot}^{\rho}\gamma_{5}N_{\bar{n}}\right]  _{3}-\frac{1}{x_{3}}\left[
\Dsl k  \gamma_{\bot}^{\sigma}C\right]  _{12}~\left[\Dsl  k^{\prime}\gamma_{\bot
}^{\alpha}\gamma_{\bot}^{\rho}\gamma_{5}N_{\bar{n}}\right]  _{3}\right)
\frac{\partial}{\partial k_{2\bot}^{\beta}}%
\nonumber \\
  +\frac{1}{4(kk^{\prime})^{2}}\left(  -\frac{1}{x_{1}x_{2}}~\left[
\Dsl k^{\prime}\gamma_{\bot}^{\alpha}\Dsl k  \gamma_{\bot}^{\sigma}\gamma_{\bot
}^{\beta}\Dsl k^{\prime}C\right]  _{12}~\left[  \gamma_{\bot}^{\rho}\gamma
_{5}N_{\bar{n}}\right]  _{3}
\right.  
\nonumber \\ 
\left.
+\frac{1}{x_{1}x_{3}}\left[ \Dsl k^{\prime
}\gamma_{\bot}^{\alpha}\Dsl k  \gamma_{\bot}^{\sigma}C\right]  _{12}~\left[
\Dsl k^{\prime}\gamma_{\bot}^{\beta}\gamma_{\bot}^{\rho}\gamma_{5}N_{\bar{n}
}\right]  _{3}
    +\frac{1}{x_{2}x_{3}}~\left[  \Dsl k  \gamma_{\bot}^{\sigma}%
\gamma_{\bot}^{\beta}  \Dsl k^{\prime}C\right]  _{12}~\left[ \Dsl k^{\prime}\gamma_{\bot
}^{\alpha}\gamma_{\bot}^{\rho}\gamma_{5}N_{\bar{n}}\right]  _{3}\right)  .\
\end{align}

The described  projectors depend on the derivatives with respect to quark momenta $\partial/\partial k_{1,2} $
and also include  the terms   with derivative with respect to charmonium momentum $\partial/\partial P$.  
These terms only depend on the twist-3 DAs  $V_{1}^{B}$, $A_{1}^{B}$ and $T_{1}^{B}$ and they are related to the matrix
element of the leading-twist operator with the total derivative%
\begin{equation}
(z_{3}n)\left\langle 0\right\vert (\bar{n}\partial)\left[  \xi_{1}\xi_{2}\xi_{3}\right]  \left\vert B(k)\right\rangle
=-i(\bar nk)(z_{3}n)\left\langle 0\right\vert \xi_{1}\xi_{2}\xi_{3}\left\vert B(k)\right\rangle , 
\label{totDme}%
\end{equation}
where $(\bar n k)\equiv k_{-}=2m_{B}^{2}/k_{+}$. This contribution does not vanish, in contrast
to an operator with the total transverse derivative. It is clear that such
contributions generate power corrections associated with the power suppressed component of
the total baryon momentum $k$. Notice, that  the calculation of the contributions with
derivatives $\partial/\partial k_{1,2}$  implies  that
$k_{3}=k-k_{1}-k_{2}$ as it follows from the  momentum conservation. 
The  contributions with the matrix elements as in Eq.(\ref{totDme})  can be rewritten 
in a special way. This is related with the details of the Fourier
transformation  and with the derivation  of the momentum
conservation $\delta$-function.

Let us consider, as example,  the contribution with
$V_{1}(x)$ and $V_{1}(y_{i})$ and the diagram as in Fig.\ref{figure}$(a)$.  The following
discussion is also applicable for other cases.  Performing the standard steps of
the calculation  one obtains the following result
\begin{equation}
\delta(k+k^{\prime}-P)\ iM[V_{1}V_{1}]_{\text{tw5}}=~\bar{V}_{n}\gamma_{\bot
}^{\rho}N_{\bar{n}}\int Dy_{i}V_{1}(y_{i})\int Dx_{i}~V_{1}(x_{i}%
)\ T_{V1V1}(x_{i},y_{i}),
\end{equation}
where
\begin{align}
T_{V1V1}(x_{i},y_{i})  &  =\int dz_{1}^{\prime}dz_{2}^{\prime}dz_{3}^{\prime
}e^{iP\cdot(z_{1}^{\prime}+z_{2}^{\prime})/2}~e^{-i(k_{1}^{\prime}%
z_{1})-i(k_{2}^{\prime}z_{2})-i(k_{3}^{\prime}z_{3})}~\nonumber\\
&  \int dz_{1}dz_{2}dz_{3}~k_{-}(-iz_{3}\cdot n)~e^{-i\left(  k_{1}%
z_{1}\right)  -i\left(  k_{2}z_{2}\right)  -i\left(  k_{3}z_{3}\right)
}~\tilde{D}[z_{123},z_{123}^{\prime}], \label{TV1V1}%
\end{align}
where $\tilde{D}[z_{123},z_{123}^{\prime}]$ denotes the analytical expression
for the diagram in the position space with the requiered operator vertices,
$z_{i}^{\prime}$ and $z_{i}$ describe the coordinates of the QCD vertices, see
Fig.\ref{figure2}.  The various Fourier exponents in the \textit{lhs} appear from the
hadronic matrix elements.  The quark momenta $k_{i}^{\prime}$ and $k_{i}$
 in Eq.(\ref{TV1V1})  the same as in Eq.(\ref{def:ki}).

In order to extract the momentum conservation $\delta$-function from the
expression for $T_{V1V1}$ one uses that $x_{3}=1-x_{1}-x_{2}$, $y_{3}%
=1-y_{1}-y_{2}$ and the translation invariance of the diagram $\tilde
{D}[z_{123},z_{123}^{\prime}]$. Performing  redefinition $z_{i}\rightarrow z_{i}+z_{3}$ and
$z_{i}^{\prime}\rightarrow z_{i}^{\prime}+z_{3}^{\prime}$ for $i=1,2$  one 
 defines  the Fourier exponent with the coordinate $z_3$%
\begin{align}
e^{iP\cdot(z_{1}^{\prime}+z_{2}^{\prime})/2}~e^{-i(k_{1}^{\prime}%
z_{1})-i(k_{2}^{\prime}z_{2})-i(k_{3}^{\prime}z_{3})}~e^{-i\left(  k_{1}%
z_{1}\right)  -i\left(  k_{2}z_{2}\right)  -i\left(  k_{3}z_{3}\right)  }  
\nonumber \\
 \rightarrow \left\{ e^{i(Pz_{3})-i(k_{-}^{\prime}z_{3+})/2-i(k_{+}z_{3-})/2}\right\}~ e^{iP\cdot(z_{1}^{\prime}+z_{2}^{\prime})/2}~e^{-i(k_{1}^{\prime
}z_{1})-i(k_{2}^{\prime}z_{2})}~e^{-i\left(  k_{1}z_{1}\right)  -i\left(
k_{2}z_{2}\right)  },
\end{align}
and
\begin{equation}
\tilde{D}[z_{123},z_{123}^{\prime}]\rightarrow\tilde{D}[z_{12},z_{12}^{\prime
}].
\end{equation}
Therefore
\begin{align}
T_{V1V1}(x_{i},y_{i})  &  =\int dz_{3}~k_{-}(-iz_{3}\cdot n)e^{i(Pz_{3}%
)-i(k_{-}^{\prime}z_{3+})/2-i(k_{+}z_{3-})/2}\label{TV1V1-2}
\nonumber \\
&  \times\int dz_{1}^{\prime}dz_{2}^{\prime}dz_{3}^{\prime}e^{iP\cdot
(z_{1}^{\prime}+z_{2}^{\prime})/2}~e^{-i(k_{1}^{\prime}z_{1})-i(k_{2}^{\prime}z_{2})}
\nonumber  \\
&\times \int dz_{1}dz_{2}~e^{-i\left(  k_{1}z_{1}\right)  -i\left(  k_{2}%
z_{2}\right)  }~\tilde{D}[z_{12},z_{12}^{\prime}].
\end{align}
The integrals over $dz'_i dz_j$ in this expression can be understood as FT of the diagram
$\tilde{D}$%
\begin{align}
\int dz_{1}^{\prime}dz_{2}^{\prime}dz_{3}^{\prime}e^{iP\cdot(z_{1}^{\prime
}+z_{2}^{\prime})/2}~e^{-i(k_{1}^{\prime}z_{1})-i(k_{2}^{\prime}z_{2})}\int
dz_{1}dz_{2}~e^{-i\left(  k_{1}z_{1}\right)  -i\left(  k_{2}z_{2}\right)
}~\tilde{D}[z_{12},z_{12}^{\prime}]
\nonumber \\
=D(k_{12},k_{12}^{\prime},P).
\end{align}
The first line in Eq.(\ref{TV1V1-2}) gives the $\delta$-function if the
factor \ $k_{-}(-iz_{3}\cdot n)$ is rewritten as%
\begin{equation}
k_{-}(-iz_{3}\cdot n)=k_{-}\left.  n_{\alpha}\frac{\partial}{\partial
l^{\alpha}}e^{-i\left(  l\cdot z_{3}\right)  }\right\vert _{l=0},
\end{equation}
where $l$ is just an auxiliary momentum. \ This gives
\begin{equation}
T_{V1V1}(x_{i},y_{i})=k_{-}\left.  \left[  n_{\alpha}\frac{\partial}{\partial
l^{\alpha}}\delta(P-k_{-}^{\prime}n/2-k_{+}\bar{n}/2-l)\right]  D(k_{12}%
,k_{12}^{\prime},P)\right\vert _{l=0}%
\end{equation}%
\begin{equation}
=k_{-}\left.  \left[  -n_{\alpha}\frac{\partial}{\partial P^{\alpha}}%
\delta(P-k_{-}^{\prime}n/2-k_{+}\bar{n}/2)\right]  D(k_{12},k_{12}^{\prime
},P)\right\vert _{P=2m_{c}\omega}%
\end{equation}%
\begin{equation}
=\delta(P-k_{-}^{\prime}n/2-k_{+}\bar{n}/2)\ \frac{2m_{B}^{2}}{k_{+}}\left.
n_{\alpha}\frac{\partial}{\partial P^{\alpha}}D(k_{12},k_{12}^{\prime
},P)\right\vert _{P=2m_{c}\omega}%
\end{equation}%
\begin{equation}
\simeq\delta(P-k^{\prime}-k)\ \frac{2m_{B}^{2}}{(kk^{\prime})}\left.
k_{\alpha}^{\prime}\frac{\partial}{\partial P^{\alpha}}D(k_{12},k_{12}%
^{\prime},P)\right\vert _{P=2m_{c}\omega},
\end{equation}
where one assumes that $P^{2}$ is not fixed before the differentiation. This
result explains  the computational prescription for the terms with derivatives $\partial/\partial P$ .

 Let us also to note  that such terms also include the endpoint singularities,
which cancel in the sum of the all twist-5 contributions. Therefore, it is
important to take them into account properly.

\section{Analytical expressions for the hard kernels $K_{XY}$}

\label{app:d}

In these section we provide the analytical expressions for the hard kernels
$K_{XY}(x_{i},y_{i})$ in the integrals $J[X,Y]$ defined in Eq.(\ref{def:J}).
In the following we use convenient notation%
\begin{align}
D_{i}  &  =x_{i}(1-y_{i})+y_{i}(1-x_{i}),\\
f(x_{i})  &  \equiv f(x_{123})\equiv f(x_{1},x_{2},x_{3}).
\end{align}

The integrals, which describe twist-4$\times$twist-4 contribution read
\begin{equation}
\left.  J_{44}^{B}\right\vert _{\text{even}}=J[\mathcal{V}^B_{1},\mathcal{V}^B%
_{1}]+J[\mathcal{V}^B_{2},\mathcal{V}^B_{1}]+J[\mathcal{A}^B_{1},\mathcal{A}^B%
_{1}]+J[\mathcal{A}^B_{2},\mathcal{A}^B_{1}]+J[\mathcal{A}^B_{1},\mathcal{V}^B%
_{1}]+J[\mathcal{A}^B_{2},\mathcal{V}^B_{1}],
\end{equation}
where \
\begin{equation}
K_{\mathcal{V}_{1}\mathcal{V}_{1}}(x_{i},y_{i})=K_{\mathcal{A}_{1}%
\mathcal{A}_{1}}(x_{i},y_{i})=\frac{1}{2}\frac{x_{2}y_{2}}{D_{1}D_{2}D_{3}},
\end{equation}%
\begin{equation}
K_{\mathcal{V}_{2}\mathcal{V}_{1}}(x_{i},y_{i})=K_{\mathcal{A}_{2}%
\mathcal{A}_{1}}(x_{i},y_{i})=\frac{1}{4}\frac{x_{1}y_{1+}x_{2}y_{2}}%
{D_{1}D_{2}D_{3}},
\end{equation}%
\begin{equation}
K_{\mathcal{A}_{1}\mathcal{V}_{1}}(x_{i},y_{i})=\frac{x_{2}y_{2}}{D_{1}%
D_{2}D_{3}},
\end{equation}%
\begin{equation}
K_{\mathcal{A}_{2}\mathcal{V}_{1}}(x_{i},y_{i})=\frac{1}{2}\frac{x_{2}%
y_{2}-x_{1}y_{1}}{D_{1}D_{2}D_{3}}.
\end{equation}

\begin{equation}
\left.  J_{44}^{B}\right\vert _{\text{odd}}=J[\mathcal{T}^B_{21}+\mathcal{T}^B%
_{41},\mathcal{T}^B_{21}+\mathcal{T}^B_{41}]+J[\mathcal{T}^B_{22}-\mathcal{T}^B%
_{42},\mathcal{T}^B_{21}-\mathcal{T}^B_{41}]+J[\mathcal{T}^B_{22}+\mathcal{T}^B%
_{42},\mathcal{T}^B_{21}+\mathcal{T}^B_{41}],
\end{equation}
where%
\begin{equation}
K_{\mathcal{T}_{21}+\mathcal{T}_{41},\mathcal{T}_{21}+\mathcal{T}_{41}}%
(x_{i},y_{i})=-\frac{1}{2}\frac{x_{2}+y_{2}-3x_{2}y_{2}}{D_{1}D_{2}D_{3}},
\end{equation}%
\begin{equation}
K_{\mathcal{T}_{22}+\mathcal{T}_{42},\mathcal{T}_{21}+\mathcal{T}_{41}}%
(x_{i},y_{i})=-\frac{1}{4}\frac{1-2x_{1}-2y_{2}+6x_{1}y_{2} }{D_{1}D_{2}D_{3}},
\end{equation}%
\begin{equation}
K_{\mathcal{T}_{22}-\mathcal{T}_{42},\mathcal{T}_{21}-\mathcal{T}_{41}}%
(x_{i},y_{i})=-\frac{1}{4}\frac{ x_{3}y_{3} }{D_{1}D_{2}D_{3} }.
\end{equation}

 The  twist-3$\times$twist-5  integrals have the following kernels. 
\begin{eqnarray}
&&\left.  J_{35}^{B}\right\vert _{\text{even}}= J[V^B_{5},V^B_{1}]+J[V^B_{5},A^B_{1}]+J[A^B_{5},A^B_{1}]+J[A^B_{5},V^B_{1}]
\\ &&
\phantom{J[V_{5},V_{1}]]}
+J_{B}[V^B_{Z},V^B_{1}]+J_{B}[A^B_{Z},V^B_{1}]+J_{B}[A^B_{Z},A^B_{1}]+J_{B}[V^B_{Z},A^B_{1}].
\end{eqnarray}
The first four integrals have relatively simple structure%
\begin{equation}
K_{V5V1}(x_{i},y_{i})=K_{A5A1}(x_{i},y_{i})=\frac{1}{16}\frac{x_{1}x_{2}}{D_{1}D_{2}D_{3}}%
(-1+2y_3+x_3 y_3+x_1y_2+x_2y_1),
\end{equation}%
\begin{equation}
K_{A5V1}(x_{i},y_{i})=K_{V5A1}(x_{i},y_{i})=\frac{1}{16}\frac{x_{1}x_{2}}{D_{1}D_{2}D_{3}}
\left(
y_1-y_2+x_1(y_2-y_3)-x_2(y_1-y_3)
\right)  .
\end{equation}

The integrals $J_{Z}$ defined in Eqs.(\ref{JVZV1})-(\ref{JAZA1}) are also
well defined, but they are given by the sum of the singular terms $J[X,Y]$.
The analytical expression for the total sum is quite complicated and it is
more constructive  to describe  the expressions for the individual kernels in $J[X,Y] $.
Then
\begin{equation}
J_{Z}[V^B_{Z},V^B_{1}]=J[V^B_{1xx},V^B_{1}]+J[V^B_{1yy},V^B_{1}]+J[V^B_{1xy},V^B_{1}%
]+(1-\delta_{B\Lambda})J[V^B_{1},V^B_{1}]. \label{JBVZV1A}%
\end{equation}
Each term in Eq.(\ref{JBVZV1A}) has the following kernel%
\begin{equation}
K_{V1xxV1}(x_{i},y_{i})=\frac{1}{32}\left(  \frac{A}{D_{1}D_{3}}+\frac
{B}{D_{1}D_{2}}+\frac{C}{D_{2}D_{3}}\right)  ,
\end{equation}
where
\begin{equation}
A=-\frac{1}{x_{1}}\left(  \frac{1}{D_{1}}+\frac{1-2x_{2}}{D_{3}}\right)
(x_{3}y_{1}+x_{1}y_{3})+1-\frac{1}{2x_{1}}+\frac{1}{2x_{3}}-\frac{1}%
{2x_{1}x_{3}}+\frac{y_{1}}{y_{3}}+\frac{x_{3}y_{1}}{x_{1}y_{3}},
\end{equation}

\begin{equation}
B=\frac{x_{2}(1-2x_{1})}{2x_{1}x_{3}},\,\,
C=-\frac{(1-2x_{2})}{x_{1}D_{3}}(x_{2}y_{3}+x_{3}y_{2})+\frac{y_{2}}{y_{3}%
}+\frac{x_{2}}{2x_{1}x_{3}}+\frac{x_{3}y_{2}}{x_{1}y_{3}}.
\end{equation}%
\begin{equation}
K_{V1yyV1}(x_{i},y_{i})=K_{V1xxV1}(x_{213},y_{213}).
\end{equation}%
\begin{equation}
K_{V1xyV1}(x_{i},y_{i})=\frac{1}{8}\left(  \frac{A}{D_{1}D_{3}}+\frac{C}%
{D_{2}D_{3}}\right)  ,
\end{equation}
where%
\begin{equation}
A=-\frac{1}{D_{3}}(x_{1}y_{3}+x_{3}y_{1})+\frac{y_{1}}{2y_{3}},\ \ C=-\frac
{1}{D_{3}}(x_{3}y_{2}+x_{2}y_{3})+\frac{y_{2}}{2y_{3}}.
\end{equation}%
\begin{equation}
K_{V1V1}(x_{i},y_{i})=\frac{1}{64}\left(  \frac{A}{D_{1}D_{3}}+\frac{C}%
{D_{2}D_{3}}\right)  ,
\end{equation}
where%
\begin{equation}
A=\left(  \frac{1-2x_{1}}{D_{1}}-\frac{1-2x_{3}}{D_{3}}\right)  (y_{1}%
x_{3}+y_{3}x_{1})-\bar{x}_{2}-\frac{2y_{1}x_{3}}{y_{3}},
\end{equation}%
\begin{equation}
C=\left(  \frac{1-2x_{2}}{D_{2}}-\frac{1-2x_{3}}{D_{3}}\right)  (y_{2}%
x_{3}+y_{3}x_{2})-\bar{x}_{1}-\frac{2y_{2}x_{3}}{y_{3}}.
\end{equation}

The next integral reads
\begin{equation}
J_{Z}[V^B_{Z},A^B_{1}]=J[V^B_{1xx},A^B_{1}]+J[V^B_{1yy},A^B_{1}]+J[V^B_{1xy},A^B_{1}]+(1-\delta_{B\Lambda})J[V^B_{1},A^B_{1}].
\end{equation}
The kernels of the individual integrals read
\begin{equation}
K_{V1xxA1}(x_{i},y_{i})=\frac{1}{32}\left(  \frac{A}{D_{1}D_{3}}+\frac
{B}{D_{1}D_{2}}+\frac{C}{D_{2}D_{3}}\right)  ,
\end{equation}
where
\begin{equation}
A=\frac{1}{x_1}\left(  \frac{1}{D_{1}}+\frac{1-2x_{2}}{D_{3}}\right)
(x_{3}y_{1}+x_{1}y_{3})+1-\frac{1}{2x_{1}}+\frac{1}{2x_{3}}-\frac{y_{1}}%
{y_{3}}-\frac{1}{2x_{1}x_{3}}-\frac{x_{3}y_{1}}{x_{1}y_{3}},
\end{equation}%
\begin{equation}
B=\frac{2 x_{2}(1-2x_{1})}{x_{1}x_{3}},\,\, 
C=-\frac{1-2x_{2}}{x_{1}D_{3}}(x_{3}y_{2}+x_{2}y_{3})+\frac{\bar{x}%
_{2}y_{2}}{x_{1}y_{3}}+\frac{x_{2}}{2x_{1}x_{3}}.
\end{equation}%
\begin{equation}
K_{V1yyA1}(x_{i},y_{i})=K_{V1xxA1}(x_{213},y_{213}).
\end{equation}%
\begin{equation}
K_{V1xyA1}(x_{i},y_{i})=\frac{1}{8}\left(  \frac{A}{D_{1}D_{3}}+\frac{C}%
{D_{2}D_{3}}\right)  ,
\end{equation}
where%
\begin{equation}
A=\frac{1}{D_{3}}(x_{1}y_{3}+x_{3}y_{1})-\frac{y_{1}}{2y_{3}},\ \ \ C=-\frac
{1}{D_{3}}(x_{2}y_{3}+x_{3}y_{2})+\frac{y_{2}}{2y_{3}}.
\end{equation}%
\begin{equation}
K_{V1A1}(x_{i},y_{i})=\frac{1}{64}\left(  \frac{A}{D_{1}D_{3}}+\frac{C}%
{D_{2}D_{3}}\right)  ,
\end{equation}
where%
\begin{equation}
A=-\left(  \frac{1-2x_{1}}{D_{1}}-\frac{1-2x_{3}}{D_{3}}\right)  (x_{3}%
y_{1}+x_{1}y_{3})+3x_{1}-x_{3}+2\frac{y_{1}x_{3}}{y_{3}},
\end{equation}%
\begin{equation}
C=\left(  \frac{1-2x_{2}}{D_{2}}-\frac{1-2x_{3}}{D_{3}}\right)  (x_{3}%
y_{2}+x_{2}y_{3})-3x_{2}+x_{3}-2\frac{y_{2}x_{3}}{y_{3}}.
\end{equation}
The next integral reads%
\begin{equation}
J_{Z}[A^B_{Z},V^B_{1}]=J[A^B_{1xx},V^B_{1}]+J[A^B_{1yy},V^B_{1}]+J[A^B_{1xy},V^B_{1}%
]+\delta_{B\Lambda}J[A^B_{1},V^B_{1}].
\end{equation}
The kernels of the individual integrals read%
\begin{equation}
K_{A1xxV1}(x_{i},y_{i})=K_{V1xxA1}(x_{i},y_{i}).
\end{equation}%
\begin{equation}
K_{A1yyV1}(x_{i},y_{i})=K_{A1xxV1}(x_{213},y_{213}).
\end{equation}%
\begin{equation}
K_{A1xyV1}(x_{i},y_{i})=K_{V1xyA1}(x_{i},y_{i}).
\end{equation}%
\begin{equation}
K_{A1V1}(x_{i},y_{i})=K_{V1A1}(x_{i},y_{i}).
\end{equation}

The next integral reads
\begin{equation}
J_{Z}[A^B_{Z},A^B_{1}]=J[A^B_{1xx},A^B_{1}]+J[A^B_{1yy},V^B_{1}]+J[A^B_{1xy},V^B_{1}]+\delta_{B\Lambda}J[A^B_{1},A^B_{1}].
\end{equation}
The kernels of the individual integrals read%
\begin{equation}
K_{A1xxA1}(x_{i},y_{i})=K_{V1xxV1}(x_{i},y_{i}).
\end{equation}%
\begin{equation}
K_{A1yyA1}(x_{i},y_{i})=K_{A1xxA1}(x_{213},y_{213}).
\end{equation}%
\begin{equation}
K_{A1xyA1}(x_{i},y_{i})=K_{V1xyV1}(x_{i},y_{i}).
\end{equation}%
\begin{equation}
K_{A1A1}(x_{i},y_{i})=K_{V1V1}(x_{i},y_{i}).
\end{equation}

The last integral reads%
\begin{eqnarray}
J_Z[T^B_Z,T^B_1]=  J[T^B_{0xx},T^B_{1}]+J[T^B_{0yy},T^B_{1}]+J[T^B_{0xy},T^B_{1}]+J[\tilde{T}^B_{2xy},T^B_{1}]
\nonumber \\
+(1-\delta_{B\Lambda})J[T^B_{1},T^B_{1}].
\end{eqnarray}
The kernels of the individual integrals read
\begin{equation}
K_{T0xxT1}(x_{i},y_{i})=\frac{1}{16}\left(  \frac{A}{D_{1}D_{3}}+\frac{B}%
{D_{1}D_{2}}+\frac{C}{D_{2}D_{3}}\right)  ,
\end{equation}
where%
\begin{equation}
A=1-\frac{1}{x_{3}}+\frac{1}{2x_{1}}+\frac{1}{2x_{3}}-\frac{1}{2x_{1}x_{3}},
\end{equation}%
\begin{equation}
B=(x_{2}y_{1}+x_{1}y_{2})\left(  \frac{1}{x_{3}y_{3}}+\frac{1}{x_{1}y_{3}%
}-\frac{1}{x_{1}D_{1}}\right)  +\frac{x_{2}}{2x_{1}x_{3}},\,\,
C=-\frac{x_{2}}{x_{1}}+\frac{x_{2}}{2x_{1}x_{3}}.
\end{equation}%
\begin{equation}
K_{T0yyT1}(x_{i},y_{i})=K_{T0xxT1}(x_{213},y_{213}).
\end{equation}%
\begin{equation}
K_{T0xyT1}(x_{i},y_{i})=\frac{1}{16}\left(  \frac{A}{D_{1}D_{3}}+\frac{B}%
{D_{1}D_{2}}+\frac{C}{D_{2}D_{3}}\right)  ,
\end{equation}
where%
\begin{equation}
A=1-\frac{1}{2x_{3}}+\frac{1}{2x_{1}}-\frac{x_{1}}{x_{2}}-\frac{1}{2x_{1}%
x_{3}}+\frac{x_{1}}{2x_{2}x_{3}},
\end{equation}

\begin{equation}
B=\frac{x_{1}}{2x_{2}x_{3}}+\frac{x_{2}}{2x_{1}x_{3}}+\frac{2x_{2}y_{1}}%
{x_{3}y_{3}}+\frac{2x_{1}y_{2}}{x_{3}y_{3}},
\end{equation}%
\begin{equation}
C=1-\frac{1}{2x_{3}}+\frac{1}{2x_{2}}-\frac{x_{2}}{x_{1}}-\frac{1}{2x_{2}%
x_{3}}+\frac{x_{2}}{2x_{1}x_{3}}.
\end{equation}%
\begin{equation}
K_{\tilde{T}2xyT1}(x_{i},y_{i})=\frac{1}{16}\left(  \frac{A}{D_{1}D_{3}}%
+\frac{B}{D_{1}D_{2}}+\frac{C}{D_{2}D_{3}}\right)  ,
\end{equation}
where%
\begin{equation}
A=-1+\frac{1}{2x_{3}}-\frac{1}{2x_{1}}-\frac{x_{1}}{x_{2}}+\frac{1}%
{2x_{1}x_{3}}+\frac{x_{1}}{2x_{2}x_{3}},\,\,
B=\frac{x_{1}}{2x_{2}x_{3}}-\frac{x_{2}}{2x_{1}x_{3}}%
\end{equation}%
\begin{equation}
C=1-\frac{1}{2x_{3}}+\frac{1}{2x_{2}}+\frac{x_{2}}{x_{1}}-\frac{1}{2x_{2}%
x_{3}}-\frac{x_{2}}{2x_{1}x_{3}}.
\end{equation}%
\begin{align}
K_{T1T1}(x_{i},y_{i})  =&\frac{1}{16}\frac{1}{D_{1}D_{2}}\left\{  \left(
\frac{1-2x_{1}}{D_{1}}+\frac{1-2x_{2}}{D_{2}}\right)  (x_{2}y_{1}+x_{1}%
y_{2})\right. 
\nonumber  \\
&\phantom{\frac{1}{4}\frac{1}{D_{1}D_{2}} 
\frac{1-2x_{1}}{D_{1}}}  \left.  -\frac{1}{y_{3}}\left(  2y_{1}x_{2}+2x_{1}y_{2}+x_{2}y_{3}%
+x_{1}y_{3}\right)  \right\}  .
\end{align}

\end{document}